\documentstyle[epsfig,psfig,twoside,cite,11pt]{article}
\def\Hab{\footnote{Habilitationsschrift im Sinne von \S~3 Abs.~2 der
Habilitationsordnung des Fachbereichs Naturwissenschaften I der Bergischen
Universit\"at--Gesamthochschule Wuppertal.}}
\begin{document}
\newcommand{\xx}{{\bf x}}
\newcommand{\yy}{{\bf y}}
\newcommand{\Ss}{{\bf s}}
\newcommand{\Eepsilon}{{\mbox{\boldmath $\epsilon$}}}
\newcommand{\Eeta}{{\mbox{\boldmath $\eta$}}}
\newcommand{\FF}{{\bf F}}
\newcommand{\ff}{{\bf f}}
\newcommand{\Gg}{{\bf g}}
\newcommand{\be}{\begin{equation}}
\newcommand{\ee}{\end{equation}}
\newcommand{\bes}{\begin{eqnarray}}
\newcommand{\ees}{\end{eqnarray}}
\newcommand{\NN}{\nonumber\\}
\newcommand{\av}[1]{\langle #1 \rangle}
\newcommand{\XXX}[1]{{\em (Material to be added: #1)}}
\def\BIB{\vfill\eject\begin{thebibliography}{100}\small}

\thispagestyle{empty}
\section*{}
\def\thefootnote{\fnsymbol{footnote}}
\noindent
{\Large Interdisciplinary application of nonlinear time series methods}%
\Hab
\\[1cm]

\noindent
{\large Thomas Schreiber}
\\
\noindent
{\em Physics Department, University of Wuppertal, D-42097 Wuppertal,
Germany}
\\[1cm]

\noindent
{\em Abstract:}

This paper reports on the application to field measurements of time series
methods developed on the basis of the theory of deterministic chaos. The major
difficulties are pointed out that arise when the data cannot be assumed to be
purely deterministic and the potential that remains in this situation is
discussed.  For signals with weakly nonlinear structure, the presence of
nonlinearity in a general sense has to be inferred statistically. The paper
reviews the relevant methods and discusses the implications for deterministic
modeling. Most field measurements yield nonstationary time series, which poses
a severe problem for their analysis. Recent progress in the detection and
understanding of nonstationarity is reported.  If a clear signature of
approximate determinism is found, the notions of phase space, attractors,
invariant manifolds etc.\ provide a convenient framework for time series
analysis. Although the results have to be interpreted with great care, superior
performance can be achieved for typical signal processing tasks. In particular,
prediction and filtering of signals are discussed, as well as the
classification of system states by means of time series recordings. 
\vfill
\def\thefootnote{\arabic{footnote}}
\setcounter{footnote}{0}
\eject

\renewcommand{\contentsname}{}
{\footnotesize\tableofcontents}
\eject

\section{Introduction} \label{sec:intro}
The most direct link between chaos theory and the real world is the analysis of
time series data in terms of nonlinear dynamics. Most of the
fundamental properties of nonlinear dynamical systems have by now been observed
in the laboratory. However, the usefulness of chaos theory in cases where the
system is not manifestly deterministic is much more controversial. In
particular, evidence for chaotic behaviour in field measurements has been
claimed --- and disputed --- in many areas of science, including biology,
physiology, and medicine; geo- and astrophysics, as well as the social sciences
and finance.

This article will take a critical look at the published literature, evaluating
the perspectives and the limitations of the approach. While common
misconceptions will be elucidated, I will try to adopt a constructive point of
view by highlighting those cases where information in fact has been gained by
the application of methods from chaos theory.

Along with the treatment of conceptual issues, instructive practical examples
from the literature and from work done in our group will be evaluated. I will
try to depict the state of the art of the application of chaos theory to real
time series. Neither naive enthusiasm to explain all kinds of unsolved time
series problems by nonlinear determinism is justified, nor is the pessimistic
view that no real system is ever sufficiently deterministic and thus out of
reach for analysis.  At least, chaos theory has inspired a new set of useful
time series tools and provides a new language to formulate time series problems
--- and to find their solutions.

Previous works of review character are Grassberger et al.~\cite{gss}, Abarbanel
et al.~\cite{abarbanel}, as well as Kugiumtzis et
al.~\cite{kugiumtzis_rev1,kugiumtzis_rev2}. Apart from a collection of research
articles by Ott et al.~\cite{coping}, two books on nonlinear time series from
the point of view of chaos theory are available so far, one by
Abarbanel~\cite{abarbook} and one by Kantz and Schreiber~\cite{ourbook}. While
in the former volume chaoticity is usually assumed --- as already reflected in
the title --- the latter book puts some emphasis on practical applications to
time series that are not manifestly found, nor simply assumed, to be
deterministic and chaotic. Apart from these works, a number of conference
proceedings volumes are devoted to chaotic time series, including
Refs.~\cite{Mayer-Kress,casdagli,SFI,dyndis,freital}. Nonlinear time
series methods that arise as extensions and generalisations of linear tools can
be mostly found in the statistical literature. Major references are the books
by Tong~\cite{tong} and by Priestley~\cite{priestley}.

The application of nonlinear time series methods to field measurements has been
accompanied by considerable controversy in the literature. Early enthusiasm has
led to straightforward attempts to find, and even quantify, deterministic chaos
in many types of systems, ranging from atmospheric
dynamics~\cite{nico,fraedrich,essex} and financial
markets~\cite{hsieh,peters,decoster} to heart~\cite{goldberger,chialvo} and
brain activity~\cite{bablo}. In Ref.~\cite{prit} it is even claimed that
cigarette smoking optimises the ``dimensional complexity'' of an indicator of
brain function. This wave of publications has been followed by a number of
critical papers pointing out the methodological deficiencies of the
former. Some of these will be cited below in their proper contexts. This review
is however not the place to repeat the known arguments or the discussion.  In
my opinion, part of the controversy and the resulting frustration is due to the
misconception that low-dimensional chaos is such an appealing theory that it
can be {\em expected} to be present generically in nature. Of course, most
researchers would deny that they have made such an a priori
assumption. Nevertheless, the amount of evidence we require for a ``climate
attractor'' etc. does depend on how likely, that is, how convincing, we find
such a concept. For example in medical research it is extremely tempting to
have a means of measuring the ``complexity'' of the cardiac rhythm or even the
brain function.

Now, if we {\em assume} chaoticity in the sense of low-dimensional determinism
as a starting point of our analysis, we can directly justify the use of delay
coordinates by Takens' theorem. The number of degrees of freedom in the system
is readily estimated as the embedding dimension where the number of false
neighbours drops below the noise floor, the rate of increase of uncertainty is
identified with the Lyapunov exponent and so forth. This rationale is pursued
for example in Ref.~\cite{abarbook}. An experience I share with many other
researchers is that this way to proceed is quite dangerous since the
possibilities for spurious results and wrong conclusions are overwhelming. As
an example, take the analysis of the Salt Lake area data in
Ref.~\cite{abarbook}. Taking for granted that nonlinear dynamics is at work,
locally linear phase space predictions seem most appropriate to forecast future
values. The predictions shown in Ref.~\cite{abarbook} seem fair enough but a
closer inspection of the data shows that already the simple linear rule to
follow the trend of the last two observations, $x_{n+1}=2x_n-x_{n-1}$, is more
appropriate than the local linear approach in that it gives forecast errors of
about half the rms magnitude.

A second possibility is to try to {\em establish} low-dimensional chaos by
positive evidence. We could for example look for self-similar geometry over a
reasonable range of length scales and demonstrate that uncertainties indeed
grow exponentially over a certain period of time. Eventually, we should be able
to extract empirical deterministic models that predict future values and that
can be iterated to yield time series with statistical properties comparable to
the data. This approach is preferred by most theoretical researchers and has
been emphasised for example in Ref.~\cite{ourbook}. The major drawback is that
only exceptional time series show such a clear signature, all of which are from
laboratory experiments set up specifically for the study of chaotic phenomena.
If this were the only way to go, applications to real world problems would have
to be largely abandoned.

Finally, we can move the focus of our study from the question of whether
deterministic chaos is really present to the question of whether deterministic
chaos provides a useful language for the evaluation of a given signal. The
concept that superior performance alone is a valid argument for the use of a
particular method is not as surprising for an engineer or clinician as it may
be for a physicist. In particular in time series analysis, very few people
actually believe that the stock market or the brain actually are linear
autoregressive machines. Nevertheless, linear time series methods have been
applied to time series from these systems with considerable success. Evidence
for the practical superiority of chaotic time series methods has so far been
rather scarce.

In the following I will at no point {\em assume} deterministic chaos.
However, I will first review in Section~\ref{sec:theory} the fundamentals of
dynamical systems theory as the theoretical basis for nonlinear time series
methods. In many cases this will only amount to a theoretical motivation; very
few facts are rigorously proven for finite, noisy time series. I will briefly
review the concepts of dynamical systems, strange attractors, phase space
embedding, and the invariant characteristics of a process.

Next I will try to give an understanding of the signatures of determinism in
finite observations.  Section~\ref{sec:fromdata} will discuss what happens to
the theoretical concepts when they are applied to real data of finite
resolution and length. Some limitations are known rigorously, others can be
understood heuristically. Some problems seem to be of purely technical nature
but nevertheless may prove to be serious in practice. In particular, extended
embedding theorems and amendments of the embedding procedure will be
discussed. Estimators for characteristic quantities like dimension, entropy,
and Lyapunov exponents are studied with respect to their practical viability.
This material will allow us to gauge for a given time series problem how far we
are from the linear case and how close we are to a nonlinear deterministic
situation. Accordingly, we will choose either linear methods, coarse but robust
nonlinear tools, or more refined phase space methods.

One formal requirement for almost all time series methods is stationarity.
Specific tests for nonstationarity in a nonlinear context that have been
proposed in the literature are discussed in Section~\ref{sec:statio}, where
also hints will be given what can be done in the presence of nonstationarity
apart from choosing a different time series problem. In
Section~\ref{sec:surro}, formal statistical tests for nonlinearity in a time
series will be set up, with particular emphasis on possible nonlinear
determinism in the data. The section will state what has to be done in order to
perform such a test correctly, but it will also discuss what can (and what
cannot) be learned from such a test.  While standard solutions for the
forecasting and filtering of linearly correlated but otherwise random sequences
exist, and methods for strongly deterministic but chaotic systems are also well
established, signals of a mixed character are more difficult to deal with.
Section~\ref{sec:process} will discuss the problems that arise in practice and
give some specific successful applications, including medical data analysis.
While most estimates of invariants based on short and noisy data are dubious as
absolute numbers, many authors have found the comparison of such numbers, or of
nonlinear qualitative features, across an ensemble of systems (e.g.  sick and
healthy patients) quite promising. As discussed in Section~\ref{sec:classify},
some of these works are of little use since the discrimination task could have
been solved equally well by standard methods, other examples seem to give
results that are by far superior to previous approaches.

Pure low-dimensional determinism is quite special and can be found in nature
only to a crude approximation. The range of potential practical applications of
nonlinear theory can only by increased significantly if the underlying
paradigm is generalised in some respects. Current efforts concerning the
analysis of data from extensively chaotic (e.g. spatio-temporal) systems, as
well as from mixed, nonlinear and stochastic, sources are discussed in
Section~\ref{sec:perspectives}.

In this paper, I will not give many technical details of the practical
implementation of the methods. These will be reviewed in a forthcoming
article~\cite{tisean} which is accompanied by the publicly available software
package TISEAN%
\footnote{The TISEAN software package is publicly available for download from 
   either
      {\tt http://\linebreak[1]%
      www.\linebreak[1]%
      mpipks-dresden.\linebreak[1]%
      mpg.\linebreak[1]%
      de/\linebreak[1]% 
      $\tilde{~}$tsa/\linebreak[1]%
      TISEAN/\linebreak[1]%
      docs/\linebreak[1]%
      welcome.html}
      or
      {\tt http://\linebreak[1]%
      wptu38.\linebreak[1]%
      physik.\linebreak[1]%
      uni-\linebreak[1]%
      wuppertal.\linebreak[1]%
      de/\linebreak[1]%
      Chaos/\linebreak[1]%
      DOCS/\linebreak[1]%
      welcome.html}. 
   The distribution includes an on-line documentation system.}
which contains many of the methods discussed here.

\section{Theoretical foundation} \label{sec:theory}
This section will briefly recall the definitions and properties of some
concepts of chaos theory, insofar as they are relevant to applied time series
problems. Most of the basic concepts are usually formulated in a purely
deterministic setting; that is, without any external noise. The time evolution
is then given by a dynamical system in phase space. Since usually the state
points cannot be observed directly but only through a measurement function,
typically involving a projection onto fewer variables than phase space
dimensions, we have to recover the missing information in some way.  This can
be done by time delay embeddings and related methods. We can then quantify
properties of the system through measurements made on the embedded time
series. Since it is eventually the underlying system we want to characterise,
these properties should ideally be unaffected by the measurement and the
embedding procedure. The presentation here serves the main purpose of fixing
the notation for the following section and is therefore extremely brief.
Theoretical issues that are directly related to time series methods are
discussed in more detail in the monographs Refs.~\cite{abarbook,ourbook}.
More general references on the theory of deterministic dynamical systems
include the volumes by Ott~\cite{Ott}, as well as the older books by Berg\'e,
Pomeau, and Vidal~\cite{Berge} and by Schuster~\cite{Schuster}. More advanced
material is contained in the work by Katok and
Hasselblatt~\cite{KatokHasselblatt}. A gentle introduction to dynamics
is given by Kaplan and Glass~\cite{KaplanGlass}. The volume by
Tsonis~\cite{tsonis} puts more emphasis on the applied side.

\subsection{Dynamical systems and predictability}
When we are trying to understand an irregular (which essentially here means
non-periodic) sequence of measurements, an immediate question is what kind of
process can generate such a series. In the deterministic picture, irregularity
can be autonomously generated by the nonlinearity of the intrinsic
dynamics. Let the possible states of a system be represented by points in a
finite dimensional phase space, say some ${\cal R}^d$. The transition from the
system's state $\xx (t_1)$ at time $t_1$ to its state at time $t_2$ is then
governed by a deterministic rule: $\xx(t_2) = T_{t_2-t_1}(\xx (t_1))$.  This
can be realised either in continuous time by a set of ordinary differential
equations:
\be\label{eq:ode}
   \dot\xx(t)=\FF(\xx(t))
\,,\ee
or in discrete time $t=n\Delta t$ by a map of ${\cal R}^d$ onto itself:
\be\label{eq:map}
   \xx_{n+1}=\ff(\xx_n)
\,.\ee
The family of transition rules $T_t$, or its realisation in the forms
(\ref{eq:ode}) or (\ref{eq:map}), are refered to as a {\em dynamical system}.
The particular choice of $\FF$ (resp. $\ff$) allows for many types of
solutions, ranging from fixed points and limit cycles to irregular behaviour.
If the dynamics is dissipative (area contracting, the case assumed throughout
this work), the points visited by the system after transient behaviour has died
out will be concentrated on a subset of Lebesgue measure zero of phase space.
This set is referred to as an {\em attractor}, the set of points that are
mapped onto it for $t\to\infty$ as its {\em basin of attraction}. Since not all
points on an attractor are visited with the same frequency, one defines a
measure $\mu(\xx)d\xx$, the average fraction of time a typical trajectory
spends in the phase space element $d\xx$. In an {\em ergodic} system,
$\mu(\xx)$ is the same for almost all initial conditions. Phase space averages
taken with respect to $\mu(\xx)d\xx$ are then equal to time averages taken over
a typical trajectory.

In real world systems, pure determinism is rather unlikely to be realised since
all systems somehow interact with their surroundings. Thus the deterministic
picture should be regarded only as a limiting case of a more general framework
involving fluctuations in the environment and in the system itself. However, it
is the limiting case that is best studied theoretically and that is expected to
show the clearest signatures in observations.

\begin{figure}
   \centerline{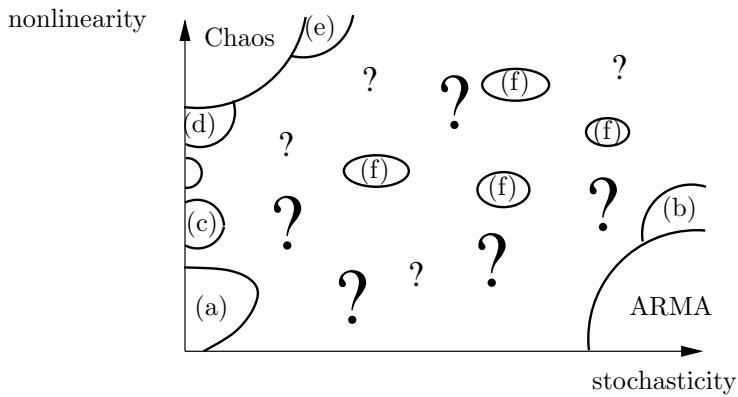} 
   \caption[]{\label{fig:world}Sketch of the
   variety of systems spanned by the properties ``nonlinearity'' and
   ``stochasticity''. Areas where theoretical knowledge and technology for the
   analysis of time series are available are outlined.  Besides the (possibly
   noisy) periodic oscillations (a), these are mainly the deterministic chaotic
   and the linear stochastic areas. Also the common routes to chaos (c) and
   (d), and extensions for small nonlinearity (b) or small noise (e) are
   marked. There are a few ``islands'' (f), like {\em hidden Markov models} and
   a few others, where a connection can be made between a nonlinear stochastic
   model approach and particular real world phenomenon.}
\end{figure}

The deterministic approach is not the most common way to explain irregularity
in a time series. The traditional answer given by the time series literature is
that external random influences may be acting on the system. The external
randomness explains the irregularity, while linear dynamical rules may be
sufficient to explain structure found in the sequence. The most general linear
(univariate) model is the {\em autoregressive moving average} (ARMA) process,
given by
\be\label{eq:arma}
   x_n=\sum_{i=1}^M a_ix_{n-i} + \sum_{i=0}^N b_i\eta_{n-i}
\,,\ee
where, $\{\eta_n\}$ are Gaussian uncorrelated random increments. The linear
stochastic description is attractive mainly because many rigorous results are
available, including the properties of finite sample estimators.

Most sources of irregular signals, take for example the brain or the
atmosphere, are known to be nonlinear. Nevertheless, if many weakly coupled
degrees of freedom are active, their evolution may be averaged to quantities
that are to a good approximation Gaussian random variables. If this
approximation is valid, it is also reasonable to assume that an observed degree
of freedom interacts with the averaged variable in a mean field way, justifying
the linear dynamics in Eq.(\ref{eq:arma}). However, there are many situations
where this approximation fails, for example if the degrees of freedom of a
system act in a coherent way, which can happen in nonlinear systems even when
the coupling among the degrees of freedom is weak.

The two paradigms, nonlinear deterministic and linear stochastic behaviour, are
the extreme positions in the area spanned by the properties ``nonlinearity''
and ``stochasticity''. They are singled out not because they are particular
realistic for most situations, but rather because of their paradigmatic role
and their solid mathematical background. Since the literature abounds with
premature conclusions like that a system that is not found to be linear must be
deterministic instead, let us emphasise that we are dealing with rather narrow
limiting cases by drawing a sketch of the world of time series models that
also contains all kinds of mixtures, see Fig.~\ref{fig:world}.

The description of a particular time series by an empirical model will of
course be guided by the paradigm adopted for the study. The idea is that one
cannot possibly do a better modeling job than recovering the equations that
actually govern the system under observation. One should note, however, that a
description of a system which includes external influences is only complete if
the input sequence (for example in Eq.(\ref{eq:arma}) the sequence of
increments $\{\eta_n\}$) is known.

One can ask what constitutes a complete description in the case of a
deterministic system. In the mathematical sense, the system equations
(\ref{eq:ode}) or (\ref{eq:map}) together with the initial conditions are
sufficient. For this to be true, both parts must be known with infinite
precision, which is unphysical. If the system is chaotic, then even in the
noise free case, errors in the initial condition and in the specification of
the model will grow in time and will have to be corrected. Of course, as soon
as noise is present, even if only in the measurement procedure, the situation
becomes worse. If we could recover $\FF$ (resp. $\ff$) from the observations
correctly, we still could not simply generate future values by applying $\ff$
to the observed present state because we could not account for the noise.
Thus we may face the confusing situation that the original equations of motion
do not necessarily give the best model in terms of predictions. (See
Refs.~\cite{lop2,jk} for illustrative examples.)

Since there is not much that can be done by way of modeling the noise in the
system, modeling deterministic systems will still attempt to fit a function
$\ff$ (or $\FF$) to the data such that Eq.(\ref{eq:map}) (or (\ref{eq:ode}),
resp.) will hold to the best available approximation. The most widespread
approach is to solve Eq.(\ref{eq:map}) in the least squares sense, minimising
\be
   \sum_{n=1}^{N-1} [\xx_{n+1}-\hat{\ff}(\xx_n)]^2
\,.\ee
Historical references on the prediction problem include the papers by Farmer
and Sidorowich~\cite{fsid0}, and by Casdagli~\cite{Casdagli_pred}. Advanced
procedures are available that take better care of the noise problem, see
Refs.~\cite{kost,jaeger}, as well as Sec.~\ref{sec:finite}. As an alternative
to explicitly fitting equations of motion to the data, it has been
proposed~\cite{brown_sync} to {\em synchronise} a model system with the
observed phenomenon (maybe given a posteriori by a time series) as a means of
identifying the correct model equations.  This approach is conceptually less
unequivocal since there exist examples where systems lock into a generalised
synchronised mode although the systems are quite different in structure.

\subsection{Phase space, embedding, Poincar\'e sections} \label{sec:the_embed}
An immediate consequence of the formulation of the dynamics in a vector space
is that when analysing time series, we will almost always have only incomplete
information. Although more and more multi-probe measurements are being carried
out, still the vast majority of time series taken outside the laboratory are
single-valued. But even if multiple simultaneous measurements are available,
they will not typically cover all the degrees of freedom of the system.
Fortunately, however, the missing information can be recovered from time
delayed copies of the available signal, if certain requirements are fulfilled.
The theoretical framework for this approach is set by a number of theorems,
all of which specify the precise conditions when an attractor in delay
coordinate space is equivalent to the original attractor of a dynamical system
in phase space.

Let $\{\xx(t)\}$ be a trajectory of a dynamical system in ${\cal R}^d$
and $\{s(t)=s(\xx(t))\}$ the result of a scalar measurement on it. Then a delay
reconstruction with {\em delay time} $\tau$ and {\em embedding dimension} $m$
is given by
\be\label{eq:delay}
   \Ss(t)=(s(t-(m-1)\tau),s(t-(m-2)\tau), \ldots, s(t))
\,.\ee
The celebrated delay embedding theorem by Takens~\cite{Takens} states that
among all delay maps of dimension $m=2d+1$, those that form an embedding of a
compact manifold with dimension $d$ are dense, provided that the measurement
function $s:{\cal R}^d\to {\cal R}$ is $C^2$ and that either the dynamics or
the measurement function is generic in the sense that it couples all degrees of
freedom.  In the original version by Takens, $d$ is the integer dimension of a
smooth manifold, the phase space containing the attractor. Thus $d$ can be much
larger than the attractor dimension.

Sauer, Yorke, and Casdagli~\cite{embed} were able to generalise the theorem to
what they call the {\em fractal delay embedding prevalence theorem}. It states
that under certain genericity conditions, the embedding property is already
given when $m>2d_{\rm box}$ where $d_{\rm box}$ is the {\em box counting
  dimension} of the attractor of the dynamical system. Further generalisations
in Ref.~\cite{embed} assert that, provided sufficiently many coordinates
are used, also more general schemes than simple delay embeddings are allowed.
For practical purposes, filtered and SVD embeddings have interesting
properties. Generalisations to periodically or stochastically driven systems
(where the driving force is assumed to be known) are heuristically
straightforward but the relevant genericity requirements are rather involved.
The proofs have been worked out by Stark and coworkers~\cite{stark}.
Depending on the application, a reconstruction of the state space up to
ambiguities on sets of measure zero may be tolerated. For example, for the
determination of the correlation dimension, events of measure zero can be
neglected and thus any embedding with a dimension larger than the (box
counting) attractor dimension is sufficient~\cite{saueryorke}.

Although the embedding theorems provide an important means of understanding the
reconstruction procedure, none of them is formally applicable in practice. The
reason is that they all deal with infinite, noise free trajectories of a
dynamical system.%
\footnote{A finite piece of trajectory is a one-dimensional curve which is
   generically and trivially embedable in three dimensions.}
It is not obvious that the theorems should be ``approximately
valid'' if the requirements are ``approximately fulfilled'', for example, if
the data sequence is long but finite and reasonably clean but not
noise free --- the best we can hope for in time series analysis. Some of these
issues will be discussed in Section~\ref{sec:finite}.

When analysing time continuous systems, Poincar\'e sections are an attractive
alternative to  the reconstruction with fixed delay times. Instead of the time
continuous trajectory, only its intersection points with a fixed {\em surface
of section} are regarded. Generically, the resulting set has a dimension which
is exactly one less than the attractor dimension. This concept is particularly
useful if the system is driven periodically and the surface of section can be
taken as the hyperplane defined by a fixed phase of the driving force. In this
case, in the intersection points are equally spaced in time. Reducing the
dimensionality of the problem can be an advantage but it comes at the price of
reducing the number of available points for a statistical analysis.

\subsection{Quantitative description} \label{sec:quant}
A time series is usually not a very compact representation of a time evolving
phenomenon. It is necessary to condense the information and find a
parametrisation that contains the features that are most relevant for the
underlying system. Most ways to quantitatively describe a time series are
derived from methods to describe an assumed underlying process.  Thus, for
example, measures of chaoticity in a time series are usually derived from
measures of chaoticity in a dynamical system. The rationale is that a certain
class of processes is assumed to have generated the time series and then the
measure quantifying that process is {\em estimated} from the data. Therefore it
is often necessary to distinguish between the abstract quantity, for example
the power spectrum of a stochastic process, and its estimate from a time
series, for example the periodogram.

Since the underlying process is only observed through some measurement
procedure, it is most useful to attempt to estimate quantities that are {\em
invariant} under reasonable changes in the measurement procedure. As will be
seen below (in Section~\ref{sec:invar}), the finite resolution and duration of
time series recordings damage the invariance properties of quantities which are
formally invariant for infinite data. If the value of an observable depends on
the observation procedure it looses its value as an absolute characteristic.
While in some cases we can still make approximate statements, the
interpretation of results has to be undertaken with great care. If we want to
compare the results between different experiments, at least we have to unify
the measurement procedure and the details of the analysis. If in a realistic
situation invariance has been given up anyway, the quantities discussed in the
following are no longer singled out that strongly among all possible ways of
turning a time series into a number. Consequently, there is no lack of ad hoc
definitions and characteristics that have been used in the literature. Since
they are invariably defined for time series, rather than the underlying
processes, some of them will be discussed later in Sec.~\ref{sec:fromdata}.

\subsubsection{Linear observables}
In the linear approach to time series analysis, a quantitative characterisation
of a process is done on the basis of either the two-point autocovariance
function or the power spectrum. If only a finite time series $\{s_n, n=1\ldots
N\}$ is available, the autocovariance function can be estimated e.g. by
\be\label{eq:autocor}
   C(\tau)={1\over N-\tau}\sum_{n=\tau+1}^{N} s_n s_{n-\tau}
\,.\ee
Depending on the circumstances, other estimators may be preferable.  A whole
branch of research is devoted to the proper estimation of the power spectrum
from a time series. The simplest estimator, known as the {\em periodogram}
$P_k$, is based on the Fourier transform of $\{s_n\}$,
\be\label{eq:fourier}
   S_k=\sum_{n=0}^{N-1}s_n e^{i2\pi kn/N}
\ee
through $P_k=|S_k^2|$. Issues of spectral estimation will not be discussed
here.  An introduction and pointers to the literature can be found for example
in {\em Numerical Recipes}~\cite{numrec}.

According to the Wiener-Khinchin theorem, the power spectrum of a process
equals the Fourier transform of its autocovariance function. For finite time
series this is only true if either $C(\tau)$ is computed on a periodically
continued version of $\{s_n\}$, or $P_k$ is computed on a version of $\{s_n\}$
that is extended to $n=-N,\ldots,N$ by padding with $N$ zeroes. Nevertheless,
both descriptions contain basically the same information, only that it is
presented in different forms. Furthermore, there is a direct connection between
the power spectrum and the coefficients of an ARMA model, Eq.(\ref{eq:arma}),
yielding a third possible representation.

The power spectrum of a process (and its autocovariance function) is unchanged
by the time evolution of the system (if it is stationary, see
Section~\ref{sec:statio} below). However, it is affected by smooth coordinate
changes, e.g. by the characteristics of a measurement device. Usually, the
non-invariance of the power spectrum is not a serious drawback. The power
spectrum is most useful for the study of oscillatory signals with sharp
frequency peaks. The location of these peaks {\em is} conserved, only their
relative magnitude may be affected by the change of coordinates.

Sharp peaks in the power spectrum indicate oscillatory behaviour and are useful
indicators in linear as well as in nonlinear signals. Broad band contributions,
however, have a less clear interpretation since they can be either due to
deterministic or stochastic irregularity. Therefore, the power spectrum is only
of limited use for the study of signals with possible nonlinear deterministic
structure.

\subsubsection{Lyapunov exponents}
The hallmark of deterministic chaos is the sensitive dependence of future
states on the initial conditions. An initial infinitesimal perturbation will
typically grow exponentially, the growth rate is called the Lyapunov exponent.
Let $\xx_{n_1}$ and $\xx_{n_2}$ be two points in state space with distance
$\|\xx_{n_1}-\xx_{n_2}\|=\delta_0\ll 1$. Denote by $\delta_{\Delta n}$ the
distance after a time $\Delta n$ between the two trajectories emerging from
these points, $\delta_{\Delta n}=\|\xx_{n_1+\Delta n}-\xx_{n_2+\Delta n}\|$.
Then the Lyapunov exponent $\lambda$ is determined by
\be\label{eq:lyapu}
   \delta_{\Delta n} \simeq \delta_0 e^{\lambda\Delta n},
   \quad \delta_{\Delta n}\ll 1, \quad \Delta n \gg 1
\,.\ee
A positive, finite, value of $\lambda$ means an exponential divergence of
nearby trajectories, which defines {\em chaos}. A mathematically more rigorous
definition will have to involve a first limit $\delta_0\to 0$ such that a
second limit $\Delta n\to \infty$ can be performed without saturation due to
the finite size of the attractor.

Here, only the single (maximal) Lyapunov exponent will be discussed.  Lyapunov
{\em spectra} can be defined that take into account the different growth rates
in different local directions of phase space. However, the non-leading
exponents are notoriously difficult to estimate from time series data. Only in
very few cases of clean laboratory time series trustworthy results have been
obtained so far.  (See~\cite{ourbook} for a discussion of the arising
problems.) For field data, Lyapunov spectra beyond the first exponent have not
so far been demonstrated to be a useful concept.

There have been a number of attempts to generalise the Lyapunov exponent to
systems which are not purely deterministic. For the usual definition, an
arbitrarily small amount of noise leads to a diffusive separation of initially
close trajectories and a divergent Lyapunov exponent (mind the order of the
two limits involved). For very small noise levels, there may still be a range
of length scales where the separation proceeds exponentially, until the finite
size saturation is reached. This is the behaviour that is probed by the real
space methods of estimating Lyapunov exponents from data, in particular the
two very similar algorithms introduced independently by Rosenstein et al.
\cite{rosenstein} and by Kantz~\cite{Holger}. From the theoretical point of
view, intermediate length scale definitions are less attractive since the
resulting quantities are no longer invariant under smooth coordinate
transformations.

An alternative way to introduce noise into the definition of Lyapunov exponents
is to study the separation of initially close trajectories of two identical
copies of a system which are evolving subject to the same noise
realisation. Then the Lyapunov exponent quantifies the contribution to the
divergence that originates in the intrinsic instability of the deterministic
part of the system. This is essentially the kind of instability probed by the
tangent space methods to obtain Lyapunov exponents from data, most prominently
Refs.~\cite{Eckmann,sano,Ruedi1,Ruedi2}.

Lyapunov exponents quantify the average exponential growth rate of
infinitesimal initial errors. Their natural units are therefore inverse times.
However, this does not justify to quote inverse Lyapunov exponents as average
predictability horizons or predictability times. (The two processes of
averaging and taking the inverse of a quantity do not commute.) In fact, the
degree of instability and predictability can vary considerably throughout phase
space, as it has been pointed out for example by Abarbanel, Brown, and
Kennel~\cite{abar_locallyap} and by Smith~\cite{lennylyap}. The Lyapunov
exponents constitute a particular way of averaging over these variations. They
are constructed in a way such that the average becomes independent of the
initial condition and invariant under smooth coordinate changes. For typical
prediction times, one has to form different averages which cannot be expected
to be invariant. It has been argued that the loss of information about the
system by averaging in a specific way over local variations of the instability
or predictability is too severe. Several
people~\cite{abar_locallyap2,bruno_locallyap} have therefore proposed concepts
of {\em local} Lyapunov exponents and predictabilities. Local Lyapunov
exponents are defined in a quite similar way as the usual exponents, except
that the limit $\Delta n\to \infty$ in omitted, whence they become position
dependent, or local. In particular, Bailey, Ellner, and Nychka~\cite{Ellner}
have studied the statistical properties of these exponents. They consider the
case that dynamical noise is perturbing the system. In that case they can prove
a central limit theorem about the existence and convergence of finite time
Lyapunov exponents. Unfortunately, local quantities are almost never invariant
in any useful sense. Quite trivially, they will change whenever the positions
are transformed. These changes may easily been kept track of. But as soon as
the coordinate changes are not isometries, the statistical weights of different
areas in phase space are changed. Thus the values, and distributions of values,
of local Lyapunov exponents and local predictability times are manifestly
non-invariant.

\subsubsection{Dimension and entropy}
Besides the exponential divergence of trajectories, the most striking feature
of chaotic dynamical systems is the irregular geometry of the sets in phase
space visited by the system state point in the course of time. This {\em
fractal} geometry is a natural consequence of the divergence of trajectories
which can be realised in a finite phase space only through some folding
mechanism. Stretching, folding, and volume contraction lead to statistically
self-similar structure on small length scales.

While the average stretching rate is quantified by the Lyapunov exponent, the
loss of {\em information} due to the folding is reflected by the {\em entropy}
of the process. The self-similar character of the resulting point sets and
measures defined on them can be characterised by fractal dimensions.  Several
definitions of non-integer dimensions have been proposed in the literature.
Most well known is the Hausdorff dimension of a set and the more easily
computable {\em box counting} (or {\em capacity}) dimension. We can also weight
the points in the set by the frequency with which they are visited on
average. Then we need a definition of the dimension in terms of the natural
measure $\mu(\xx)d\xx$ defined on the set.

One way to proceed is to take weighted averages of the number of points
contained in the elements of a partition of phase space and study their
dependence on the refinement of the partition. The translation of this scheme
into a time series context leads to the box-counting methods of dimension
estimation. The practical problems that arise when a space of moderate
dimensionality must be covered by boxes of small length can be overcome by
sophisticated bookkeeping algorithms. However, these methods make rather
inefficient use of the statistics available and suffer from severe finite size
effects on the larger length scales. They are therefore not recommended for the
study of invariant properties of real world time series.

An alternative way to define the dimension of a measure $\mu(\xx)d\xx$ is by
means of correlation integrals $C_q(\epsilon)$. Let us define the locally
averaged density $\rho_{\epsilon}$ to be the convolution of $\mu$ with a kernel
function $K_{\epsilon}(r)=K(r/\epsilon)$ of {\em bandwidth} $\epsilon$ that
falls off sufficiently fast for the convolution to exist:
\be\label{eq:cqrho}
   \rho_{\epsilon}(\xx)= \int_{\yy}d\yy\mu(\yy) K_{\epsilon}(\|\xx-\yy\|)
\,.\ee
Most commonly, the kernel is chosen to be $K_{\epsilon}(r) =
\Theta(1-r/\epsilon)$ where $\Theta(\cdot)$ is the Heaviside step function,
$\Theta(x)=0$ if $x\le 0$ and $\Theta(x)=1$ for $x>0$. Other kernels are
popular in statistical density estimation. The  correlation integral of order
$q$ is given by the order $q$ average of $\rho_{\epsilon}$:
\be\label{eq:cq}
   C_q(\epsilon) = \int_{\xx}d\xx\,\mu(\xx)\;
         [\rho_{\epsilon}(x)]^{q-1}
\,.\ee
For a self-similar measure we have
\be\label{eq:cqscale}
   C_q(\epsilon)\propto \epsilon^{(q-1)D_q}, \qquad \epsilon\to 0
\,.\ee
In the literature, $D_q$ is called the order-$q$ dimension. This definition
includes the dimension $D_0$ that has been shown to coincide with the Hausdorff
dimension in many cases, and the {\em information dimension} $D_1$ through
l'Hospital's rule.  Although $D_1$ is the most relevant because of its
information theoretic meaning --- it quantifies the scaling of the amount of
information needed to specify the state of the system with the required
accuracy --- we will usually at most be able to estimate a lower bound on it,
the {\em correlation dimension} $D_2$. The correlation dimension as a means of
quantifying the ``strangeness'' of an attractor has been introduced by
Grassberger and Procaccia~\cite{GP}. For finite samples, the double integral in
$C_2$ can be evaluated down to much smaller scales than the other
$C_q$'s. Although generic attractors are expected to be {\em multifractal},
that is, $D_q$ depends on $q$, this property is difficult to study in real time
series.  Only for exceptionally long, clean signals, $D_q$ can be obtained for
$q\neq 1,2$. For real world recordings it is already an ambitious goal to
establish a possible fractal nature by finding a scaling region of $C_2$.

When analysing time series we are usually dealing with distributions of delay
vectors with delay $\tau$ in an $m$-dimensional reconstructed phase space. The
$m$ dependence of $C_q$ in the limit of large $m$ can then be expressed as
\be\label{eq:cqscalem}
   C_q(m,\epsilon)=\alpha(m) e^{-(q-1)h_q\tau m} \epsilon^{(q-1)D_q},
      \qquad \epsilon\to 0, m\to\infty
\ee
which defines the order $q$ {\em entropy} $h_q$. The pre-factor $\alpha(m)$
depends on the norm $\|\cdot\|$ and the kernel function. Although $\alpha(m)$
does not affect the asymptotic value of the {\em entropy} $h_q$, the
convergence for finite $m$ can be dramatically different, as demonstrated in
Ref.~\cite{frank}. (See also Ref.~\cite{dimi_norm}.)  Again, the case $q=1$ is
singled out since the (Shannon, or Kolmogorov) entropy $h_1$ is additive when
independent processes are joined. Also, $h_1$ is related to the Lyapunov
exponents via Pesin's identity, a fact that can be used for consistency
checks. However, as with the dimensions, the case $q=2$ is much more accessible
with time series data. See for example Ref.~\cite{pt}. An algorithm for the
determination of the Kolmogorov entropy is given in Cohen and
Procaccia~\cite{CP}. 

Equivalent scaling behaviour to that of Eq.(\ref{eq:cqscalem}) is valid for a
large class of kernel functions in the average Eq.(\ref{eq:cqrho}), see
Refs.~\cite{ghez1,ghez2}. Besides the hard kernel given by the Heaviside
function, the most natural choice is a Gaussian, so that for example $C_2$
reads:
\be\label{eq:c2g}
   C_2^G(\epsilon) = \int\!\!\!\int_{\xx,\yy}d\xx\,d\yy\,\mu(\xx)\mu(\yy)\;
      e^{-{\|\xx-\yy\|^2 \over 4\epsilon^2}}
\,.\ee
Apart from yielding smoother curves for finite sample estimates, Gaussian
kernel correlation integrals have some other attractive properties. For
example, $\log C_2^G$ is additive under the pointwise summation of independent
variables, in particular, a deterministic signal and measurement noise. It is
not so much in use mainly since its numerical implementation seems quite
awkward when the definition, Eq.(\ref{eq:c2g}) is used
straightforwardly. However, it is quite easily obtained from the usual (step
kernel) correlation integral by%
\footnote{It can be easily seen that
   $$
      C_2^G(\epsilon) = \int_0^{\infty}d \tilde\epsilon \;
         e^{- {\tilde\epsilon^2\over 4\epsilon^2}} 
         {d\over d\tilde\epsilon} C_2(\tilde\epsilon)
   $$
   from which the result follows by partial integration.}
\be\label{eq:cg}
   C_2^G(\epsilon) = {1\over 2\epsilon^2}\int_0^{\infty}d \tilde\epsilon \;
      e^{- {\tilde\epsilon^2\over 4\epsilon^2}} h\, C_2(\tilde\epsilon)
\,.\ee

\subsection{Comparing dynamics and attractors}\label{sec:compare}
The quantities considered so far were all meant to characterise a single
process. Different processes can of course be compared by comparing these
numbers. It may however be interesting to have some means to answer the
question of how different two processes are directly, without going through the
reduction to a small number of characteristics. Recently, several authors
independently have begun to use relative measures for the classification of
systems through time series~\cite{EEG,clust} and for the study of
nonstationary signals~\cite{B1,casEEG,statio}. Therefore, some theoretical
background will be given for the less ad hoc measures of
dissimilarity. Practical aspects as well as some more informal but useful
quantities will be taken up in Sec.~\ref{sec:dissim}.

Let us first make a clear distinction between the problem of defining a
measure of dissimilarity between {\em attractors} (dynamics, measures,
probability distributions) and between {\em trajectories}. The latter is
related to the question if two systems are dynamically synchronised in a
general sense. {\em Generalised synchronisation} means that there is a smooth
mapping that relates the states of the two systems at any time. The present
work does not address the latter question. Relevant references include
Refs.~\cite{lou,koc1,koc2,rulkov,rosenblum}.

Information theory provides a measure of distance between two probability
densities $\mu(\xx)$ and $\nu(\xx)$ which is based on the Kullback
entropy~(Ref.~\cite{Kull}).  Let
\be
   H_K(\mu,\nu)=\int_{\xx}d\xx\,\nu(\xx) \log {\mu(\xx) \over\nu(\xx) }
\,.\ee
Then
\be\label{eq:Kull}
   \gamma_K(\mu,\nu) = H_K(\mu,\nu)+H_K(\nu,\mu)
        = \int_{\xx}d\xx\, (\mu(\xx)-\nu(\xx))(\log\mu(\xx)-\log\nu(\xx))
\ee
is positive definite, symmetric, and fulfills the triangle inequality. Thus
$\gamma_K(\cdot,\cdot)$ is a metric. For the same reasons discussed above,
expression Eq.(\ref{eq:Kull}) will be replaced by its analog based on second
order correlation integrals. A distance $\gamma_2(\mu,\nu)$ can then be defined
by:%
\footnote{Strictly speaking, this formula is only valid for smooth
   distributions $\mu,\nu$. For measured data we can safely assume that
   smoothness is imposed on fractal distributions by measurement errors.}
\be\label{eq:dist}
   \gamma_2(\mu,\nu)^2 = \int_{\xx}d\xx\, [\mu(\xx)-\nu(\xx)]^2
       = \lim_{\epsilon\to 0}\, [C_2(\epsilon; \nu) + C_2(\epsilon; \mu) -
       2C_2(\epsilon; \mu,\nu)]
\,\ee
where $C_2(\epsilon; \mu,\nu)$ is the {\em cross-correlation integral}
\be\label{eq:x2}
   C_2(\epsilon; \mu,\nu) = \int\!\!\!\int_{\xx,\yy}d\xx\,d\yy\,
      \mu(\xx)\nu(\yy)\; K_{\epsilon}(\|\xx-\yy\|)
\,.\ee
The case that $K_{\epsilon}(r)=\Theta(\epsilon-r)$ as the generalisation of the
Grassberger-Procaccia correlation integral (Ref.~\cite{GP}) has been introduced
by Kantz~\cite{holger}. The Gaussian kernel case
$K_{\epsilon}(r)=e^{-r^2/4\epsilon^2}$ together with its finite sample
properties has been studied by Diks and coworkers~\cite{diks_dist}.  If the
proper limit $\epsilon\to 0$ is taken, $\gamma_2(\mu,\nu)$ is nothing but the
$L^2$ distance of the two probability densities. For finite $\epsilon$,
$C_2(\epsilon; \nu) + C_2(\epsilon; \mu) - 2C_2(\epsilon; \mu,\nu)$ is no
longer formally a distance, except for the Gaussian kernel case.

One drawback of the second order distance $\gamma_2(\mu,\nu)$ as compared to
the Kullback distance $\gamma_K(\mu,\nu)$ is that it is no longer invariant
under smooth coordinate transformations acting on both distributions.  Other
measures of distance, in particular for discrete point sets, have been
discussed by Moeckel and Murray~\cite{moeckel}. One could further ask what
happens if two distributions are observed but they may have been obtained with
different measurement functions. This amounts to the question if the two
distributions are absolutely continuous with respect to each other. The hope
for a time series based answer seems unrealistic at this stage.

The quantities mentioned so far are all based on geometrical ideas. Apart from
asking for similar phase space geometry one can also ask for similar dynamical
evolution laws. If approximate predictive models can be established for the
time series, one can derive measures of dissimilarity that often allow stable
estimates for rather short sequences. Kadtke~\cite{Kadtke} uses
global models of the form
\be
    s_{n+1}=\ff(\Ss_n)=\sum_{i=1}^M a_i \ff_i(\Ss_n)
\ee
to fit several time series or segments individually. Then changes and
differences in the dynamics are monitored by changes in the model parameters
$a_i$. Technically, it is important to choose a model class with as few basis
function $\ff_i$ as possible. Otherwise the values of individual coefficients
$a_i$ may strongly depend on unimportant details of the signal. Two numerically
distinct sets of parameters may equally well model the same data. This approach
requires the dynamical models to involve adjustable parameters, excluding
locally constant or locally linear methods.

There are at least two other ways to compare the dynamics of two predictive
models $\ff$ and $\Gg$. One approach that has been taken by Hern\'andez and
coworkers~\cite{EEG} is to use both models to make predictions on a time
series $\{s_n\}$ and compare them for each time step $n$:
\be
   \gamma_P(\ff,\Gg;\{\Ss_n\})^2 =
      {1 \over N} \sum_{n=1}^N [\ff(\Ss_n)-\Gg(\Ss_n)]^2
\,.\ee
Since $\gamma_P(\ff,\Gg)$ is nothing but the $L^2$ distance of the vectors
formed by the individual predictions, it is a distance measure in the
mathematical sense. In general its value depends on the choice of time series
$\{\Ss_n\}$. If $\ff$ has been obtained by a fit to the signal
$\{\Ss^{(1)}_n\}$ and $\Gg$ by a fit to $\{\Ss^{(2)}_n\}$, a symmetric distance
measure between $\{\Ss^{(1)}_n\}$ and $\{\Ss^{(2)}_n\}$ is given by
\be
   \gamma_P(\ff,\Gg)^2 = \gamma_P(\ff,\Gg;\{\Ss^{(1)}_n\})^2
       + \gamma_P(\ff,\Gg;\{\Ss^{(2)}_n\})^2
\,.\ee

The {\em cross-prediction error} used in Refs.~\cite{statio,clust} is quite
similar, but it compares the individual predictions to the observed values
rather than to each other:
\be
   \gamma_C(\ff;\{\Ss_n\})^2 =
      {1 \over N-1} \sum_{n=1}^{N-1} [s_{n+1}-\ff(\Ss_n)]^2
\,.\ee
A small value of $\gamma_C(\ff;\{\Ss^{(2)}_n\})$ indicates that the dynamics
on $\{\Ss^{(2)}_n\}$ is a subset of the dynamics found on $\{\Ss^{(1)}_n\}$ and
modeled by $\ff$. A symmetric measure of dissimilarity (not a formal distance
measure in general) is given by
\be\label{eq:xpred}
   \gamma_C(\ff,\Gg)^2 = \gamma_C(\ff;\{\Ss^{(2)}_n\})^2
      + \gamma_C(\Gg;\{\Ss^{(1)}_n\})^2
\,.\ee
In the last two schemes, in principle any method of prediction can be used.  In
Refs.~\cite{statio,clust} stable results have been obtained with simple locally
constant phase space predictors. Prediction errors will be discussed in
Sec.~\ref{sec:noninvar}. Section~\ref{sec:dissim} will discuss a few practical
aspects of the above approaches.

\section{Nonlinear analysis of limited data} \label{sec:fromdata}
In the previous section, some definitions and theoretical motivation were given
for a number of concepts which now have to be adapted to the case that instead
of a measure, an attractor, a dynamical system, all we have is a finite, noisy
time series. The way to proceed crucially depends on the point of view we want
to assume about the nature of the system. As said earlier, we cannot {\em
assume} deterministic chaos for any measured time series. If we want to use the
theoretical results available, we need to {\em establish} it from the data,
maybe backed up by additional considerations. Often, we will not be able to
find low-dimensional structure, but we may still borrow some concepts just
because they give a convenient framework for certain problems.

\subsection{Embedding finite, noisy time series}\label{sec:embed_finite}
How much information can be recovered from time delayed copies of finite sets
of noisy measurements is quite a complicated question and a general answer is
not available. The embedding theorems mentioned previously all assume that
the observations are available with arbitrary precision. For some results, in
particular those concerning the attractor dimension, it is also assumed that
arbitrarily small length scales can be accessed which implies that an infinite
amount of data is available. A mathematical theorem cannot simply be expected
to be almost valid if the conditions are almost fulfilled. Consequently,
several authors have investigated what happens to the embedding procedure when
noise is present and the sequence is of finite length. For the embedding
procedure, noise seems to be the dominant limiting factor.

Only a few theoretical results relevant for practical work are available on
the embedding of noisy signals. First of all, we have to make a fundamental
distinction between noise due to measurement error and noise that is intrinsic
to the dynamics. In the first case, we suppose that there is a deterministic
dynamical system underlying the signal. Thus it is clear what we want to
reconstruct by the embedding procedure. If the noise is coupled to the system
we have to specify in what sense we want to use an embedding in the first
place.  Unfortunately, the nature of the noise is usually not known
independently. There is no general straightforward way to infer its properties
from a time series without making strong assumptions about the dynamical
system or the spectral properties of the noise.

One remarkable paper about the effect of measurement noise on the embedding
procedure is that by Casdagli and coworkers~\cite{reconstruct}. Their main
result is that a reconstruction technique that leads to a formally valid
embedding with noise free data can nevertheless amplify noise even in a
singular way. That means that in such a case not all degrees of freedom of the
system can be recovered from a scalar time series even for arbitrarily small
amounts of noise.  The examples studied in Ref.~\cite{reconstruct} suggest that
this situation is quite typical and not just found in constructed pathological
examples.  Thus bold interpretations of Takens' theorem, for example, that we
can recover the full dynamics of the human body from a recording of a single
variable, is not only in contradiction with common sense but also disproven by
mathematical arguments.

Some results are available on the embedding of noise driven signals. One line
of thought supposes that the driving noise sequence is known. The dynamical
system then becomes a nonlinear input-output device.
Casdagli~\cite{Casdagli-inout} and Stark et al.~\cite{stark} formalise the
idea that the observations of the output of such a system can be embedded in
the sense that time delayed copies of the observation sequence together with
the state of the input variable specify the state of the system equally well
as the full output together with the input state. These results are maybe more
useful for time series analysis than they may seem, given the fact that we
almost never know the noise sequence. There are certain signals where the
external influence can be inferred to some extent from the observed output.
Consider for example a recording of the cardiac cycle, for example an
electrocardiogram (ECG).  The cycle itself is fairly regular but the
initiation of a new cycle seems not to be fully determined by degrees of
freedom of the heart itself. But even if the beat times were random, we could
always infer a posteriori that triggering must have occurred once we observe a
new cycle. An illustration of this point will be given in an example below.

There are other theoretical works that also follow the idea that dynamical
noise can be isolated in certain cases where the observations contain
sufficient redundancy. Muldoon and coworkers~\cite{HeggusMuldoon} study the
case that more probes are available than necessary to cover the degrees of
freedom of the system. They demonstrate in a number of examples that 
sufficient redundancy in the  measurements allows for a distinction between the
detreministic part of the signal and the dynamical noise. This allows also to
recover missing variables by an embedding procedure.

\subsection{Practical aspects of embedding} \label{sec:embedp}
In most interdisciplinary applications we do not know much about the nature of
the noise. For example, biological systems are almost never isolated, and
measurements are always of finita accuracy. Observational noise is not always
white and Gaussian, although this is often the case.  If we make any assumption
about the statistical properties of the noise, we have to carefully check the
consistency of the results.

One of the most immediate restrictions of the embedding theorems for finite
data is that the information contained in a time delay representation of real
data is influenced by the choice of embedding parameters.  While the theorems
do not restrict the delay time $\tau$ (only a few exceptional cases must be
excluded), the proper choice of $\tau$ does matter for practical
applications. Also, there are many cases where the theoretically sufficient
embedding dimension $m$ is not optimal for a certain purpose. Larger (but
sometimes also smaller) values may give superior results. The literature on
this issue is quite confusing and at times contradictory.  Part of the
confusion is due to the fact that optimality can only be assessed with respect
to a specific application. When fitting the dynamics by a global polynomial
model, the embedding dimension should be as small as possible in order to limit
the number of coefficients in the model.  On the other hand, for local
projective noise reduction, the redundancy of an embedding with small $\tau$
and large $m$ allows for better noise averaging.  For signal classification we
do not even need a formal embedding since the difference between states may be
statistically better defined in a low-dimensional projection where small
neighbourhoods tend to contain more points.

The discussion will therefore be cut short by giving some pointers to the
literature and by proposing simply to carry out each study with several
embedding strategies and to compare the results. Theoretical work on the
embedability of noisy sequences is found in
Refs.~\cite{reconstruct,ding,malinetskij}. More heuristic studies are
Refs.~\cite{fraser,ls,lps,kennelisabelle,buzpfist,buzpfist2,kugi}.  One general
remark is that if one attempts to formally optimise the performance of an
embedding, one should use a large enough class of possible embeddings. There is
no theoretical reason to restrict the study to time delay embeddings with
equally weighted lags that are integer multiples of a common lag time. In fact
it was already reported in Ref.~\cite{gss} that minimising the redundancy in a
reconstruction $(s(t-\tau_{m-1}),s(t-\tau_{m-2}),\ldots,s(t))$ does not
necessarily yield $\tau_n=n\tau_1$ but for example
$\tau_1<\tau_2<2\tau_1$. More general functions of $s(t'), t-w\le t'\le t$ in a
time window of length $w$ may be considered.  The {\em singular value
decomposition} constitutes the special case of maximising the variance among
all linear combinations within a time window.  But neither is it necessary to
consider linear functions only, nor is the variance always the most interesting
characteristic.

Let us finish this section with a particular 
type of signal where a time delay embedding proves useful even though the
signal has a strong stochastic (or high-dimensional, in any case unpredictable)
component, the electrocardiogram (ECG). The ECG records the electro-chemical
activity of the heart which is essential for its pumping mechanism. The cardiac
muscle can be regarded as a spatially extended excitable medium with an
excitable, an excited, and a refractory phase. At the onset of a cardiac cycle,
a stimulus is initiated at the {\em sino-atrial (SA) node}, a specialised
collection of muscle cells. The resulting depolarisation wave proceeds along a
well defined pathway, first through the atria and then to the ventricles. The
excited tissue contracts and thereby ejects blood to the body and the lungs.
Eventually, all cardiac tissue has been excited and is refractory whence the
stimulus dies out. (If this condition fails, then the {\em re-entry} phenomenon
can occur which is the cause of serious arrhythmiae.) The pathway of the
depolarisation wave is quite similar from cycle to cycle, the variation over a
few cycles can be parametrised approximately by a one or two-parameter family
of ECG curves. However, the onset of a new cycle fluctuates from beat to beat
in a way that is not well understood. Certainly, the interbeat fluctuations
cannot be modeled successfully by a low-dimensional deterministic
approach. Coupling to the breath activity, blood pressure, as well as more
complex control signals from the central nervous system have to be taken into
account.  With this picture in mind, it could not be expected from the
embedding theorems that a delay coordinate representation of the ECG is useful
at all.

Let us consider a stochastically driven, damped harmonic oscillator as a toy 
model for such an input-output system with an unknown, fluctuating input 
sequence: 
\be\label{eq:toy}
   \ddot{x} + \dot{x} + x = a(t)
\,.\ee
The driving term is taken to be zero except for kicks of random strength at
times $t_i$ such that the inter-beat intervals, $t_i-t_{i-1}$ are random in the
interval $[p,q]$. Figure~\ref{fig:kirech} shows two trajectories of such a
system with different choices of the inter-beat time interval $[p,q]$. The
kicks are realised by finite jumps by a random amount in the interval
$[0,1]$. To the left, solutions of Eq.(\ref{eq:toy}) are plotted versus
time. In the middle, the true phase space spanned by $x(t)$ and $\dot{x}(t)$ is
shown while to the right a delay representation is used with a delay of one
time unit. In the upper row, beats are initiated with a time separation of
$p=0$ to $q=T/2$ where $T=4\pi/\sqrt{3}\approx 7.26$~time units is the period
of oscillation.  This does not allow the system to relax sufficiently between
kicks in order to form characteristic structure in phase space. Consequently an
embedding provides no clear picture and additional information would be needed
for an analysis of such a time series.  In the lower row, no kicks were closer
in time than $p=T$, the maximal separation being $q=3T$. The inter-beat parts
of the trajectory are distinct because of the different kick strength, but
since this is the only fluctuating parameter except for the inter-beat
interval, they are essentially restricted to a two-dimensional manifold. This
manifold is preserved under time delay embedding although neither the sequence
of beat times nor the beat amplitudes are used in any way. The randomness acts
locally around the origin in the indeterminacy as to when the next beat will
occur and how far the system will be taken by the kick.

\begin{figure}
\centerline{\hspace{1cm}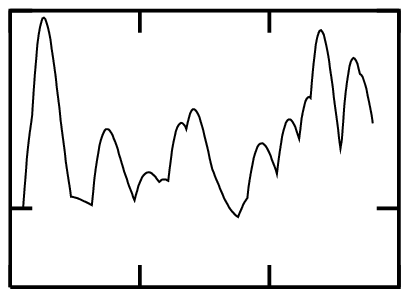\hspace{-0.4cm}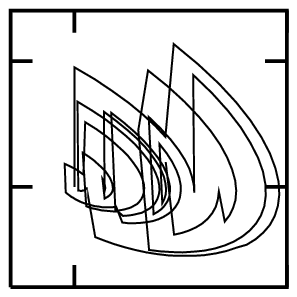\hspace{-2.7cm}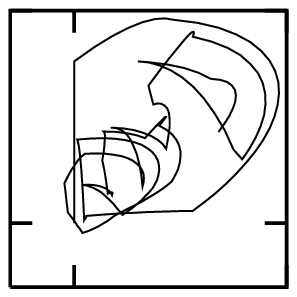}
\centerline{\hspace{1cm}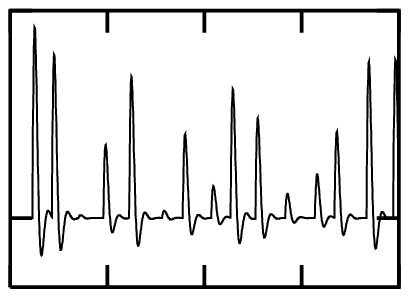\hspace{-0.4cm}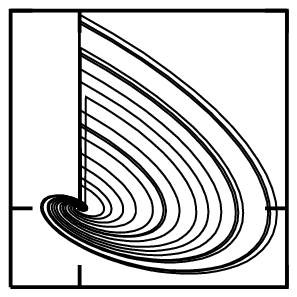\hspace{-2.7cm}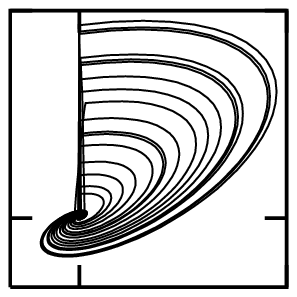}
\caption[]{Trajectories of a kicked, damped harmonic oscillator. Left: signal
   plotted versus time. Middle: true two-dimensional
   phase space. Right: delay embedding. Upper: kicks occur close in
   time. Lower: kicks are well separated in time and the system can relax
   between kicks. See text for discussion.\label{fig:kirech}}
\end{figure}

\begin{figure}
\centerline{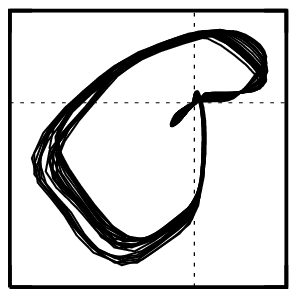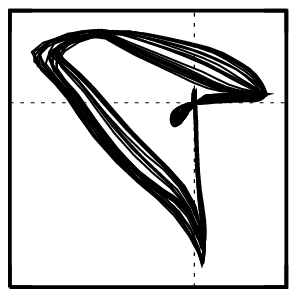}
\caption[]{Delay coordinate embeddings of a human electrocardiogram. The delay
   time is 12~ms resp. 24~ms at an (interpolated) sampling rate of 500~Hz. 
   Note that
   trajectories spend fluctuating stretches of time near the origin, where
   therefore an indeterminacy occurs. (The ECG voltages are in
   $\mu$V.)\label{fig:delay}}  
\end{figure}

Similarly to this toy example, most of the lacking information in the ECG, that
is, the times at which a new beat is triggered and possible other parameters of
the new cycle, can be deduced from the recording itself. Once the cycle is on
its way, we can find its origin quite easily. Thus, the redundancy in the ECG
trace explains why delay representations of ECGs are found to be approximately
confined to low-dimensional manifolds, see for example Fig.~\ref{fig:delay}. In
the left panel delay has been set to 12~ms in order to resolve best the large
spike (the QRS-complex) that corresponds to the depolarisation of the ventricle
(the large loop in Fig.~\ref{fig:delay}).  In the right panel, a 
longer delay of 24~ms has been used in order to unfold the smaller structures
around the origin which represent the atrial depolarisation (P-wave) and
ventricular re-polarisation (T-wave) phases.  After each beat, the signal
retires to the baseline, or the origin in delay space, where it may spend some
time until a new beat occurs. Since at that point the future is indeterminate,
the set visited by the trajectory is essentially finite dimensional, without
being formally deterministic.

\subsection{Estimating dynamics and predicting}\label{sec:finite}
Perhaps the most fundamental idea behind the approach to time series analysis
taken in this paper is that an irregular signal may have been generated
by a nonlinear dynamical system with only a few active degrees of
freedom. Therefore one of the most important goals should be to establish
effective equations of motion that follow this principle and are consistent
with the data. The ability to generate a time series that is equivalent to
the measured one can be taken as evidence for the validity of the approach, and
is therefore interesting in its own right. But there are many other situations
where effective model equations can be of great value.

Most properties of chaotic systems are much more easily determined from
equations than from a time series. Thus, if a time series can be well
represented by model equations, one might even abandon the analysis of the
series in favour of an analysis of the model. But this situation is rather
rare.  Except for well controlled laboratory experiments, dynamical modeling is
seldom faithful enough to justify such an approach. Nevertheless, analysing an
empirical model, and maybe synthetic time series data generated from it, can
provide a valuable consistency test for the results of time series analysis.
The best we can hope for when fitting a model to data is that the result comes
close to the real underlying dynamics. However, chaotic dynamical systems
generically show the phenomenon of {\em structural instability}. This means
that models with very similar parameters may exhibit qualitatively different
global dynamics, for example close to an attractor {\em crisis} (see for
example Ref.~\cite{Ott}). Therefore, if we simply iterate fitted model
equations, we may see substantially different behaviour from the actual system
even if the model in itself is faithful. One way to moderate this danger is to
introduce a small amount of dynamical noise comparable to the modeling error
when iterating the equations. Dynamical noise softens the sensitive dependence
on parameters to some extent. Alternatively, or additionally, one could study
{\em ensembles} of models which are compatible with the data. The ensemble
variation of statistical properties can then be taken as an indicator for the
expected effect of the remaining modeling error.

The most obvious reason to reconstruct model equations from a time series is
that one may be interested in predictions of future values. This task is of
course quite common in meteorology and finance and several other
fields. Moreover, in many situations the average error when predicting a time
series can be taken as an indicator of the structure present in the signal.
Thus, we will often use a nonlinear prediction error as a quantifier for the
comparison of signals. The use of prediction errors will be discussed in
Sec.~\ref{sec:noninvar}. A choice of models that can be employed will be
described below.

In the context of nonlinear dynamics, the modeling task consists in estimating
the function $\FF$ or $\ff$ that is supposed to generate the data via
Eqs.(\ref{eq:ode}) or (\ref{eq:map}) respectively. This may look like the
common problem of estimating a nonlinear function, and there is indeed a close
relation. However, the information available is not quite what one would like
to have for that purpose. All we usually have is a noisy scalar time series:
\be
    s_n=s(\xx_n)+\xi_n,\qquad \xx_n=\ff(\xx_{n-1})+\Eeta_n
\,.\ee
Here, also an intrinsic noise term $\Eeta$ has been included, since no real
system is ever really isolated. Since we cannot completely recover $\{\xx_n\}$
from $\{s_n\}$, the best we can do is to use some kind of embedding of
$\{s_n\}$ and look for a mapping $\ff_s$ that acts on the embedding vectors.
If the original phase space is $d$-dimensional this mapping may have to be
defined in up to $2d$ dimensions according to the theory of embeddings.
However, information about $\ff_s$ is only given through the data, that is, on
the attractor. This may render the estimation problem singular, depending on
the model class from which $\ff$ shall be estimated.

Even if the embedding problem can be solved (or avoided, if multiple
simultaneous measurements are available) the estimation problem remains
difficult. The standard approach would be to choose some parameter dependent
model for $\hat{\ff}_s$ and optimise the parameters using a maximum likelihood
or least squares procedure. This however implies that the value $y=\ff_s(\xx)$
is known at a number of locations, perhaps with some uncertainty. But for the
usual procedure to work, the locations $\xx$ have to be given without error.
This cannot be assumed in time series analysis because $\ff_s$ is sampled only
at the noisy data points. This and other practical problem in estimating
dynamics from a time series are discussed in Kostelich~\cite{kost}.  In
Refs.~\cite{lop2,Lop} illustrative material and a partial solution can be
found. The most thorough discussion, the one that also comes closest to a
satisfactory solution of the problem, is offered by Jaeger and
Kantz~\cite{jaeger}. The solution involves two ingredients. The first is to
replace the ordinary least squares procedure by a procedure that also optimises
the positions $\{\xx\}$ (sometimes called {\em total least squares}).
Unfortunately this renders the fitting problem nonlinear, even if the model
class is a linear combination of basis functions. The second part requires to
minimise the one-step error $(y-\hat{\ff}_s(\xx))^2$ but simultaneously
optimising the precision of $\ff_s(\xx)$, $\ff_s(\ff_s(\xx))$, etc, a difficult
nonlinear minimisation problem which is quite computer time intensive.

As for the model class from which $\hat{\ff}$ is to be determined, a number of
different propositions have been made. One possibility favoured by many
authors is to expand the dynamics in Taylor series locally in phase space.
This has first been proposed in the context of Lyapunov exponent estimation by
Eckmann and coworkers \cite{Eckmann}. They perform local linearisation on time
series data to obtain the dynamics in tangent space. In a classical paper
\cite{fsid0}, Farmer and Sidorowich port the idea to the prediction problem.
In practice, the expansion is carried out up to at most linear order.  Since
one has to work in several dimensions, the number of coefficients in higher
order approximations becomes too large for a local treatment.
In $m$-dimensional delay coordinates, the local model is then quite simply:
\be \label{eq:loclin}
    s_{n+\Delta n} = a_0^{(n)} + \sum_{j=1}^m a_j^{(n)} s_{n-(j-1)\tau}
\,,\ee
where $\Delta n$ is the time over which predictions are being made and $\tau$
is the time delay as usual. The coefficients $a_j^{(n)}, j=0,\ldots,m$ may be
determined by a least squares procedure, involving only points $\Ss_k$ within
a small neighbourhood around the reference point $\Ss_n$. Thus, the
coefficients will vary throughout phase space.  The fit procedure amounts to
solving $m+1$ linear equations for the $m+1$ unknowns.

When fitting the parameters $a$, several problems are encountered that seem
purely technical in the first place but are related to the nonlinear
properties of the system. If the system is low-dimensional, the data that can
be used for fitting will locally not span all the available dimensions but
only a subspace, typically. Therefore the linear system of equations to be
solved for the fit will be ill conditioned. However, we are only interested in
that part of the linear map (\ref{eq:loclin}) which relates points on the
attractor to their future. There are several ways to regularise the least
squares problem. In the presence of noise, the equations are not formally ill
conditioned but still the part of the solution that relates the noise
directions to the future point is meaningless (and uninteresting).
Equivalently to adding a small amount of noise one can add a small factor
times the unit matrix before the singular matrix is inverted.  The most
appealing approach however is to restrict the fitting procedure to the
directions spanned by the data which can locally be identified with {\em the
  principal components} or {\em singular vectors} of the data distribution.
These and a few other regularisation schemes for locally linear predictions
are discussed in great detail by Kugiumtzis and coworkers~\cite{kugiLL}. In
his contribution to the Santa Fe Institute time series contest in 1991,
Sauer~\cite{Sauer} has emphasised the close interplay between phase space
embedding and fitting of the dynamics.

The optimal degree of locality of a locally linear modeling approach has been
used by Casdagli~\cite{cas_stoch} as a measure for nonlinearity in a time
series. He compares the predictive quality of models fitted with using
different numbers of neighbours. In the absence of nonlinearity, the globally
linear fit using all available points as neighbours should give best results
since it uses the largest number of points and is structurally more robust.
For increasing degrees of nonlinearity, the tradeoff between lack of statistics
with few neighbours and curvature error with large neighbourhoods should move
the optimum towards smaller and smaller length scales. The rationale of this
paper is quite attractive -- to define the degree of nonlinearity by what is
the most useful assumption for modeling.

If one is interested in a robust, low variance measure of nonlinear
predictability without necessarily aiming at optimal forecasting power, one
should consider using locally constant approximations to the dynamics. The idea
is simply that determinism will cause similar present states to evolve into
similar future states. (This idea has been first used for predictions by
Lorenz~\cite{analog} who called it the method of {\em analogues}.)  Since we
expect the signal to be noisy, it is advantageous to consider a collection of
similar states rather than the single most similar state observed so far
(Lorenz' analogue). Thus, in order to make a prediction on the point $\Ss_n$,
we form a neighbourhood ${\cal U}_n$, either with a fixed radius or a fixed
number of elements. The prediction model is that $s_{n+\Delta n} = a^{(n)}$
where $a^{(n)}$ may vary throughout phase space. The fitting problem then
degenerates to finding the local (in phase space) average over the future
points of $\Ss_k, k \in {\cal U}_n$:
\be \label{eq:fclazy}
   s_{n+\Delta n} =
      \frac{1}{|{\cal U}_n|} \sum_{k\in {\cal U}_n} s_{k+\Delta n}
\,.\ee
Here, $|{\cal U}_n|$ denotes the number of points in the neighbourhood.  In
its simplest implementation, all that has to be set are the embedding
parameters and a length scale, the radius of phase space neighbourhoods.
Usually, none of these parameters need to be determined by a fit. Therefore
the locally constant model can be regarded as almost parameter free.  Very
similar algorithms has first been used by Pikovsky~\cite{arkady} and later by
Sugihara and May~\cite{SM}.  It is quite popular in the context of
nonlinearity testing~\cite{KI}, see also Sec.~\ref{sec:surro}. Extensions to
the locally constant or linear multivariate function interpolation procedure
by spline smoothing and adaptive parameter optimisation are implemented in the
MARS (multivariate adaptive regression splines) package which is described in
Friedman~\cite{MARS}.

An approach that is quite different from those mentioned so far attempts to fit
the dynamics by a nonlinear function that is globally defined in phase space.
The price for having a single expression for the model function is that it has
to accommodate the nonlinear structure appropriately. The most straightforward
generalisation of the linear autoregressive models is to include higher order
polynomial terms in the ansatz for $\hat{\ff}_s$. This is sometimes called a
Volterra series expansion. Another popular model class are {\em radial basis
functions}. Their use is documented by a large body of papers, for example
Powell~\cite{powell}, Broomhead and Lowe~\cite{rbf},
Casdagli~\cite{Casdagli_pred}, and Smith~\cite{lenny_rbf}.  As usual, the model
is given by a linear combination of orthogonal functions $\Phi_i(\Ss)$.  Here,
the functions are essentially of the same form, radially symmetric about some
centre point $\Ss_i^{(c)}$: $\Phi_i(\Ss)=\Phi(\|\Ss-\Ss_i^{(c)}\|)$.  This
yields
\be \label{eq:rbf}
   \hat{\ff}_s(\Ss)=a_0 + \sum_{i=1}^M a_i \Phi(\|\Ss-\Ss_i^{(c)}\|)
\,.\ee
Quite a variety of functional forms of $\Phi(r)$ can be used, including
Gaussians or powers. For the art of choosing the centre points $\Ss_i^{(c)}$
(which do not have to be points from the data set), reference is made to the
literature --- individual experimentation is also recommend.

After giving examples of model classes which are most popular in the nonlinear
time series community, it should be stressed again that multivariate function
estimation is a common problem in statistics and has a rich literature.
Neural networks have been very fashionable over the past few years. They have
proven to have remarkable capabilities, but, as the huge literature indicates,
they need a lot of expertise and experience to be used reliably. Another branch
of research pursues the idea of a regression tree in order to organise
multidimensional structure for prediction, see Breiman~\cite{brei} for a
classical reference.

If any parameter dependent model is used, for example one like
Eq.(\ref{eq:rbf}), or a neural network, care has to be taken to avoid
overfitting. Overfitting means that a larger model class can always increase
the accuracy of a fit, even though at some point only the details of the
particular realisation of the process that is available for fitting are
accommodated. This problem can be avoided by limiting the number of adjustable
coefficients in the model.  There are at least two ways to look at the
problem. Both result in a penalty for the number of coefficients in the cost
function that is to be minimised, but the exact form of this penalty differs
for the two approaches. Akaike~\cite{akaike} observes that if we are planning
to use the model for making predictions, then what we want to minimise is not
the least squares error of the model {\em on the data} but the {\em
expectation} of that error for the case that the model is applied to new data
from the same source.  This can mean that a model obtained from such a fit may
outperform the original equations when prediction errors are compared. The
second approach is due to Rissanen~\cite{rissanen}. The idea here is that
modeling provides a way to represent a data set in a more compact way than
storing the measured data. If the data can be reproduced by a simple model, one
can store the model and the errors instead, which is enough to recover all the
information. In such a context, fitting is a tradeoff between reducing errors
and increasing the size of the model.  Practically usable formulas have only
been derived in this context for linear (AR) models. One of the reasons is that
it is difficult to compare the importance of parameters across different model
classes.

In order to avoid overfitting in practical problems, one has to validate the
predictive power of a model on yet unused data. Simply splitting the available
data into two parts, one for fitting and one for testing, is the cleanest
possibility, but unfortunately is quite wasteful in terms of data
usage. Alternatively, one can use a cross-validation technique, as they are
common in the statistical literature. For $k$-fold cross-validation, the data
set is split into two segments in $k$ different ways and the fitting and
testing is repeated $k$ times. At each time, a different part of the data is 
used as a test set and the remaining points are used for fitting. The errors 
of the $k$ tests may then be averaged together. (For the correct averaging, in
particular if the different test sets overlap, consult the statistical
literature.) The advantage is that testing is done
on all available points, making best use of the data base. At the same time,
each fit is based on a large part of the time series. If the expense
in computer time that is necessary to repeat the fit $k$ times is feasible,
$N$-fold cross-validation can be used on $N$ data points, thereby optimising
the available statistics for the fits. This case is sometimes called {\em
take-one-out} statistics for obvious reasons.  Strictly speaking, $k$-fold
cross-validation assumes that there are no serial dependencies between data in
the different segments, which may or may not be true. If in doubt, on can
ensure that the training section and the test section are sufficiently far
apart in time. With this restriction, locally constant and locally linear
predictors provide take-one-out out-of-sample errors automatically since the
current point has to be excluded from the neighbourhoods. In the presence of
serial correlations, one should also exclude temporally close points that may
still be dynamically related.

It should be stressed that cross-validation is a technique for model
verification and not for model optimisation. If several models are proposed and
their out-of-sample errors are compared, the error quoted for the best of these
models can no longer be regarded as an out-of-sample statistic since it is
obtained by optimisation over a known training set.

Sec.~\ref{sec:noninvar} will discuss a few issues that arise when the
error with respect to a nonlinear prediction scheme is to be used as a
quantifier for the predictability in a system, or a measure of
``complexity''. If such a quantifier is only used in a relative way to compare
different signals, it is not formally necessary to use out-of-sample errors.
In fact, unless cross-validation at high $k$ is carried out, in-sample errors
usually have lower sample variance and may therefore give better discriminative
power. For the case of nonlinearity testing, this has been discussed by Theiler
and Prichard~\cite{fields}.

\subsection{Estimating invariants} \label{sec:invar}
All quantitative indicators of chaos involve in their definitions some kind of
limit. If these indicators are to be estimated from a finite time series
measurement, none of these limits can actually be carried out. Formally, the
desirable theoretical properties of these indicators, in particular their
invariance under smooth coordinate transformations, will be lost. If the
indicator is to be computed for the purpose of a comparative study, lack of
invariance may be compensated for by standardising the characterisation
procedure. Nonlinear indicators which are not necessarily invariant but are
optimised for their power to discriminate between different dynamical states
are discussed in Sec.~\ref{sec:noninvar} below.

With the typical data quality in non-laboratory experiments, and given that
pure low-dimensional determinism is quite a particular phenomenon in nature, we
will very seldom be able to reliably estimate the proper dimension or Lyapunov
exponent of a real world phenomenon. The issues we have to consider in such an
attempt will be discussed below -- most of them are well covered by the
literature, for example the book by Kantz and Schreiber~\cite{ourbook}. One of
the sharpest critics of naive use of the Grassberger-Procaccia correlation
dimension describes the situation roughly like that: There is little use in
computing the correlation dimension; If it is less than three one does not need
it because the structure of the attractor is obvious, if it is larger than
three it cannot be estimated reliably.%
\footnote{P. Grassberger, private communication.}
Very similar statements can be made for the other invariants from chaos theory
as well. Later, the discussion will proceed to quantities that have less
theoretical value but are easier to compute or give statistically more powerful
results.

\subsubsection{Lyapunov exponents}
Lyapunov exponents measure the rate of divergence of initially close
trajectories. A positive but finite Lyapunov exponent is therefore a sharp
criterion for the existence of deterministic chaos. The older literature on
the determination of exponents from time series can be seen as extensions of
techniques that have been developed for the analysis of systems with known
evolution equations. From these they inherit the assumption that there
actually exist such dynamical equations and trajectory separation evolves
indeed exponentially. Sano and Sawada~\cite{sano} as well as
Eckmann and coworkers~\cite{Eckmann} introduce locally
linear fits to the dynamics in order to follow the evolution in tangent space.
The algorithm by Wolf et al.~\cite{Wolf} follows several nearby trajectories to
measure the average increase of local volume. Many refinements of these
methods have been proposed, see Ref.~\cite{geist} for a comparative discussion
of all but the most recent Lyapunov algorithms. It is not wise to use these
algorithms when it cannot be taken for granted that the dynamics is
deterministic since none of them actually verifies the exponential behaviour
of trajectories.

More recently, the emphasis has shifted from the estimation of exponents under
the assumption of determinism to the verification of exponential growth of
errors. Very similar algorithms for this purpose have been proposed
independently by Rosenstein et al.~\cite{rosenstein} and by
Kantz~\cite{Holger}. We follow the latter reference here. The key idea is that
initially close trajectories do not necessarily diverge exactly exponentially,
but only on average. In order to cancel fluctuations around the general
exponential growth, one has to average appropriately over many trajectory
segments. Let $\cal W$ denote a set of delay reconstructed points $\Ss_n$
selected at random from a long trajectory such that they approximate the true
probability distribution. Let $|\cal W|$ denote the number of members in $\cal
W$.  The set of points in an $\epsilon$-neighbourhood of $\Ss_n$ is denoted by
${\cal U}_n$. Now define
\be \label{eq:lyap}
   S(\Delta n) = {1 \over |{\cal W}|} \sum_{n\in {\cal W}}
       \ln \left( {1 \over |{\cal U}_n|} \sum_{k\in {\cal U}_n}
           | s_{n+\Delta_n} - s_{k+\Delta_n} | \right)
\,.\ee
If the distances $|s_{n+\Delta_n} - s_{k+\Delta_n}|$ grow like $e^{\lambda
\Delta n}$, then so does $\exp S(\Delta n)$, but with less
fluctuations. Kantz~\cite{Holger} discusses why this is the correct (that is,
unbiased) way to average. A plot of $S(\Delta n)$ versus $\Delta n$ must show a
reasonably straight line over a range of length scales before we accept its
slope as an estimate of the Lyapunov exponent $\lambda$. In order to find such
a scaling range one has to choose the radius $\epsilon$ of the neighbourhoods
$\cal U$ as small as possible, but not so small that too few neighbours are
found or that distances are dominated by noise.

Examples for the successful use of this approach for computer generated
sequences and for time series from low-dimensional systems in laboratory
experiments have been given in the original articles by Rosenstein et
al.~\cite{rosenstein} and by Kantz~\cite{Holger}, as well as in
Ref.~\cite{ourbook}. In many time series from field measurements, initially
close trajectories are found to diverge rapidly. Algorithms that {\em assume}
that this divergence is due to an intrinsic instability of the dynamics will
then issue a positive Lyapunov exponent. However, an intrinsic instability of a
chaotic dynamical system should result in {\em exponential} growth of the
discrepancies, which is difficult to establish. Regard, for example, the
divergence rate plot shown in Fig.~\ref{fig:rrlyap}. For a sequence of 2000
time intervals between heartbeats in a normal human, the function $S(\Delta n)$
defined in Eq.(\ref{eq:lyap}) has been computed using unit delay, initial
neighbourhoods ${\cal U}$ of diameter 0.015~s around each point were formed in
2-10 dimensions. Indeed, the trajectories diverge quite fast. However, they
reach a saturation soon. The curves are not fitted well by straight lines which
would be the case if the divergence was exponential. If one were to assign a
slope to the lines, the result would strongly depend on the length scale,
embedding dimension etc., and would be quite useless as an estimator of the
Lyapunov exponent. It should, however, be remarked that the growth is not simply
diffusive either.  A non-invariant parameter, like the time of growth from
$S(0)$ to twice its value for given fixed embedding and neighbourhood
parameters may well be useful for the comparison of different subjects but may
not have much relation to a possible intrinsic instability of the system.
\begin{figure}
   \centerline{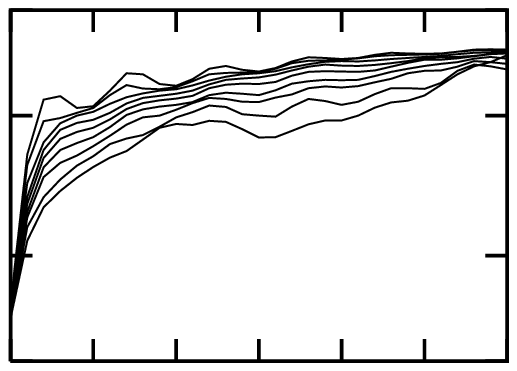}
   \caption[]{\label{fig:rrlyap}Divergence of initially close trajectories for
      a series of time intervals between heartbeats in a normal human (semilog
      scale). $S(\Delta n)$ is shown for $m=2,\ldots,10$. Trajectories do
      separate, but no straight line indicating exponential growth can be
      established. }
\end{figure}

\subsubsection{Correlation dimension} \label{sec:c2data}
The problems that arise when correlation integrals and the correlation
dimension are estimated from finite time series have been discussed extensively
in the literature. Statistical estimators for fractal dimensions and their
theoretical properties are studied in
Refs.~\cite{dimerr,smith,takens_est,cutler1,cutler2}. Original contributions
pointing out potential sources for spurious results are found for example in
Refs.~%
\cite{oneoverf,oneoverft,grass_nature1,grass_nature2,ruelle_fiction}. Some of
the material has been reviewed for example in
Refs.~\cite{gss,theiler_dim,dim,ourbook}. I will therefore only briefly state
the main points.

If the probability distribution implied by the natural measure is approximated
by a sum of delta functions at $N$ points $\{\xx\}$ independently drawn from
it, we can estimate the correlation integral $C_2$ by the {\em correlation sum}
\be\label{eq:c2}
   \hat{C}_2(\epsilon) =
      {2\over N(N-1)} \sum_{i=1}^N\sum_{j=i+1}^N
      \Theta(\epsilon-\|\xx_i-\xx_j\|)
\,.\ee
Thus, $\hat{C}_2(\epsilon)$ is just the fraction of all pairs of points that
are closer than $\epsilon$. It has been shown for example by
Grassberger~\cite{grass_finite} that $\hat{C}_2$ is an unbiased estimator of
$C_2$.  The hat will henceforth be suppressed. In time series applications, the
assumption that the $\{\xx\}$ are independently drawn from the underlying
distribution is usually violated due to serial, also called temporal
correlations.

It has been pointed out by several authors that serial correlations can lead to
spurious results for the correlation dimension, see for example
Refs.~\cite{theiler_dim,oneoverf,oneoverft,gss}. The necessary correction is
also well known (it has been proposed by Grassberger~\cite{grass_nature2} and
by Theiler~\cite{theiler_dim}): pairs of points $i,j$ that are closer than some
correlation time $t_{\rm min}$ have to be excluded from the double sum in
Eq.(\ref{eq:c2}). The loss of statistics is not dramatic since the total of
pairs grows like $N^2$ while only a number of terms $\propto N$ is
suppressed. Therefore it is advisable to be generous when choosing $t_{\rm
min}$, the time scale given by the decay of the linear autocorrelation function
is often not sufficient. A useful tool to determine the decay of nonlinear 
correlations is the {\em space-time separation plot} introduced by Provenzale
et al.~\cite{provenzale}, see Sec.~\ref{sec:statio}, Eq.(\ref{eq:stp}) below.
The effect of failure to exclude serially correlated pairs from the correlation
sum can be seen by comparing Fig.~\ref{fig:eeg} of the present section and 
Fig.~\ref{fig:eegip} of Sec.~\ref{sec:statio}.

It should be remarked that the literature might give the wrong impression that
the sensitivity to serial correlations is a flaw specific to the
Grassberger-Procaccia dimension algorithm. In fact, linear correlations and
nonlinear determinism are sources of predictability which are detected by any
algorithm that does not explicitly exclude the structure imposed by one of
these sources. This is the reason why also for example prediction errors or
false nearest neighbours techniques have to be augmented by a comparison to
linearly correlated random surrogates in tests for nonlinearity, unless
similar corrections are carried out as for the correlation dimension.  For the
false nearest neighbours approach~\cite{FNN} this has been pointed out for
example in Ref.~\cite{kennel-unpub}.

If one wants to estimate a correlation dimension, plotting $C_2(\epsilon)$ in a
log-log plot is not always the best thing to do since deviations from the
desired power law scaling do not appear very pronounced in this
representation. A better method might be to plot the local slopes of the
log-log plot,
\be\label{eq:d2}
   D_2(\epsilon) = {d \log C_2(\epsilon) \over d \log \epsilon}
\,,\ee
versus $\log\epsilon$. These slopes can for example be obtained by a straight
line fit over a small range of values of $\epsilon$.

Theiler~\cite{takens_theiler} gives a maximum likelihood estimator of the
Grassberger--Procaccia correlation dimension.  Since maximum likelihood
estimation of the correlation dimension goes back to Takens~\cite{takens_est},
such quantities are often referred to as {\em Takens' estimator}. The estimator
is given by
\be\label{eq:d2t}
   D_{\mbox{ML}}(\epsilon)=
     {C_2(\epsilon) \over
     \int_0^{\epsilon}{C_2(\epsilon')\over\epsilon'}d\epsilon'}
\,.\ee
This quantity can also be plotted against $\log\epsilon$ for different values
of the embedding dimensions. Its advantage over $D_2(\epsilon)$ is that it
incorporates all the statistical information that is available below the length
scale $\epsilon$. It is however implied that the dominating contamination at
the small length scale is given by the effect of the finite size of the data
set. In many practical situations this is not quite the case since measurement
errors destroy the self-similarity as well. The effect of noise on the
correlation integral has been studied in a number of
papers~\cite{smith,diks,kugig,olt,level} which are reviewed and compared
in~Ref.~\cite{comment}. Olofsen and coworkers~\cite{olof} as well as Schouten
and coworkers~\cite{schouten} derive maximum likelihood estimators of the
correlation dimension for data which are contaminated with noise. However, the
noise amplitude enters the analysis as an unknown parameter which complicates
the application of their results in practical situations.

\begin{figure}
\centerline{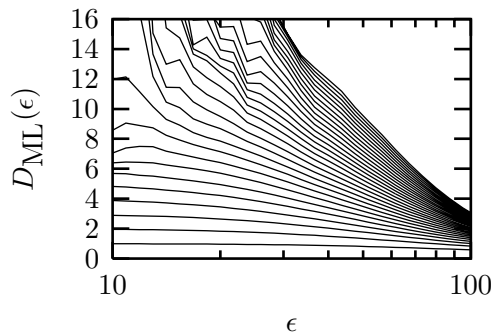}
\caption[]{Maximum likelihood estimator of the correlation dimension as a
   function of the cutoff length scale and the embedding dimension for an
   intracranial recording~\cite{elger} of the neural electric potential in a
   human.  No scaling region of approximately constant $D_{\mbox{ML}}$ can be
   found.  The time series was provided by Lehnertz and
   coworkers.\label{fig:eeg}}
\end{figure}

If all precautions are taken and a dimension estimate is attempted on a complex
data set, one should not be surprised if the result is negative in the sense
that no scaling region and no proper saturation can be found. In fact, few real
systems are sufficiently low-dimensional for this kind of analysis.  As an
example that illustrates this statement but which also shows that estimates of
the correlation {\em sum} can be useful nevertheless, let us study an
intracranial recording of the electric field in the brain of a human.  The data
and its analysis is described thoroughly in Elger and Lehnertz~\cite{elger}.
The data has been taken in an epilepsy patient and one of the questions is
whether individual seizures can be anticipated from these recordings a few
minutes before they occur. I will report on this application in more detail in
Sec.~\ref{sec:EEG}. Huge amounts of data are available since multiple channel
recordings are taken at 173~Hz continuously over several days for pre-surgical
screening purposes. Not surprisingly, however, the data are quite nonstationary
which is in fact essential for seizure anticipation to be possible. If one
wants to assign an attractor dimension to such a time series, one first has to
select windows in time which are short enough so that the dynamics can be
considered to be effectively stationary within each window. A reasonable
tradeoff between approximate stationarity and time series length is obtained
with windows of 30~s duration. Figure~\ref{fig:eeg} shows the result of an
attempt to calculate the correlation dimension of such a segment. After
low-pass filtering (40~Hz cutoff), the correlation sum $C_2(\epsilon)$ has been
computed with a delay of 5 sampling time units and embedding dimensions 1
to~30.  Dynamical correlations have been excluded by setting the minimal
temporal separation of neighbours to 50 samples. From this data, the maximum
likelihood estimator of the correlation dimension, Eq.(\ref{eq:d2t}), is
obtained as a function of the upper cutoff length scale $\epsilon$ and plotted
versus $\log\epsilon$.  Clearly, there are no scaling regions where
$D_{\mbox{ML}}$ becomes independent of $\epsilon$ and $m$. Thus it may be
concluded that a low-dimensional attractor is not a good model for this data
set. However, it will discussed in Sec.~\ref{sec:EEG} below how the correlation
sum, although not suitable for a dimension estimate, could be used for
monitoring changes in the brain state between different segments of a long
recording.

\subsection{Non-invariant characterisation}\label{sec:noninvar}
So far, invariance of an observable has been emphasized as a requirement for
the objective characterisation of time series data. However, we have also seen
that estimation of truly invariant quantities is an ambitious goal that is
worth pursuing only with sufficiently high data quality and for systems from
the appropriate class. This does not imply that we have to give up a
quantitative description in all other cases. An absolute, portable
characterisation is not always indispensable. Non-invariant characterisation of
time series data can for example be useful for the comparative study of
multiple data sets.  Of course, if we want to use an observable for comparisons
that is not invariant under changes in the measurement procedure, we have to
standardize the measurement procedure. Practical issues concerning the
comparison of time series are discussed in Sec.~\ref{sec:classify}. Let us here
only discuss some nonlinear time series measures which are not invariant under
coordinate changes but which have been used in the literature for various
reasons, for example because of their robustness to noise and to small
sample sizes.

Both noise and limited numbers of data points constitute severe problems
whenever properties at small length scales in phase space have to be probed.
In the presence of noise, length scales below the noise amplitude cannot be
accessed without an explicit noise reduction step. Also, on small length scales
the discrepancy between the finite collection of points and an underlying
probability distribution becomes most pronounced. The obvious way to get away
from these problems is to use coarse-grained quantities which are defined on
intermediate length scales. While giving up invariance, the statistical
properties of these quantities are often favourable. The most extreme step in
this direction is to encode the time series as a symbol string and then analyse
this sequence of discrete values. This approach is often called {\em symbolic
dynamics}, although in the dynamical systems literature this name is reserved
to a symbolic description that results from an encoding according to a {\em
generating partition}. (See for example Bowen~\cite{bowen}, Christiansen and
Politi~\cite{freddy}, or Grassberger and Kantz~\cite{GK}.) In the latter case,
the symbol string has the same entropy as the full time series and, in fact,
the full trajectory of a dynamical system can be in principle recovered from
the (bi-infinite) symbol sequence. The partitions which are commonly used for
symbolic encodings of time series data are almost never generating in that
sense. Since any further refinement of a partition that is generating preserves
this property, one often replaces a generating partition by a very fine ad hoc
partition. Upon further refinement, the partition becomes ``approximately
generating'' to a higher and higher degree. However, if fine partitions are
used, the main advantage of symbolic dynamics, to reduce the information in a
signal to the essential part, is lost.

To obtain a symbolic encoding, one can directly partition the measurements into
a small number of classes by defining suitable thresholds.  Alternatively, a
more general partitioning can be defined after a phase space reconstruction
step in multidimensional space. The ergodic properties of symbol sequences are
traditionally studied much more than continuous state dynamical
systems. References from the dynamical systems context are for example the
works by Herzel~\cite{herzel}, Ebeling and coworkers~\cite{ebeling},
Sch\"urmann and Grassberger~\cite{symbol}, and Hao~\cite{hao}.  Applications
are discussed in Refs.~\cite{wacker,k} and many others. Symbolic encoding
constitutes a severe selection among the available information.  This may be a
desirable property in cases with high noise levels.

At a moderate level of coarse graining, any finite length scale version of the
invariant quantities discussed above can in principle be used for comparative
purposes. Most popular seem to be intermediate length scale estimates of the
correlation dimension and of the maximal Lyapunov exponent. One example of such
an estimate is the maximum likelihood estimator given by
Theiler~\cite{takens_theiler}, see Eq.(\ref{eq:d2t}). In general, its value
depends on the upper cutoff length scale $\epsilon$, and the embedding
parameters.  Another popular statistic based on the correlation integral goes
back to a paper by Brock et al.~\cite{brockpaper1} and is usually referred to
as the BDS statistic. The authors of Refs~\cite{brockpaper1,brockpaper2} make
use of the fact that for a sequence of independent random numbers,
$C_m(\epsilon)=C_1(\epsilon)^m$ holds, where $m$ is the embedding dimension.
In these papers, also a formal test for this property is introduced. The
original BDS statistic is specifically designed so that one can derive the
asymptotic distribution analytically. A simpler expression that contains the
same information is $C_m(\epsilon)/C_1(\epsilon)^m$.

A whole class of measures for nonlinearity is given by constructive measures of
predictability that use a specific modeling approach to make forecasts of time
series. If an estimate of the dynamics $\hat{\ff}$ has been produced (see
Sec.~\ref{sec:finite}), one can define an average prediction error for example
by
\be\label{eq:pred}
    e^2 = {1 \over N-1-(m-1)\tau}
          \sum_{n=(m-1)\tau+1}^{N-1}
           (s_{n+1}-\hat{\ff}(\Ss_n))^2
\,.\ee
This, or some differently averaged error of prediction, is then used as an
indicator for the unpredictability of the signal. If such an interpretation is
put forth, it is essential to use some cross-validation technique to ensure
that $e$ is an out-of-sample error. An out-of-sample error is obtained if the
data set that is used for the estimation of $e$ is a different one from that
used to fit $\hat{\ff}$. See also Sec.~\ref{sec:finite}.

One of the simplest nonlinear predictive models is the locally constant
approximation given by Eq.(\ref{eq:fclazy}).  Since it does not involve the
numerical optimisation of parameters, the danger of overfitting is rather
small. Of course, the predictions obtained are often not optimal but on the
other hand statistically quite stable. A similar non-parametric prediction
scheme has been used by Sugihara et al.~\cite{B1b} and by Kennel and
Isabelle~\cite{KI}. Barahona and Poon~\cite{volterra} (among others) have used
a global polynomial model (a Volterra series) for nonlinearity testing.  In
Ref.~\cite{power}, a number of popular measures for nonlinearity are compared
quantitatively for the task of discriminating noisy chaotic data from
randomised surrogates with the same linear properties. The finding there was
that in this particular setting at the edge of detectability, the most stable
statistics outperform more subtle measures.

A number of test statistics for the detection and quantification of
nonlinearity have been used in the literature which, while not explicitly
called prediction errors, can be seen as specific ways to quantify
nonlinear predictability in the sense used here. Among these, the test
statistic proposed by Kaplan and Glass~\cite{KG} is particularly suited to
quantify deterministic structure in densely sampled data which permit the 
estimation of local flow vectors. Also the technique of {\em false nearest
  neighbours} advocated by Kennel, Brown, and Abarbanel~\cite{FNN} can be
regarded in this way. 

Pompe~\cite{pompe}, and Palu\v{s}~\cite{milan2,milan1} advocate the use of
coarse-grained redundancies, generalisations of the time-delayed mutual
information. Prichard and Theiler~\cite{pt} (among others) have pointed out
that it can be computationally advantageous to estimate information theoretic
quantities like redundancy and mutual information by their second order
generalisations. The latter can be obtained from correlation integrals, thus
avoiding the common problems with box-counting approaches. Correlation
integrals are also much easier to compute than the adaptive partitionings used
for example by Fraser and Swinney~\cite{fraser}. The drawback of using
second order quantities is that generalised entropies lack the additivity
property. The generalised mutual information is therefore no longer positive
definite. This is however unproblematic as long as it is used only as a
relative measure.

With measured data we will never be able to carry out the proper limit of 
small length scales. As a rule, the necessary coarse graining leads 
to a loss of invariance properties. There is one notable exception to this
rule. Unstable periodic orbits embedded in a strange attractor define a family
of invariant quantities which are accessible at finite length scales. In
particular, the existence, length, and stability of each orbit are such
invariants.%
\footnote{Confusingly, the purely topological properties which one would expect
   to be invariants in the first place, are not in general. The topological
   length of a cycle in a flow system depends on the choice of Poincar\'e
   section and winding numbers etc. are only invariant under families of
   transformations where the family depends smoothly on its parameters. For
   example the knot structure changes under reflection. Reflection is smooth
   but cannot be connected with the identity by a smooth family of
   transformations.} 
Consequently, many people have pursued the analysis of unstable periodic orbits
from time series. Periodic orbit expansions~\cite{A0,AAC,AAC2,PObook} are of
great theoretical appeal but they require knowledge of the dynamical system or
data of exceptional quality for useful results. See Ref.~\cite{NMR-RMP} for a
review. Similar data requirements are valid for the topological analysis of
time series and the extraction of templates~\cite{NBT}. Recently, the emergence
of methods for the stabilisation of chaotic systems~\cite{OGY,coping} has
attracted renewed interest in the detection of unstable fixed points or low
order periodic orbits~\cite{PM,soso}. Unstable fixed points have been found and
stabilised in a number of real world systems. Controlability with methods from
chaos theory has been often taken as an indication of the presence of chaos in
these systems. However, Christini and Collins~\cite{sonicht} have shown that
also stochasitc, non-chaotic systems can be successfully controlled by such
methods. If the detection and analysis of unstable fixed points is used as a
means to detect nonlinearity and chaos, the same issues of significance and
possible spurious results have been considered as for other quantifiers of
nonlinearity.

\subsection{Measures of dissimilarity} \label{sec:dissim}
The idea to use relative measures between time series or segments of a long
sequence for signal classification and nonstationarity testing has been brought
up independently in a number of recent publications~\cite{EEG,B1,statio,clust}.
In principle, it is desirable to use relative measures that can be interpreted
as a distance or a dissimilarity. As we have seen in Section~\ref{sec:compare},
one such measure can be derived from the cross-correlation integral,
Eq.(\ref{eq:x2}).  For the practical estimation of the cross-correlation
integral, refer to what has been said about the correlation sum
(Section~\ref{sec:c2data}).  There are at least two other ways to construct an
informal measure of dissimilarity from an estimator $C_{XY}(\epsilon)$ of the
cross-correlation integral $C_2(\epsilon; \mu,\nu)$. (This notation implies
that $\mu(x)$ (resp. $\mu(y)$) are the probability distributions of the random
variables $X$ and $Y$, respectively.)  Kantz~\cite{holger} defines an informal
distance between attractors by the minimal length scale $\epsilon_0$ above
which the attractors are indistinguishable up to an accuracy $\delta$:
\be
   \max(|\log C_{XX}(\epsilon)-\log C_{XY}(\epsilon)|, \,
       |\log C_{YY}(\epsilon)-\log C_{XY}(\epsilon)|) < \rho
       \quad \forall \epsilon>\epsilon_0
\,.\ee
This approach is particularly useful when comparing clean model attractors and
noisy measurements. In that case the length scale at which the two attractors
start to differ indicates the noise level. Another possible definition of a
dissimilarity based on the cross-correlation integral is $1 - C_{XY}(\epsilon)
/ \sqrt{C_{XX}(\epsilon) C_{YY}(\epsilon)}$. Further, Albano et
al.~\cite{albano} use a Kolmogorov-Smirnov test to detect dissimilarity of
two correlation integrals.

The different measures of dissimilarity based on predictive models (see
Section~\ref{sec:compare}) have been introduced in a more a hoc way. They have
less of a theoretical foundation and weaker invariance properties than
cross-correlation integrals. In practical work however, this is often the price
to be paid for statistical robustness and modest data requirements.
Locally constant phase space predictors can yield stable results with a few
hundred points and global polynomial or radial basis function models with even
less. It should be stressed here that it is not essential for the present
purpose that the predictions are optimal in the usual sense, as long as the
predictions are sensitive to differences in the dynamics. In other words, for
comparative purposes it may be advantageous to trade a possible bias for 
a lower variance.

\section{Nonstationarity} \label{sec:statio}
Almost all methods of time series analysis, traditional linear, or nonlinear,
require some kind of stationarity. Therefore, changes in the dynamics during
the measurement usually constitute an undesired complication of the
analysis. There are however situations where such changes represent the most
interesting structure in the recording. For example, electro-encephalographic
(EEG) recordings are often taken with the main purpose of identifying changes
in the dynamical state of the brain. Such changes occur e.g. between different
sleep stages, or between epileptic seizures and normal brain activity.

In the past, emphasis has been put on the question how stationarity can be
established. If nonstationarity was detected, often the time series was
discarded as unsuitable for a detailed analysis, or it was split into segments
that were short enough to be regarded as stationary. More recently, authors
have begun to exploit the information contained in time-variable dynamics as an
essential part of the underlying process. Thus this section will discuss tests
for stationarity but also report on the steps that have been taken towards a
time resolved study of nonstationary signals.

The most common definition of a stationary process found in textbooks (often
called {\em strong stationarity}) is that all conditional probabilities are
constant in time. Note that this definition is only applicable to the abstract
generating process, and not to a realisation that produces a time series.  If
we regard a deterministic system as the limiting case of a stochastic process
where the conditional probability density for a transition from state $\xx$ to
state $\xx'$ is given by $\delta(\xx'-f(\xx))$, the definition requires
$f(\cdot)$ to be unchanged with time. In the study of time series, the
transition probabilities are unknown and have to be estimated from the data,
subject to statistical fluctuations. In some cases, for example in intermittent
systems, these fluctuations are large and the properties of measured time
series can change dramatically, even though the underlying process is formally
stationary after the above definition.  There is no agreement on a definition
of stationarity for time series. It seems reasonable to require that the
duration of the measurement is long compared to the time scales of the
systems. If this is the case, all temporal changes can be modeled as part of
the dynamics. For this reason, processes with power law correlations are often
considered nonstationary since no length of measurement could ever cover all
time scales. On the other hand, processes with very well seperated time scales
can lead to time series which are stationary for practical purposes. The heart
beat of a resting person is often homogeneous over several minutes. Longer
recordings, however, cover new elements due to slower biological cycles. Since
the common 24~h ECG recordings cover just a single cycle of the circadian
rhythm, they are more problematic with respect to stationarity than shorter or
longer sequences.

\subsection{Moving windows}
A number of statistical tests for stationarity in a time series have been
proposed in the literature.  Most of the tests I am aware of are based on ideas
similar to the following: Estimate a certain parameter using different parts of
the sequence. If the observed variations are found to be significant, that is,
outside the expected statistical fluctuations, the time series is regarded as
nonstationary. In many applications of linear (frequency based) time series
analysis, stationarity has to be valid only up to the second moments (``weak
stationarity''). Then, the obvious approach is to test for changes in
quantities up to second order, like the mean, the variance, or the power
spectrum. See e.g. Priestley~\cite{priestley} and references therein.

In a nonlinear dynamical framework, weak stationarity is not an interesting
property. Quite often, the linear properties of the processes do not carry much
information anyway. It is therefore desirable to use some nonlinear quantifier
in order to trace nonstationarity. In particular, Isliker and
Kurths~\cite{kurths} use a binned probability distribution. The method proposed
there, however, suffers from a problem that arises with most nonlinear
quantifiers. Unless quite narrow assumptions are made, the probability
distribution of these quantities is not known exactly. Therefore we cannot
usually assess the significance of changes in these quantities in a rigorous
way.  Also, a signal might be considered stationary for some purpose, but not
for another. A typical case is dimension estimation which requires stationarity
in the probability of close recurrences.  The authors of Ref.~\cite{kurths} use
a $\chi^2$ test for the difference of histograms on sections of the data. This
test, however, assumes that the histogram is formed by independent draws from
some probability distribution. In the presence of serial correlations or
deterministic structure, this is usually not justified. A possible remedy is to
exclude points close in time form the histograms, thereby however losing
statistical stability.

Computing nonlinear indicators for moving windows of data is attractive because
it allows for a time resolved study of possible changes. As we have seen,
however, there is a tradeoff between time resolution and statistical accuracy.
A different way to proceed therefore is to completely give up the detailed time
information and concentrate on testing the null hypothesis that the sequence is
stationary. Let us remark that stationarity is an awkward concept to test for.
What we would like to have is the assertion that a given time series is
stationary. The failure of some test to reject the hypothesis of stationarity
is not sufficient -- the test might just have no power against the particular
kind of nonstationarity present. Quite generally, a statistical test can never
prove the null hypothesis. Thus we would rather like to test against the null
hypothesis that the data is {\em non}-stationary. Unfortunately, this is such a
hopelessly composite hypothesis that we do not know how to devise a statistical
test for it.

Therefore, formal tests against stationarity, like the one set up by
Kennel~\cite{kennel} have to be understood with a particular alternative
hypothesis in mind. The alternative in Ref.~\cite{kennel} is that the phase
space geometry of the time series, reflected by the nearest neighbour
structure, is changing in time.  The basic idea is that the expectation value
of the number of reconstructed phase space points that have their nearest
neighbour in the same half of the sequence is minimal for a stationary
sequence. When thinking of geometry in phase space, nonstationarity introduces
a tendency that points close in space are also close in time.

\subsection{Recurrence plots}
The relation between closeness in time and in phase space is the most relevant
manifestation of nonstationarity in a nonlinear dynamical setting. The basic
graphical tool that evaluates temporal and phase space distance of states is
the classical recurrence plot of Eckmann et al.~\cite{recurr}. In its original
version, a pair of times $n_1,n_2$ is called a recurrence if $\Ss_{n_1}$ is
one of the $k$-th nearest neighbours of $\Ss_{n_2}$, for some predefined value
of $k$. An alternative given by Koebbe~\cite{Koebbe} is to define a recurrence
to occur at times $n_1,n_2$ at resolution $\epsilon$ if $n_1\neq n_2$ and
$\|\Ss_{n_1}-\Ss_{n_2}\|\le \epsilon$. Usually, the $\Ss_n$ are delay
embedding vectors and the results depend on the embedding parameters. A
recurrence {\em plot} is generated by marking all recurrences at a given
neighbour order $k$ or resolution $\epsilon$ in a graph with coordinates $n_1$
and $n_2$. In the second form, a recurrence plot can be simply identified with
the expression \be\label{eq:recurr} r_{\epsilon}(n_1,n_2) = (1-\delta_{n_1
  n_2})\, K_{\epsilon}(\|\Ss_{n_1}-\Ss_{n_2}\|) \ee where the kernel function
is usually taken to be the Heaviside step function $K_{\epsilon}(r)=\Theta
(\epsilon-r)$. The full recurrence structure is contained in the {\em
recurrence matrix}, which is simply defined by $R_{n_1n_2} =
\|\Ss_{n_1}-\Ss_{n_2}\|$. Obviously, $R$ is invariant under isometries
(translations, rotations, and reflections) in phase space. Mc~Guire and
coworkers~\cite{mcguire} give an algorithm to explicitly reconstruct an
attractor up to isometries from a recurrence matrix. Of course, since an
$(N\times N)$ symmetric matrix has $N(N-1)$ independent entries, a recurrence
matrix is not a very economical representation of $N$ vectors, and the
requirement that the entries are distances in a space of dimension $m$ poses a
strong constraint. 

Since recurrence plots are rather difficult to read they have not gained much
popularity beyond the admiration of the intriguing patterns they
exhibit~\cite{zbilut}. Zbilut and coworkers~\cite{zbilut2} propose different
parameters for the statistical quantification of recurrence plots but they give
little clues on how to interpret these numbers. Nevertheless, the recurrence
plot can be a useful starting point for the analysis of nonstationary sequences
if the relevant information is extracted in a suitable way. The most detailed
account of these techniques has been given by Casdagli~\cite{cas_recurr}, where
also the interrelations to other methods are discussed thoroughly. Let us use a
very simple nonstationary dynamical system as an illustration in the
following. Consider a one-parameter family of sawtooth maps $[0,1]\mapsto
[0,1]$:
\be\label{eq:saw}
   x_{n+1} = f_{\phi_n}(x_n)=f(x_n+\phi_n \bmod 1),\qquad
      f(x) = \left\{\begin{array}{ll}
         2x       & x< 1/2      \\
         2-2x     & 1/2 \le x <1 
      \end{array}\right.
\ee
Take a time series of length $N=20000$ and let $\phi_n$ vary with time
such that it covers two oscillations of a damped sine function within the
measurement period: $\phi_n=(1+e^{-n/N}\sin 4\pi n/N)/2$. Recurrence
plots of (a) the time series for two-dimensional embeddings, $\epsilon=0.1$ and
0.01, for a three-dimensional embedding, $\epsilon=0.005$, and (b) for the
parameter $\phi_n$ at $\epsilon=0.001$ are shown in Fig.~\ref{fig:tent_recurr}.
Casdagli~\cite{cas_recurr} has pointed out that for a faithful embedding and
in the limit $\epsilon\to 0, \quad N\to \infty$, the recurrence plot of the
time series from a system approaches that of the varying parameter. For the
above example, this can be verified from Fig.~\ref{fig:tent_recurr}.

\begin{figure}
   \centerline{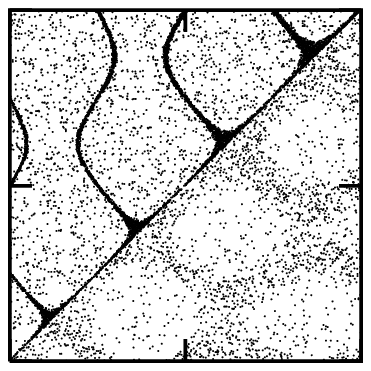\hspace*{-1cm}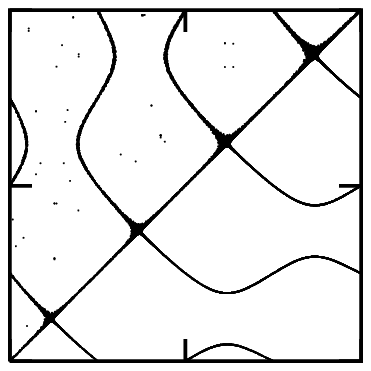}
   \caption[]{\label{fig:tent_recurr} Recurrence plots of a tent map time
   series subject to a parameter drift (see text for details). The left panel
   was obtained with two-dimensional embeddings. Above the diagonal:
   $\epsilon=0.01$ (50\% of all points chosen at random for better
   contrast). Below the diagonal: $\epsilon=0.1$ (0.2\% shown). In the right
   panel, embedding was done in three dimensions. Above the diagonal:
   $\epsilon=0.005$. Below the diagonal, the recurrence plot of the parameter
   sequence $\{\phi_n\}$ is shown with $\epsilon=0.001$ (2\% shown).  For
   decreasing $\epsilon$, the recurrence plots of the signal indeed converge to
   that of the parameter sequence (right panel, below the diagonal)}
\end{figure}

Let us note that the total average $\sum_{n_1,n_2} r_{\epsilon}(n_1,n_2)$
equals the sample correlation integral $C(\epsilon)$ in
Eq.(\ref{eq:c2}). However, in the practical estimation of $d_2$ one has to
exclude terms with $|n_2-n_1|<t_{\rm min}$. One way to estimate the correlation
time $t_{\rm min}$ is the following. The partial average
\be\label{eq:stp}
   C(\epsilon,\Delta n) =      
      {1 \over N-\Delta n} \sum_{n=\Delta n+1}^N r_{\epsilon}(n,n-\Delta n)
\ee
yields the {\em space-time separation plot} introduced by Provenzale et
al.~\cite{provenzale}. Contour lines of
$C(\epsilon,\Delta n)$ should not increase with $\Delta n$ except for possible
oscillatory variation. The minimal $\Delta n$ for which this is the case
yields a guideline for the minimal time separation $t_{\rm min}$ to be used in
the correlation sum. Due to the temporal averaging, the space-time separation
plot does not allow for a time-resolved analysis. Also, the effect of
nonstationarity may average out in certain cases, as for example for the
oscillating parameter in the tent map example. Let us remark that for the same
reason also the time averaged statistic used by Kennel~\cite{kennel} fails to
reject the null hypothesis of stationarity since the nearest neighbour of each
point can have any temporal distance with about equal probability.

\subsection{Tracing parameter variation}\label{sec:trace}
A useful way to formulate nonstationary dynamics is by introducing a temporal
variation of dynamical parameters into the system, as it was already done in
the tent map example. If this variation is sufficiently slow, recurrence plots
and similar techniques can asses these changes to some extent. In
Ref.~\cite{cas_recurr}, it is shown by a scaling argument that for a dynamical
system with time varying parameters, the recurrence plot in the limit of small
$\epsilon$, large $N$, and sufficient $m$ approaches the recurrence plot of the
fluctuating parameter. This can be seen in Fig.~\ref{fig:tent_recurr} for the
time varying tent map. However, it is in general difficult to extract the time
variation of the parameter from its recurrence plot. Nevertheless, qualitative
information, like the number of fluctuating parameters and the time scales of
their fluctuations, can often be inferred from such a plot.

It can be useful to average the number of recurrences over windows in time,
in particular, if there is a stochastic component in the dynamics of the
system:
\be\label{eq:cxy}
   C_{XY}(\epsilon,w,n_1,n_2) = {1 \over \alpha}
      \sum_{i=1}^w \sum_{j=1}^w r_{\epsilon}(n_1+i,n_2+j)
\,,\ee
where $\alpha = w^2-\sum_{i=1}^w \sum_{j=1}^w \delta_{n_1+i\,n_2+j}$ is the
normalization which takes the varying number of diagonal recurrences into
account. The quantity $C_{XY}(\epsilon,w,n_1,n_2)$ defined in
Eq.(\ref{eq:cxy}) is just the cross-correlation integral between segments of
length $w$ of $\{s_n\}$, starting at $n_1$ and $n_2$. (See
Section~\ref{sec:compare}, Eq.(\ref{eq:dist}).)  The cross-correlation
integral has been introduced and discussed as a measure for the distance
between attractors by Kantz~\cite{holger}.

Relative measures between time series or segments of a long sequence for
signal classification and nonstationarity testing have been discussed in
Sections~\ref{sec:compare} and~\ref{sec:dissim} above.  As we have seen, the
quantity $C_{XY}(\epsilon,w,n_1,n_1) + C_{XY}(\epsilon,w,n_2,n_2) -
2C_{XY}(\epsilon,w,n_1,n_2)$ becomes a formal distance fulfilling the triangle
inequality in the limit $\epsilon\to 0$ and $w\to\infty$. The latter limit
causes problems with signals which can be considered to be stationary only
over short times~$w$, if at all. If we give up the formal requirement of a
distance, we can alternatively use nonlinear cross-prediction errors. The
average error of a locally constant predictor (see also Eq.(\ref{eq:xpred}))
can be written in a compact way in terms of recurrences: \be\label{eq:xpredR}
\gamma(\epsilon,w,n_1,n_2)^2 = \sum_{i=1}^w (\hat{\Ss}_{n_1+i+1} -
\Ss_{n_1+i+1})^2 \,,\ee where the prediction $\hat{\Ss}_{n_1+i+1}$ is given by
the average over an $\epsilon$-neighbourhood, 
\be\label{eq:xpredSs}
   \hat{\Ss}_{n_1+i+1} = \left\{ \begin{array}{ll} 
      {\displaystyle {1\over\alpha}
      \sum_{j=1}^w r_{\epsilon}(n_1+i,n_2+j)
      \,\Ss_{n_2+j+1} } & \alpha>0 \\
    {\displaystyle {1\over w} \sum_{j=1}^w \Ss_{n_2+j+1} } & \alpha=0 \,,
   \end{array} \right.
\ee
where $\alpha= \sum_{j=1}^w r_{\epsilon}(n_1+i,n_2+j)$ is the number of
neighbours of $\Ss_{n_1+i}$ closer than~$\epsilon$. Figure.~\ref{fig:tent_dij}
shows $\gamma(\epsilon,w,n_1,n_2)$ for the modulated tent map series with
$w=1000$ and $n_1,n_2$ in steps of 500. The variation of the prediction error
with the segment location $n_1$, $n_2$ is clearly visible.

\begin{figure}
\centerline{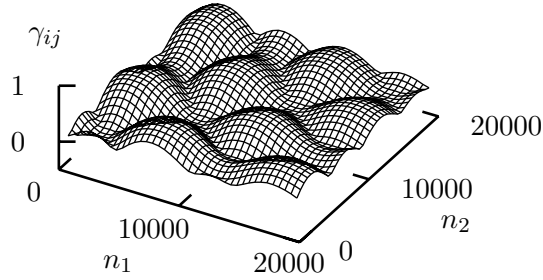}
\caption[]{\label{fig:tent_dij} Cross-prediction errors for the 
  model~(\ref{eq:xpredSs}) obtained with locally constant prediction in one
  dimension versus location in time of segments. Local neighbourhoods were
  formed with a radius of one quarter of the variance of the sequence,
  $\epsilon=0.072$. Time windows of length $w=1000$ overlapping by 500 time
  steps were used.}
\end{figure}

The information contained in Fig.~\ref{fig:tent_dij} can be processed in a way
that makes the time dependence of the parameter more clear. The quantities
$\gamma(\epsilon,w,n_1,n_2)$ can be regarded as a {\em dissimilarity matrix}
and treated by a cluster algorithm. (This technique will be discussed in more
detail below in Section~\ref{sec:classify}.) If the analysis is successful, the
clusters are localised in parameter space and can be used to define coordinates
in that space. The time varying ``distances'' of each segment $i$ to the
clusters $\nu$ ($D_i^{(\nu)}$, to be defined in Eq.(\ref{eq:dis_i_nu}) of
Section~\ref{sec:clust} below) then reflect the time variation of the
parameter(s). A successful example with two clusters and one parameter is
shown in Fig.~\ref{fig:tent}.
\begin{figure}
   \centerline{\hspace*{-0.2cm}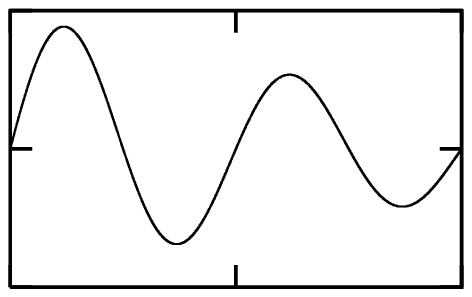%
   \hspace*{0.4cm}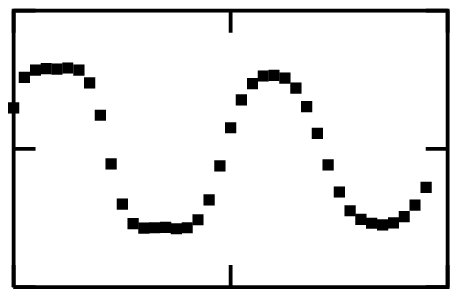}
   \caption[]{\label{fig:tent} Left panel: Time dependence of the parameter
   $\phi_n$ in the tent map example. Right panel: The information contained in
   Fig.~\ref{fig:tent_dij} was used to form two clusters of similar time
   series. For each segment, the distances $D_i^{(1)}, D_i^{(2)}$ to clusters
   $(1)$ and $(2)$ are computed. The difference $D_i^{(2)}-D_i^{(1)}$ is
   plotted for each segment versus time. Such a plot cannot only reveal that
   there is a single changing parameter but also the form of its change. That
   the units in both panels are of comparable magnitude is purely
   coincidental.  Clustering of time series is discussed in
   Sec.~\ref{sec:clust} below, where also $D_i^{(\nu)}$ is defined.}
\end{figure}

\subsection{An application}\label{sec:EEG}
Let us finish the discussion of nonstationary time series with an application
that is currently studied with considerable effort in a number of research
groups: the anticipation of epileptic seizure onset from intracranial
recordings of neural potentials. Epileptic seizures manifest themselves in
specific patterns in the neural electric field. While traditional
electro-encephalograms (EEG) with electrodes placed on the surface of the scalp
show such patterns when the epilepsy activity has reached a cortical region
that is sufficiently close to the surface, for a detailed study of {\em focal}
epilepsy in deeper regions of the brain electrodes have to be implanted in
the epileptogenic region. This is a common clinical technique in pre-surgical
screening. The specific activity during seizures is usually so pronounced that
it can be detected visually and also automatically.%
\footnote{Many authors have discussed the question if there is evidence for
   low-dimensional chaos and strange attractors in normal, or, more likely, in
   epilepsy EEG data. References include~\cite{Bablo,FEEG,TEEG,TEEG2} and many
   more, conclusions are controversial. More recently, the focus has shifted
   from the question of chaos versus noise to the quantification of changes in
   the EEG.  Most authors agree that ictal (during seizures) and inter-ictal
   EEG can be distinguished with nonlinear, but also with spectral
   methods. References for the former are for example
   Refs.~\cite{casEEG,lerner,Pijn,lehnertz,Pezard}, and Ref.~\cite{Rogo} for
   the latter.}
A far more challenging problem is to detect specific changes in the dynamics
of the recordings just {\em prior} to the actual seizure. First of all, a
reliable anticipation of a seizure several minutes ahead potentially allows
for pharmacological or electrophysiological intervention. The insights into
the mechanism that leads to the large scale pathological activity are of
equally high interest. The problem however is very intricate. At any given
time, the recorded neural activity is very rich and far from being understood.
Although usually simultaneous recordings at several positions are available,
it is not clear to what extent multivariate studies provide more insight at
this stage~\cite{Dvo}. Electrode spacings down to fractions of a millimeter
are still much larger than typical coherence lengths. The dynamics is
time-variable and only part of this variability is specific for the generation
of epileptic activity. It is not expected that the brain falls into a single
typical state prior to a seizure but rather that the pre-seizure activity
shows some characteristic yet variable behaviour. The task of time series
analysis is to find and specify such features that allow for the detection of
the critical state.

Elger and Lehnertz~\cite{elger} claim statistically significant positive
evidence for seizure predictability several minutes ahead of seizure onset.
The authors use a sliding window version of the correlation integral that has
been customised for this particular purpose.  Since the intracranial EEG
signal is nonstationary even in episodes without epileptic activity, the
window length should be short enough for the segment to be effectively
stationary but long enough to yield stable results.  Half-overlapping windows
of 30~s duration were chosen at a sampling rate of 173~Hz. The variance of the
data segments is time dependent which makes it difficult to choose a length
scale for the determination of an effective scaling index. The dominating
contribution to the variance found in pathological regions often arises from
spikes occurring at irregular intervals.  Although these spikes are
characteristic for epileptogenic tissue, they seem not to be specific
precursors of seizures and should therefore not be over-emphasised in the
analysis. In Ref.~\cite{elger} this is achieved by selecting approximate
scaling regions for each individual window. These are usually found at much
smaller scales than the spikes. Pre-seizure behaviour is found to be
accompanied by epochs of smaller effective scaling indices as compared to the
standard behaviour found in segments that are spatially and/or temporally well
separated from the seizure.

\begin{figure}
\centerline{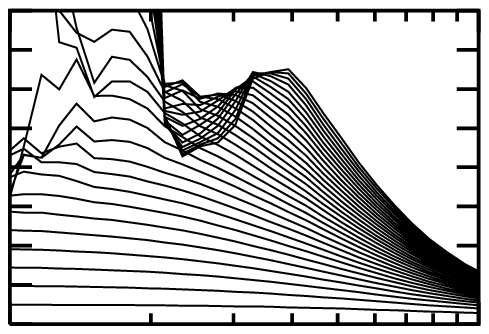\hspace{0.5cm}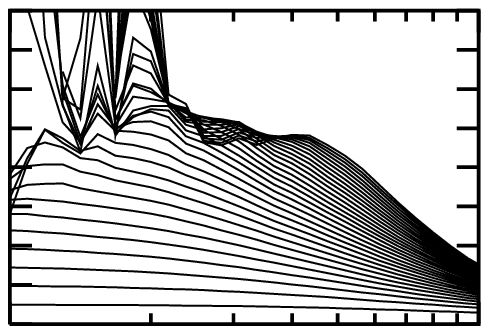}
\caption[]{Maximum likelihood estimator of the correlation dimension for an
   intracranial EEG recording~\cite{elger} of an epilepsy patient. The effect
   of dynamically correlated pairs was only incompletely corrected. Left: a
   data set measured long before the next seizure. Right: a data set measured
   about 10~min prior to seizure onset. The right panel shows the same data as
   Fig.~\ref{fig:eeg}.\label{fig:eegip}}
\end{figure}

Correlation dimension data for a time series recorded by Elger and Lehnertz
has already been shown in Sec.~\ref{sec:c2data} and it may surprise the reader
that in Ref.~\cite{elger} the authors do find small approximate scaling
regions. This finding can be reproduced by limiting the correction for the
dynamical correlations to the exclusion of pairs which are not more than three
sampling intervals apart in time. For the determination of a proper dimension,
or for the interpretation of the effect as a signature of a finite attractor
dimension, this would be disatrous, but it is fully justified by the
resulting discriminative power for the particular purpose at hand.
Figure~\ref{fig:eegip} repeats the same as was done in Fig.~\ref{fig:eeg} but
with the limited correction indicated above. The left panel shows a data
segment that was taken far away in time from any seizure. The right panel was
calculated from the same data as Fig.~\ref{fig:eeg}, a segment taken about
10~min prior to the onset of an epileptic seizure. Indeed we see tiny
``plateaus'' not present in Fig.~\ref{fig:eeg}.%
\footnote{The particular shape of the deflection that could have been 
   interpreted as a plateu resembles what is typically seen for intermittent
   systems, see Hegger and coworkers~\cite{hko}.}
Further, we see that there is a difference between the two segments in where
the pseudo-scaling is found. It is such differences that have been studied
systematically in Ref.~\cite{elger}.  The statistical material presented there
is based on 16 patients and shows that the pre-seizure states and normal epochs
follow significantly different distributions of approximate scaling indices.
Certainly, a number of ad hoc decisions have been made in devising the
algorithm and it is not fully clear from Ref.~\cite{elger} to what extent the
same sample of patients has been used to optimise parameters. Further research
will have to evaluate whether the differences are strong enough to make
reliable out-of-sample predictions for individual patients. For clinical
applicability, it will eventually be necessary to compute and interpret the
relevant quantities in real time.

The finding that the discriminative power declines upon full correction of the
dynamical correlations indicates that it is not exactly phase space geometry
that distinguishes the different states. It should however be stressed that the
shown segments are not distinguishable by their autocorrelation functions or
spectra. Thus, what is represented in Fig.~\ref{fig:eegip} is a difference in
the {\em nonlinear} dynamical correlation. This suggests that there may be
nonlinear indicators which are more sensitive to changes in the dynamics
and that might correlate even more strongly with the seizure onset.

\section{Testing for nonlinearity}\label{sec:surro}
There are two distinct motivations to use a nonlinear approach when analysing
time series data. It might be that the arsenal of linear methods has been
exploited thoroughly but all the efforts left certain structures in the time
series unaccounted for. It is also common that a system is known to include
nonlinear components and therefore a linear description seems unsatisfactory in
the first place. Such an argument is often heard for example in brain research
--- nobody expects the brain to be a linear device. In fact, there is ample
evidence for nonlinearity in particular in small assemblies of neurons.
Nevertheless, the latter reasoning is rather dangerous. That a system is known
to contain nonlinear components does not prove that this nonlinearity is also
reflected in a specific signal we measure from that system. In particular, we
do not know if it is of any practical use to go beyond the linear
approximation. After all, we do not want our data analysis to reflect our
prejudice about the underlying system but to represent a fair account of the
structures that are present in the data. Consequently, the application of
nonlinear time series methods has to be justified by establishing nonlinearity
in the time series data.

This section will discuss formal statistical tests for nonlinearity.  First, a
suitable null hypothesis for the underlying process will be formulated covering
all Gaussian linear processes or a class that is somewhat wider. We will then
attempt to reject this null hypothesis by comparing the value of a nonlinear
parameter estimated on the data with its probability distribution for the null
hypothesis. Since only exceptional cases allow for the exact or asymptotic
derivation of this distribution unless strong additional assumptions are made,
we have to estimate it by a Monte Carlo resampling technique. This procedure is
known in the nonlinear time series literature as the method of {\em surrogate
data}, see Refs.~\cite{theiler1,theiler2,fields}. Thus we have to face a
two-fold task. We have to find a nonlinear parameter that is able to actually
detect an existing deviation of the data from a given null hypothesis and we
have to provide an ensemble of randomised time series that accurately
represents the null hypothesis.

\subsection{Detecting weak nonlinearity}
In the preceding sections, several quantities have been discussed that can be
used to characterise nonlinear time series. For the purpose of nonlinearity
testing we need such quantities that are particular powerful in discriminating
linear dynamics and weakly nonlinear signatures --- strong nonlinearity is
usually more easily detectable. Quite a number of such measures has been
proposed and used in the literature. An important objective criterion that can
be used to guide the preferred choice is the discrimination {\em power} of the
resulting test. The power $\beta$ is defined as the probability that the null
hypothesis is rejected when it is indeed false.  It will obviously depend on
how and how strongly the data actually deviates from the null hypothesis.

Traditional measures of nonlinearity are derived from generalisations of the
two-point autocovariance function or the power spectrum. The use of higher
order cumulants and bi- and multi-spectra is discussed for example in
Ref.~\cite{BI}.  One particularly useful third order quantity is
\be\label{eq:skew}
   \phi^{\rm rev}(\tau) = {\sum_{n=\tau+1}^N (s_n-s_{n-\tau})^3
      \over [\sum_{n=\tau+1}^N (s_n-s_{n-\tau})^2]^{3/2} }
\,,\ee
since it measures the asymmetry of a series under time reversal. (Remember that
the statistics of linear stochastic processes is always symmetric under time
reversal. This can be most easily seen when the statistical properties are
given by the power spectrum which contains no information about the direction
of time.) Time reversibility as a criterion for discriminating time series
is discussed in detail in Ref.~\cite{diks2}.

When a nonlinearity test is performed with the question in mind if nonlinear
deterministic modeling of the signal may be useful, it seems most appropriate
to use a test statistic that is related to a nonlinear deterministic approach.
We have to keep in mind however that a positive test result only indicates
nonlinearity, not necessarily determinism. Since nonlinearity tests are usually
performed on data sets which do not show unambiguous signatures of
low-dimensional determinism (like clear scaling over several orders of
magnitude), one cannot simply estimate one of the quantitative indicators of
chaos, like dimension or Lyapunov exponent. The formal answer would almost
always be that both are probably infinite. Still, some useful test statistics
are at least inspired by these quantities. Usually, some effective value at a
finite length scale has to be computed rather than attempting to take the
proper limits.  We can largely follow the discussion in
Sec.~\ref{sec:noninvar}, considering the limiting case that the deterministic
signature to be detected is probably weak. In that case the major limiting
factor for the performance of a statistical indicator is its variance since
possible differences between two samples may be hidden among the statistical
fluctuations. In Ref.~\cite{power}, a number of popular measures of
nonlinearity are compared quantitatively. The results can be summarised by
stating that in the presence of time-reversal asymmetry, the three-point
autocorrelation (Eq.\ref{eq:skew}) gives very reliable results. However, many
nonlinear evolution equations produce little or no time-reversal asymmetry in
the statistical properties of the signal. In these cases, simple measures like
a prediction error of a locally constant phase space predictor performed best.
It was found to be advantageous to choose embedding and other parameters for
small variance, even if at these parameters no valid embedding could be
expected.

\subsection{Surrogate data tests}
All of the measures of nonlinearity discussed above have in common that their
probability distribution on finite data sets is not known analytically. Some
authors have tried to give error bars for measures like predictabilities (e.g.
Barahona and Poon~\cite{volterra}) or averages of pointwise dimensions (e.g.
Skinner et al.~\cite{skinner}) based on the observation that these quantities
are averages (or medians) of many individual terms, in which case the variance
(or quartile points) of the individual values yield an error estimate. This
reasoning is however only valid if the individual terms are independent, which
is usually not the case for time series data. In fact, it is found empirically
that nonlinearity measures often do not even follow a Gaussian distribution.
Also the standard error given by Roulston~\cite{roulston} for the mutual
information is not quite correct except for uniformly distributed data.  While
a smooth rescaling to uniformity would not do harm to his derivation, rescaling
to {\em exact} uniformity is in general non-smooth and introduces a bias in the
joint probabilities.  In order to determine the distribution of a nonlinear
statistic on realisations of the null hypothesis, it is therefore
preferable to use a Monte Carlo resampling technique. Traditional bootstrap
methods use explicit model equations that have to be extracted from the data.
This {\em typical realizations} approach can be very powerful for the
computation of confidence intervals, provided the model equations can be
extracted successfully. As discussed by Theiler and Prichard~\cite{tp}, the
alternative approach of {\em constrained realizations} is more suitable for the
purpose of hypothesis testing we are interested in here. It avoids the fitting
of model equations by directly imposing the desired structures onto the
randomised time series.  However, the choice of possible null hypothesis is
limited by the difficulty to impose arbitrary structures on otherwise random
sequences. The following section will discuss a number of null hypotheses and
algorithms to provide the adequately constrained realisations. The most general
method to generate constrained randomisations of time series is described in
Ref.~\cite{anneal}. The price for its accuracy and generality is its high
computational cost.

\subsubsection{How to make surrogate data}
It is essential for the validity of the statistical test that the surrogate
series are created properly. If they contain spurious differences to the
measured data, these may be detected by the test and interpreted as signatures
of nonlinearity. More formally, the {\em size} of a test is the actual
probability that the null hypothesis is rejected although it is in fact true.
For a valid test, the size $\alpha$ must not exceed the level of significance
$p$. The correct size crucially depends on the way the surrogates are
generated. Let us discuss a hierarchy of null hypotheses and the
issues that arise when creating the corresponding surrogate data.

A simple case is the null hypothesis that the data consists of independent
draws from a fixed probability distribution. Surrogate time series can be
simply obtained by randomly shuffling the measured data. If we find
significantly different serial correlations in the data and the shuffles, we
can reject the hypothesis of independence.%
\footnote{Independence seems not to be an interesting null hypothesis for most
  time series problems. It becomes relevant when the residual errors of a time
  series model is evaluated. For example in the BDS test for
  nonlinearity~\cite{brockpaper1}, an ARMA model is fitted to the data. If the
  data are linear, then the residuals are expected to be independent.}
The next step would be to explain the structures found by linear two-point
autocorrelations. A corresponding null hypothesis is that the data have been
generated by some linear stochastic process with Gaussian increments. The most
general univariate linear process is given by Eq.(\ref{eq:arma}). The
statistical test is complicated by the fact that we do not want to test against
one particular linear process only (one specific choice of the $a_i$ and
$b_i$), but against a whole class of processes. This is called a {\em
composite} null hypothesis. The unknown values $a_i$ and $b_i$ are sometimes
referred to as {\em nuissance parameters}. There are basically three directions
we can take in this situation. First, we could try to make the discriminating
statistic independent of the nuissance parameters. This approach has not been
demonstrated to be viable for any but some very simple statistics. Second, we
could determine which linear model is most likely realised in the data by a fit
for the coefficients $a_i$ and $b_i$, and then test against the hypothesis that
the data has been generated by this particular model. Surrogates are simply
created by running the fitted model. This {\em typical realisations} approach
is the common choice in the bootstrap literature, see e.g.  the classical book
by Efron~\cite{efron}. The main drawback is that we cannot recover the {\em
true} underlying process by any fit procedure. Apart from problems associated
with the choice of the correct model orders $M$ and $N$, the data is by
construction a very likely realisation of the fitted process.  Other
realisations will fluctuate {\em around} the data which induces a bias against
the rejection of the null hypothesis. This issue is discussed thoroughly in
Ref.~\cite{fields}, where also a calibration scheme is proposed.

The most attractive approach to testing for a composite null hypothesis seems
to be to create {\em constrained realisations}~\cite{tp}. Here it is useful to
think of the measurable properties of the time series rather than its
underlying model equations. The null hypothesis of an underlying Gaussian
linear stochastic process can also be formulated by stating that all structure
to be found in a time series is exhausted by computing first and second order
quantities, the mean, the variance and the autocovariance function.  This means
that a randomised sample can be obtained by creating sequences with the same
second order properties as the measured data, but which are otherwise
random. When the linear properties are specified by the squared amplitudes of
the Fourier transform (that is, the periodogram estimator of the power
spectrum), surrogate time series $\{s'_n\}$ are readily created by multiplying
the Fourier transform of the data by random phases and then transforming back
to the time domain:
\be\label{eq:ftsurro}
   s'_n=\sum_{k=0}^{N-1}e^{i\alpha_k}\sqrt{P_k} e^{-i2\pi kn/N}
\,,\ee
where $0\le \alpha_k<2\pi$ are independent uniform random numbers.

The two null hypotheses discussed so far (independent random numbers and
Gaussian linear processes) are not what we want to test against in most
realistic situations. In particular, the most obvious deviation from the
Gaussian linear process is usually that the data do not follow a Gaussian
single time probability distribution. This is quite obvious for data obtained
by measuring intervals between events, e.g. heart beats since intervals are
strictly positive. There is however a simple generalisation of the null
hypothesis that explains deviations from the normal distribution by the action
of a monotone, static measurement function: \be\label{eq:distort}
s_n=s(x_n),\quad x_n=\sum_{i=1}^M a_ix_{n-i} + \sum_{i=0}^N b_i\eta_{n-i}
\,.\ee We want to regard a time series from such a process as essentially
linear since the only nonlinearity is contained in the --- in principle
invertible --- measurement function $s(\cdot)$.

The most common method to create surrogate data sets for this null hypothesis
essentially attempts to invert $s(\cdot)$ by rescaling the time series
$\{s_n\}$ to conform with a Gaussian distribution. The rescaled version is
then phase randomised (conserving Gaussianity on average) and the result is
rescaled to the empirical distribution of $\{s_n\}$. These {\em amplitude
  adjusted Fourier transformed surrogates} (AAFT) yield a correct test when
$N$ is large, the correlation in the data is not too strong and $s(\cdot)$ is
close to the identity. It is argued in Ref.~\cite{surrowe} that for short and
strongly correlated sequences, the AAFT algorithm can yield an incorrect test
since it introduces a bias towards a slightly flatter spectrum. In fact, the
formal requirement the surrogates have to fulfill for the test to be correct
is that they have the same sample periodogram and the same single time sample
probability distribution as the data. Schreiber and Schmitz~\cite{surrowe}
propose a method which iteratively corrects deviations in spectrum and
distribution.  Alternatingly, the surrogate is filtered towards the correct
Fourier amplitudes and rank-ordered to the correct distribution. The accuracy
that can be reached depends on the size and structure of the data and is
generally sufficient for hypothesis testing.

The above schemes are all based on the Fourier amplitudes of the data, which is
however not exactly what we want. Remember that the autocorrelation structure
given by Eq.(\ref{eq:autocor}) corresponds to the Fourier amplitudes only
if the time series is one period of a sequence that repeats itself every $N$
time steps. Conserving the Fourier amplitudes of the data means that the {\em
periodic} autocovariance function
\be\label{eq:cp}
   C_p(\tau) = {1\over N}
      \sum_{n=1}^{N} x_n x_{(n-\tau-1)\,{\rm\scriptsize mod}\,N+1}
\ee
is reproduced, rather than $C(\tau)$. The difference can lead to serious
artefacts in the surrogates, and, consequently, spurious rejections in a test.
In particular, any mismatch between the beginning and the end of a time series
poses problems, as discussed e.g. in Ref.~\cite{theiler_sfi}. In spectral
estimation, problems caused by edge effects are dealt with by windowing and
zero padding. None of these have been successfully implemented for the phase
randomisation of surrogates.  The problem of non-matching ends can be partly
overcome by choosing a subinterval of the recording such that the end points do
match approximately.  Still, there may remain a finite phase slip at the
matching points.  The only method that has been proposed so far that strictly
implements $C(\tau)$ rather than $C_p(\tau)$ is given in
Ref.~\cite{anneal}. The method is very accurate but also rather costly in terms
of computer time. In practical situations, the matching of end points is
a simple and mostly sufficient precaution that should not be neglected.

Since the randomisation algorithm of Ref.~\cite{anneal} is of very general
applicability and conceptually quite simple, let us give a brief description.
In order to create randomised sequences with the correct distribution of
values, only permutations of the original time series are considered. The
shuffling is however carried out under the constraint that the autocovariances
of the surrogate $C'(\tau)$ are the same as those of the data, $C(\tau)$. This
is done by specifying the discrepancy as a cost function, e.g.
\be\label{eq:cost}
   E^{(q)}=\left[ \sum_{\tau=0}^{N-1} |C'(\tau)-C(\tau)|^q \right]^{1/q}
\,.\ee
The way the average over all lags $\tau$ is taken can be influenced by the
choice of $q$. Now $E^{(q)}(\{{\tilde s}_n\})$ is minimised among all
permutations $\{{\tilde s}_n\}$ of the original time series $\{s_n\}$ using the
method of simulated annealing.  Configurations are updated by exchanging pairs
in $\{{\tilde s}_n\}$. With an appropriate cooling scheme, the annealing
procedure can reach any desired accuracy. Simulated annealing has a rich
literature, classical references are Metropolis et al.~\cite{metro} and
Kirkpatrick~\cite{kirk}, more recent material can be found for example in
Vidal~\cite{annealbook}.

Constrained randomisation using combinatorial minimisation is a very flexible
method since in principle arbitrary constraints can be realised. Although it
is seldom possible to specify a formal null hypothesis for more general
constraints, it can be quite useful to be able to incorporate into the
surrogates any feature of the data that is understood already or that is
uninteresting. Let us give an example for the flexibility of the approach, a
simultaneous recording of the breath rate and the instantaneous heart rate of
a human subject during sleep. (Data set B of the Santa Fe Institute time
series contest in 1991~\cite{gold}, sample points 1800--4350.) Regarding the
heart rate recording on its own, one easily detects nonlinearity, in
particular via an asymmetry under time reversal. An interesting question
however is, how much of this structure can be explained by linear dependence
on the breath rate, the breath rate also being non-time-reversible. In order
to answer this question, we need to make surrogates that have the same
autocorrelation structure but also the same cross-correlation with respect to
the fixed input signal, the breath rate. Accordingly, the constraint is
imposed that lags 0,\ldots,500 of the auto-covariance function and lags
-500,\ldots,500 of the cross-covariance
function with the reference (breath) signal are given by the data.%
\footnote{Strictly speaking, these constraints over-specify the problem and it
  is likely that the only permutation that fulfills them exactly is the
  original time series itself. However, it can be expected that there are a
  large number of permutations which are essentially different and which
  fulfill the constraint {\em almost} exactly. In fact, it has been
  observed~\cite{anneal_unpub} for very short sequences of $N<50$ points and
  strong correlations that the annealing scheme settled on the original data.
  If this seems to happen, one can introduce a term in the cost function that
  discourages similarity to the original permutation.}
Further suppose that within the 20~minutes of observation, during one minute
the equipment spuriously recorded a constant value. In order not to interpret
this artefact as structure, the same artefact is imposed on the surrogates,
simply by excluding these data points from the permutation scheme.

Figure~\ref{fig:multi} shows the measured breath rate (upper trace) and
instantaneous heart rate (middle trace). The lower trace shows a surrogate
conserving both, auto- and cross-correlations. The visual impression from
Fig.~\ref{fig:multi} is that while the linear cross-correlation with the breath
rate explains the cyclic structure of the heart rate data, other features, in
particular the asymmetry under time reversal, remain unexplained. This finding
can be verified at the 95\% level of significance, using the time asymmetry
statistic given in Eq.(\ref{eq:skew}). Possible explanations include artefacts
due to the peculiar way of deriving heart rate from inter-beat intervals,
nonlinear coupling to the breath activity, nonlinearity in the cardiac system,
and others.

\begin{figure}
   \centerline{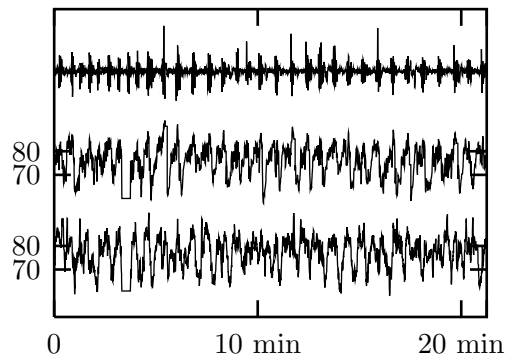}\vspace{10pt}
   \caption[]{\label{fig:multi} Simultaneous measurements of breath and heart
   rates~\cite{gold}, upper and middle traces. Lower trace: a surrogate heart
   rate series preserving the autocorrelation structure and the
   cross-correlation to the fixed breath rate series, as well as a gap in the
   data. Auto- and cross-correlation together seems to explain some, but not
   all of the structure present in the heart rate series.}
\end{figure}

Let us finish the section by giving a more exotic example, from finance. The
time series%
\footnote{The data were kindly provided by Thomas Sch\"urmann, WGZ-Bank
   D\"usseldorf}
c
consists of 1500 daily returns (until the end of 1996) of the {\em BUND
  Future}, a derived german financial instrument. As can be seen in the upper
panel of Fig.~\ref{fig:bund}, the sequence is nonstationary in the sense that
the local variance and also the local mean undergo changes on a time scale
that is long compared to the fluctuations of the series itself. This property
is known in the statistical literature as {\em heteroscedasticity} and modeled
by the so-called GARCH~\cite{garch} and related models. Here, we want to avoid
the construction of an explicit model from the data but rather ask the
question if the data is compatible with the null hypothesis of a correlated
linear stochastic process with time dependent local mean and variance. We can
answer this question in a statistical sense by creating surrogate time series
that show the same linear correlations and the same time dependence of the
running mean and running variance as the data and comparing a nonlinear
statistic between data and surrogates. The lower panel in Fig.~\ref{fig:bund}
shows a surrogate time series generated using the annealing method described
above.  The cost function was set up to match the autocorrelation function up
to five days and the moving mean and variance in sliding windows of 100 days
duration.  In Fig.~\ref{fig:bund} the running mean and variance are shown as
points and error bars, respectively, in the middle trace. The deviation of
these between data and surrogate has been minimised to such a degree that it
can no longer be resolved.  A comparison of the time-asymmetry statistic
Eq.(\ref{eq:skew}) for the data and 19 surrogates did not reveal any
discrepancy, and the null hypothesis could not be rejected.

\begin{figure}
   \centerline{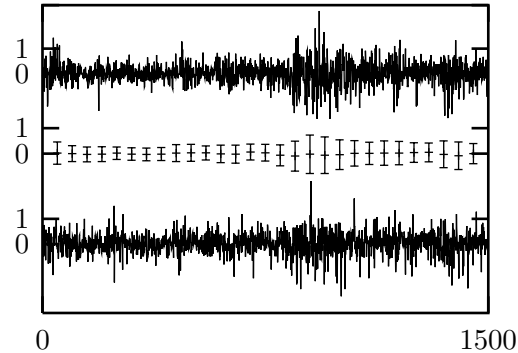}
   \caption[]{Nonstationary financial time series (BUND Future returns, top)
      and a surrogate (bottom) preserving the nonstationary structure
      quantified by running window estimates of the local mean and variance
      (middle).\label{fig:bund}}
\end{figure}

\subsection{What can be learned}
Having set up all the ingredients for a statistical hypothesis test of
nonlinearity, we may ask what we can learn from the outcome of such a test.
The formal answer is of course that we have, or have not, rejected a specific
hypothesis at a given level of significance. How interesting this information
is, however, depends on the null hypothesis we have chosen. The test is most
meaningful if the null hypothesis is plausible enough so that we are prepared
to believe it in the lack of evidence against it. If this is not the case, we
may be tempted to go beyond what is justified by the test in our
interpretation. Take as a simple example a recording of hormone concentration
in a human. We can test for the null hypothesis of a stationary Gaussian
linear random process by comparing the data to phase randomised Fourier
surrogates.  Without any test, we know that the hypothesis cannot be true
since hormone concentration, unlike Gaussian variates, is strictly
non-negative.  If we failed to reject the null hypothesis by a statistical
argument, we will therefore go ahead and reject it anyway by common sense, and
the test was pointless.  If we did reject the null hypothesis by finding a
coarse grained ``dimension'' which is significantly lower in the data than in
the surrogates, the result formally does not give any new information but we
might be tempted to speculate on the possible interpretation of the
``nonlinearity'' detected.

This example is maybe too obvious, it was meant only to illustrate that the
hypothesis we test against is often not what we would actually accept to be
true. Other, less obvious and more common, examples include signals which are
known (or found by inspection) to be nonstationary (which is not covered by
most null hypotheses), or signals which are likely to be the {\em squares} of
some fundamental quantity. An example for the latter are the celebrated
sunspot numbers. Sunspot activity is generally connected with magnetic fields
and is to first approximation proportional to the squared field strength.
Obviously, sunspot numbers are non-negative, but also the null hypothesis of a
monotonically rescaled Gaussian linear random process is to be rejected since
taking squares is not a monotonic operation. Unfortunately, the framework of
surrogate data does not currently provide a method to test against null
hypothesis involving noninvertible measurement functions. Yet another example
is given by linearly filtered time series.  Even if the null hypothesis of a
monotonically rescaled Gaussian linear random process is true for the
underlying signal, it is usually not true for filtered copies of it, in
particular sequences of first differences, see Prichard~\cite{dean} for a
discussion of this problem.

Recent efforts on the generalisation of randomisation schemes try to broaden
the repertoire of null hypotheses we can test against. The hope is that we can
eventually choose one that is general enough to be acceptable if we fail to
reject it with the methods we have. Still, we cannot prove that there is not
any structure in the data beyond what is covered by the null hypothesis.  From
a practical point of view, however, there is not much of a difference between
structure that is not there and structure that is undetectable with our
observational means.

\section{Nonlinear signal processing} \label{sec:process}
The goals of time series analysis are probably as diverse as the methods.  In
basic research, the ultimate aim is a deeper understanding of some phenomenon
in nature. In engineering, clinical research, finance etc., a better
understanding of the processes is also most welcome but the actual purpose of
the work is different, making better devices, making people healthier, making
money.  Pursuing such a goal often involves a rather specific task of time
series analysis. The interesting problem of signal classification will be dealt
with in Section~\ref{sec:classify}. One of the most well known objectives is
the prediction of future values of some quantity, for instance the price of an
entity at the stock market. The prediction problem has been discussed in
Sec.~\ref{sec:finite}. It is almost identical to the problem of
estimating the dynamics underlying a time series.  An intermediate step in most
time series studies is to filter the data in order to enhance the relevant
information. Prediction and filtering, or noise reduction, have many things in
common, but there are notable differences. In particular, for noise reduction
it is not enough to have a description of the dynamics. One also has to have a
means of finding a cleaner signal that is consistent with this dynamics.

\subsection{Nonlinear noise reduction}\label{sec:noise_reduction}
Originally, phase space methods of nonlinear noise reduction have been
developed~\cite{fsid1,ky2,sg,on} under the premise that there is a
low-dimensional dynamical system which is only observed with some observational
error that is to be suppressed. Conceptual as well as technical issues arising
in such a situation have been well discussed in the literature, see Kostelich
and Schreiber~\cite{noiserev} for a review containing the relevant
references. In interdisciplinary applications, we usually face a different
situation --- the signals themselves often contain a stochastic
component. Before we apply a filtering technique we therefore have to specify
what exactly we want to separate. Phase space projection techniques, like those
employed by Grassberger and coworkers~\cite{on}, rely on the assumption that
the signal of interest is approximately described by a manifold that has a
lower dimension than some phase space it is embedded in. This statement can be
formalised as follows. Let $\{\xx_n\}$ be the states of the system at times
$n=1,\ldots,N$, represented in some space ${\cal R}^d$.  A $(d-Q)$-dimensional
submanifold ${\cal F}$ of this space can be specified by $F_q(\yy)=0,\quad
q=1,\ldots,Q$. Even if $F_q$ is not known exactly, or if $\{\xx_n\}$ is
corrupted by noise, we can always find $\{\Eepsilon_n\}$ such that 
$\yy_n=\xx_n+\Eepsilon_n$ and
\be\label{eq:manifold} 
   F_q(\xx_n+\Eepsilon_n)=0, \qquad \forall q,n
\,.\ee 
Then $\sqrt{\av{\Eepsilon^2}}$ denotes the (root mean squared) average
error we make by approximating the points $\{\xx_n\}$ by the manifold ${\cal
  F}$. In a measurement we can only obtain noisy data $\yy_n=\xx_n+\Eeta_n$,
where $\{\Eeta_n\}$ is some random contamination. By projecting these values
onto some estimated manifold ${\cal F}$ we may be able to recover
$\xx'_n=\xx_n+\Eepsilon_n$. If we can find a suitable manifold --- and carry
out the projections --- such that $\av{\Eepsilon^2}<\av{\Eeta^2}$, then we
have reduced the observational error.  For dynamical systems embedded in delay
coordinate space there always exists a manifold for which $\Eepsilon_n\equiv0$,
but as we can see, noise reduction is possible as soon as the
$\Eepsilon_n$ are smaller than the observational error. Of course we will not
only reduce the magnitude of the errors but also alter their structure.
Therefore we will have to be careful (for example by statistically analysing
the corrections) when we are going to interpret the structure we find in the
corrected data.

Since the full, true phase space of a system is not usually accessible to time
series measurements, phase space filtering has to make heavy use of time delay
or related embedding techniques. Filtering implies that the signal is not pure
but a mixture of several components to be separated, and the use of the
embedding theorems for dynamical systems is limited. We will rather have to
take a pragmatic attitude. Please consult Sec.~\ref{sec:embed_finite} for
material and references on the embedding of finite noisy time series. 

In time series work, the most practical way to approximate data by a manifold
is by a locally linear representation. It should in principle be possible to
fit global nonlinear constraints $\hat{F}_q$ from data but the problem is
complicated by the necessity to have $Q$ locally independent equations. In the
locally linear case this is achieved by establishing local principal
components. The derivation will not be repeated here, it is carried out for
example in Refs.~\cite{ourbook,processing}. The resulting algorithm proceeds as
follows. In an embedding space of dimension $m$ we form delay vectors
$\Ss_n$. For each of these we construct small neighborhoods ${\cal U}_n$, so
that the neighbouring points are $\Ss_k, k\in{\cal U}_n$.  Within each
neighbourhood, we compute the local mean
\be
   \overline{\Ss}^{(n)}=\frac{1}{|{\cal U}_n|}\sum_{k\in {\cal U}_n} \Ss_k
\ee
and the $(m\times m)$ covariance matrix%
\footnote{It has been found advantageous~\cite{on} to introduce a diagonal
   weight matrix $R$ and define a transformed version of the covariance matrix
   $\Gamma_{ij}=R_{ii} C_{ij}R_{jj}$ for the calculation of the principal
   directions.  In order to penalise corrections based on the first and last
   coordinates in the delay window one puts $R_{00}=R_{mm}=r$ where $r$ is
   large.  The other values on the diagonal of $R$ are~1.}
\be
   C_{ij}=\frac{1}{|{\cal U}_n|}\sum_{k\in {\cal U}_n}
   (\Ss_k)_i(\Ss_k)_j-\overline{\Ss}^{(n)}_i\overline{\Ss}^{(n)}_j
\,.\ee
The eigenvectors of this matrix are the semi--axes of the best approximating
ellipsoid of this cloud of points (these are local versions of the well known
principal components, or singular vectors, see for example
Refs.~\cite{PC,BroomSVD}). If the clean data lives near a smooth manifold with
dimension $m_0<m$, and if the variance of the noise is sufficiently small for
the linearisation to be valid, then for the noisy data the covariance matrix
will have large eigenvalues spanning the smooth manifold and small eigenvalues
in all other directions.%
\footnote{For this to be valid, the neighbourhoods should be larger than the
noise level. In practice, a tradeoff between the clear definition of the noise 
directions and a good linear approximation has to be balanced.}
Therefore, we move the vector under consideration towards the manifold by
projecting onto the subspace of large eigenvectors.  The procedure is
illustrated in Fig.~\ref{fig:noise}. The correction is done for each embedding
vector, resulting in a set of corrected vectors in embedding space.  Since each
element of the scalar time series occurs in $m$ different embedding vectors, we
finally have as many different suggested corrections, of which we simply take
the average. Therefore in embedding space the corrected vectors do not
precisely lie on the local subspaces but are only moved towards them.

\begin{figure}
\centerline{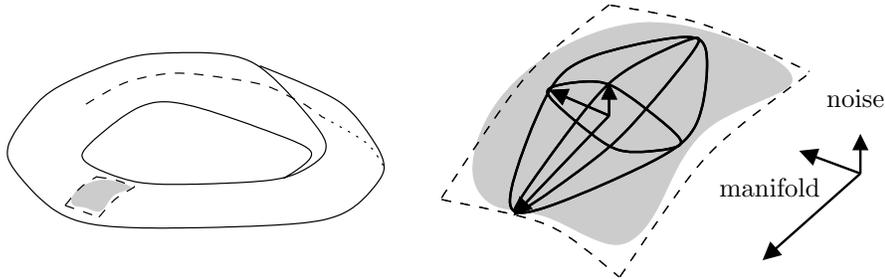}
\caption[]{Illustration of the local projection scheme. For each point to be
  corrected, a neighbourhood is formed (grey shaded area), the point cloud in
  which is then approximated by an ellipsoid. An approximately two-dimensional
  manifold embedded in a three-dimensional space could for example be cleaned
  by projecting onto the first two principal directions.\label{fig:noise}}
\end{figure}

\begin{figure}
   \centerline{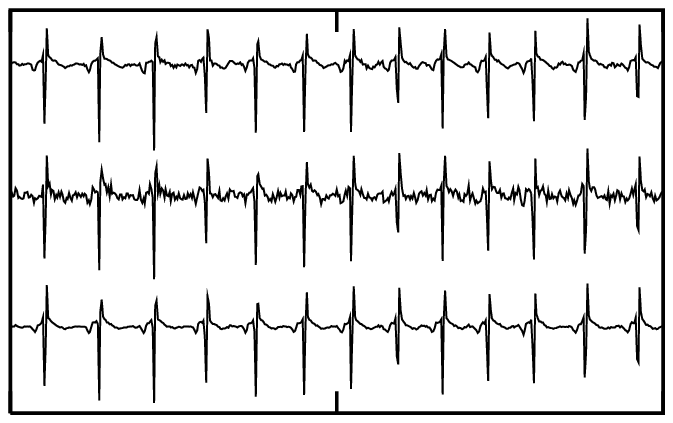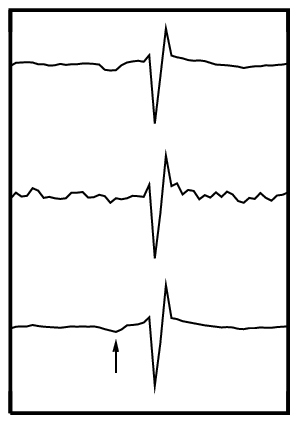}
   \caption[]{Nonlinear noise reduction applied to electrocardiogram
     data. Upper trace: original recording. Middle: the same contaminated with
     typical baseline noise. Lower: the same after nonlinear noise reduction.
     The enlargements on the right show that indeed clinically important
     features like the small downward deflection of the P-wave preceding the
     large QRS-complex (see for example Goldberger and
     Goldberger~\cite{goldbook} for an introduction to electrocardiography)
     are recovered by the procedure. Note that the noise and the signal have
     very similar spectral contents and could thus not be separated by
     Fourier methods.\label{fig:ecg}}
\end{figure}  
 
As an application, Fig.~\ref{fig:ecg} shows the result of the noise reduction
scheme applied to a noisy ECG. As discussed already in Sec.~\ref{sec:embedp}, a
delay coordinate embedding of an electrocardiogram seems to be well
approximated by a lower dimensional manifold. This is apparent already in a
two-dimensional representation (Fig.~\ref{fig:delay}), but for the purpose of
noise reduction, embeddings in higher dimensions are advantageous. The data
shown in Fig.~\ref{fig:ecg} was produced with delay windows covering 200~ms,
that is, $m=50$ at a delay time of 4~ms (equal to the sampling interval).  See
Ref.~\cite{ecg} for more details on the nonlinear projective filtering of ECG
signals. Applications of nonlinear noise reduction to chaotic laboratory data
are given in Ref.~\cite{case}. It should be noted that, as it stands, nonlinear
noise reduction is quite computer time intensive, in particular if compared to
Fourier based filters. For small and moderate noise levels, this can be
moderated by using fast neighbour search strategies, see for example
Ref.~\cite{neigh} for a review set in the context of time series
analysis. Recently, a fast version of nonlinear projective noise reduction has
been developed~\cite{stream} that cannot only be used a posteriori but also
for real time processing in a data stream.

If the data quality does not permit to use the local linear approach, one can
try to use locally constant approximations instead~\cite{lazy}. This is done
exactly in the same way as for locally constant predictions, see
Sec.~\ref{sec:finite}, Eq.(\ref{eq:fclazy}). The only difference is that
instead of predicting a future value with $\Delta n>0$, the middle coordinate
of the embedding window $\Delta n=m/2$ is estimated.

\subsection{Signal separation}
Noise reduction can be regarded as the particular case of the more general task
of signal separation where one of the signals is the noise contribution. It
turns out that the methodology developed for noise reduction can be generalised
to the separation of other types of signals. As a specific example, let us 
discuss the extraction of the fetal electrocardiogram (FECG) from non-invasive
maternal recordings. Other very similar applications include the removal of
ECG artefacts from electro-myogram (EMG) recordings (electric potentials of
muscle) and spike detection in electro-encephalogram (EEG) data~\cite{unpub}.

Fetal ECG extraction can be regarded as a three-way filtering problem since we
have to assume that a maternal abdominal ECG recording consists of three main
components, the maternal ECG, the fetal ECG, and exogenous noise, mostly from
action potentials of intervening muscle tissue. All three components have
quite similar broad band power spectra and cannot be filtered apart by
spectral methods. The fetal component is detectable from as early as the
eleventh week of pregnancy. After about the twentieth week, the signal becomes
weaker since the electric potential of the fetal heart is shielded by the {\em
  vernix caseosa} forming on the skin of the fetus. It appears again towards
delivery. In Refs.~\cite{fetal1,fetal2}, it has been proposed to use a
nonlinear phase space projection technique for the separation of the fetal
signal from maternal and noise artefacts. A typical example of output of this
procedure is shown in Fig.~\ref{fig:fecg}. The assumption made about the
nature of the data is that the maternal signal is well approximated by a
low-dimensional manifold in delay reconstruction space. After projection onto
this manifold, the maternal signal is separated from the noisy fetal
component. Now it is assumed that the fetal ECG is also approximated by a
low-dimensional manifold and the noise is removed by projection. Since both
manifolds are curved, the projections have to be made onto linear
approximations. For technical details see Refs.~\cite{fetal1,fetal2}.
\begin{figure}
   \centerline{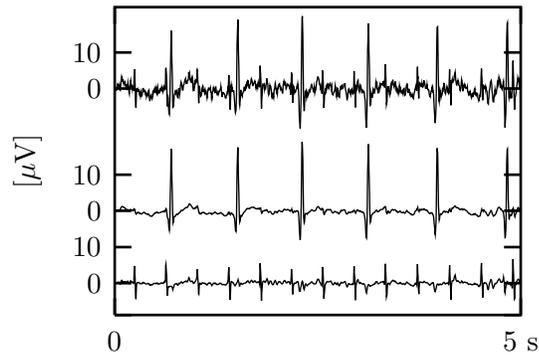}
   \caption[]{Signal separation by locally linear projections in phase space.
   The original recording (upper trace) contains the fetal ECG hidden under
   noise and the large maternal signal. Projection onto the manifold formed
   by the maternal ECG (middle) yields fetus plus noise, another projection
   yields a fairly clean fetal ECG (lower trace). The data was kindly provided
   by J.~F. Hofmeister~\cite{recording}.\label{fig:fecg}}
\end{figure}  
 
\section{Comparison and classification} \label{sec:classify}
With current methods, many real world systems cannot be fully understood on the
basis of time series measurements. Approaches aiming at an absolute analysis,
like the reliable determination of the fractal dimension of a strange attractor
have often been found to fail for various reasons, the most prominent being
that most of the systems are not low-dimensional deterministic. However, many
phenomena can still be studied in a comparative way. In that case, we do not
have to worry too much about the theoretical basis of the quantities we
use. The results are validated by the statistical significance for the
discriminative power. The classification of states can give valuable insights
into the structure of a problem, and very often, signal classification is
desirable in its own right. In clinical applications, for example, it is common
to define quantities by a standardised procedure, even if this procedure yields
an observable which has no immediate physical interpretation. Take the standard
procedure of determining the blood pressure non-invasively (the {\em
Riva-Rocci} method).  Although the measurement is indirect and does not yield
invariant results, the standardisation of the procedure ensures good
comparability of the results.  For its value as a diagnostic tool, it is
irrelevant whether the measured numbers actually represent the pressure of the
blood in a specific part of the body or not.

Comparison and classification of time series is most often done pretty much in
the same spirit as, for example, the blood pressure measurement. A complex
phenomenon is reduced in a well defined way to a single number or a small set
of numbers. Further analysis is then carried out on these numbers. In nonlinear
time series analysis these numbers can be for example nonlinear prediction
errors, coarse grained dimensions or entropies, etc. Below, an alternative
approach will be discussed which attempts to carry out the actual comparison
between the signals directly rather than between single numbers abstracted from
them.

Before we set up a classification problem, we have to decide how we want to use
the available information. We need to make economical use of the data we have
since we need them to set up and maybe optimise a classification scheme and
then to independently verify it. One way to proceed is by splitting the
available data base into two parts, a learning set where the correct
classification is known (for example, which of the patients were in the control
group), and a test set which is only used at the very end of the analysis to
verify the validity of the classification without using the correct
answer. (See the discussion of the overfitting problem in
Sec.~\ref{sec:finite}.) The advantage of this approach is that the training
phase can be supervised and directed to the desired behaviour. The disadvantage
is that we need a sufficient number of test cases to be kept apart. Another
possibility is to perform unsupervised classification of the whole data
base. One tries to find whether the whole ensemble of signals falls into
distinct groups naturally without help from a supervisor who knows the correct
answer. One can then regard the whole available data base as the test set for
the correctness and significance of the grouping. The latter approach of {\em
unsupervised learning} will be considered here mostly. Note that in both cases,
supervised and unsupervised classification, the test set cannot be used
repeatedly to optimise strategies or parameters, unless claims of significance
are modified accordingly.

\subsection{Classification by histograms}
The standard approach to classification of time series is to express the
information in each sequence by a single number or a few of them. One can
then form a histogram of these numbers, either in one or a few dimensions.
Let us give a simple example. Consider a generalised baker map
\be
  \begin{array}{lllllll}
      v_n\leq\alpha:    & u_{n+1}       & =     & \beta u_n, &
              v_{n+1}   & =     & v_n/\alpha \\
      v_n>\alpha:       & u_{n+1}       & =     & 0.5+\beta u_n,  &
              v_{n+1}   & =     & (v_n-\alpha) / (1-\alpha)
   \end{array}
\ee
with $\alpha=0.4$.  The parameter $\beta$ can be varied without changing the
positive Lyapunov exponent~\cite{Schuster}. Two groups of sequences were
generated, the first (group a) containing 50 sets with $\beta=0.6$ and the
second (group b) containing 50 with $\beta=0.8$. Each sequence has a length of
400 points. Prediction errors $\gamma_i$ are calculated from
Eqs.(\ref{eq:pred},\ref{eq:fclazy}) and collected in a histogram (see
Fig.~\ref{fig:histo}). For unsupervised learning, this histogram would not
provide enough information since the observed distribution (black bars) does
not fall into two groups naturally. In fact, in the best case we can find a
threshold value (arrow) that minimises the number of misclassifications. Still
at least 20 series will be assigned to the wrong group since the individual
distributions (white and grey bars) overlap.

\begin{figure}
\centerline{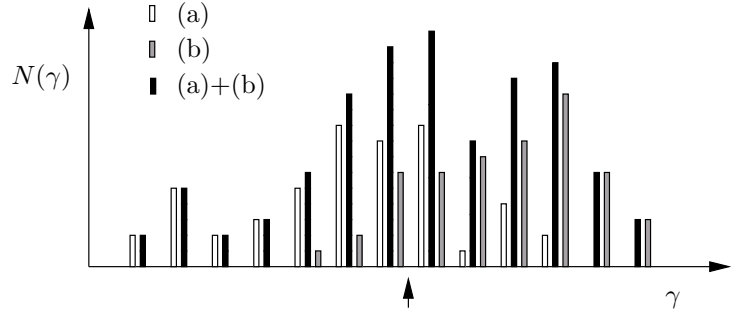}
\caption[]{Classification using a histogram. The arrow indicates the threshold
   value resulting in the fewest misclassifications possible.\label{fig:histo}}
\end{figure}

\subsection{Classification by clustering}\label{sec:clust}
The success of the usual classification schemes based on histograms or scatter
plots of a few quantities crucially depends on the right choice of observable
to characterise the signals. In general, it seems quite a loss of information
to express nonlinear dynamics by a few numbers. An alternative is to compare
the individual series directly without first extracting an observable. As we
will see in the baker map example of the preceding section, two series
produced with different parameters may be well distinguishable even though they
have comparable predictability. This motivates to generalise the usual measures
of nonlinearity discussed in Secs.~\ref{sec:quant}, \ref{sec:invar}, and
\ref{sec:noninvar} to comparative measures, or measures of similarity, as it
was done in Secs.~\ref{sec:compare} and~\ref{sec:dissim}. If we want to compare
$w$ signals, the study of symmetric dissimilarities yields $w(w+1)/2$
independent relative quantities $\gamma_{i,j}$ rather than just $w$
characteristics $\gamma_i=\gamma_{i,i}$. In this section it will be shown how
such matrices can be obtained and used for the task of classification. The
method has been proposed in Ref.~\cite{clust} where also more technical details
and further examples can be found.

The main idea is to use a cluster algorithm to find groups of data based on a
dissimilarity matrix. There are several standard methods to do so and the
choice made below is not meant to be exclusive by any means.  The task now is
to classify $w$ objects into $K$ groups or clusters.  Let us define a
membership index $u_i^{(\nu)}$ to be 1 if object $i$ is in cluster $\nu$, and 0
if not.  A cluster is given by all points with membership index 1:
$C^{(\nu)}=\{i: u_i^{(\nu)}=1\}$. The size of a cluster is
$|C^{(\nu)}|=\sum_{i=1}^w u_i^{(\nu)}$.  The average dissimilarity of object
$i$ to cluster $\nu$ (the ``distance'' of $i$ to $\nu$) is then given by
\be\label{eq:dis_i_nu}
   D_i^{(\nu)} = {1\over |C^{(\nu)}|} \sum_{j=1}^w u_j^{(\nu)} \gamma_{ij}
\,.\ee
The average dissimilarity within cluster $\nu$ is
\be \label{eq:dis_nu_nu}
   D^{(\nu)}={1\over|C^{(\nu)}|}\sum_{i=1}^w u_i^{(\nu)} D_i^{(\nu)}
\ee
and the total average intra-cluster dissimilarity:
\be
   D={1\over K}\sum_{\nu=1}^K |C^{(\nu)}|\; D^{(\nu)}
\,.\ee
This finally yields the cost function
\be
   E = K D = \sum_{\nu=1}^K {1\over |C^{(\nu)}|}
       \sum_{i,j=1}^w u_i^{(\nu)} u_j^{(\nu)} \gamma_{ij}
\ee
that quantifies the average distance within the clusters. The cost function $E$
can be minimised numerically, for example with {\em simulated annealing} (see
Ref.~\cite{clust} for details).

Let us illustrate the use of this approach with the same example studied
previously, the collection of generalised baker map data. Equation
(\ref{eq:xpred}) generalises the prediction error $\gamma_i$ used for the
histogram approach to a symmetrised cross-prediction error $\gamma_{ij}$.  With
this, two clusters are formed by minimizing~$E$.  In Fig.~\ref{fig:baker}, for
each series the average dissimilarity $D_i^{(2)}$ to cluster~2 is plotted
against that to cluster~1 ($D_i^{(1)}$). Two distinct groups can easily be seen
that coincide perfectly with the correct classes. Indeed the algorithm forms
exactly the two desired clusters.
\begin{figure}
   \centerline{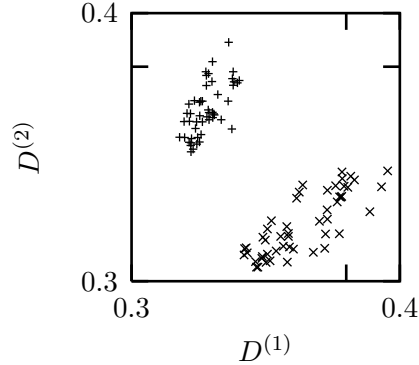}
   \caption[]{Distances of objects from two clusters generated for 100 time
   series in two groups of 50. Baker map with $\beta=0.6$ in the first group
   ({\scriptsize +}) and $\beta=0.8$ in the second 
   ({\scriptsize $\times$}). The two original groups are
   readily separated by the cluster algorithm without any misclassification, as
   compared to more than 20 mistakes using a histogram of prediction errors.
   \label{fig:baker}}
\end{figure}

Of course, to some extent the problem has only been shifted from finding a
magic characteristic number to finding a suitable (not much less magic)
dissimilarity measure. However, the approach augments the set of available
tools in a meaningful way. After all, we want to classify signals by their
dynamics, a feature that is not usually well described by a few parameters. For
any classification method, the major problem remains to separate those
differences that are relevant for the discrimination task from those that are
not. Sometimes, the calibration of the measurement is known to be of no
significance, in which case we can subtract the mean and rescale to unit
variance. But apart from such simple transformations we have so far little
means of being selective in a controlled way.  It is possible to exclude the
linear correlation structure from the analysis by normalising characteristic
parameters to values obtained with surrogate data.

\section{Conclusion and future perspectives} \label{sec:perspectives}
This review paper tries to give an impression on how useful time
series methods from chaos theory can be in practical applications.
Chaos theory has attracted researchers from many areas for various
reasons, in particular, because of its ability to explain complicated temporal
behaviour by equations with only a few degrees of freedom and without assuming
random forcing to act on the system. The attractiveness of the new paradigm
(or the desperation in fields where standard time series methods fail
miserably) has tempted people to take several steps at a time, and high
expectations have been raised.  Only now the path is being retraced step by
step. Starting from a theoretical understanding of the new class of systems,
time series methods have been tested on computer generated and well controlled
laboratory data. Some of these studies provided sobering experience as to how
fragile chaos and fractals can be and one could now be tempted to become quite
pessimistic about the usefulness of algorithms derived from these concepts.

Now that procedures have been revised, limitations and pitfalls have been
pointed out, and intuition has been gained, we can again try to expose the
algorithms to field measurements. We will do it less naively than people have
done previously --- but also with much more modest expectations.  The aim of
this paper is to get away from naive enthusiasm, but also from a roundabout
abandonment of the approach. Realistic applications will only be possible with
a pragmatic attitude --- what can we learn from the new methods even if the
assumption of determinism is not really valid for the system we study?

The obvious goal for the near future is thus to enlarge the class of time
series problems that are only, or better, or more efficiently, solvable by the
nonlinear approach. The main obstacle will probably not be the lack of good
quality data. Experimentalists have come a long way towards controlling devices
and measurement apparatuses. The major challenge lies in the nature of the
systems that are most interesting. Many outstanding time series problems in the
bio- geo- and social sciences involve multiple time scales, or put differently,
lead to nonstationary signals. Also, natural systems are never isolated and
thus are of a mixed nature, containing intrinsic and external dynamical
components.

There has been a noticable shift of focus in recent research from the mere
testing for nonstationarity and, if the result is positive, excluding a time
series from study, to the development of tools to understand the nature of the
changes in a system's dynamics. If a faithful parametric or empirical model for
the process is lacking, there is no obvious set of parameters whose changes
could be monitored. A promising approach is to define a basis to descibe
changes in the dynamics either by a number of reference sets~\cite{B1} or a
number of clusters of dynamically similar reference states
(Refs.~\cite{clust,statio}, Secs.~\ref{sec:trace} and~\ref{sec:classify}).  The
distances, or dissimilarities, of the dynamics at a given time to these refence
dynamical states constitutes then a natural set of parameters. Further work on
the problem of how to quantify the similarity of dynamical states and on how to
use that information could be rewarding.  It may then be possible to answer
questions about the number of time-dependent parameters and the time sacles and
nature of these variations.

Systems with many degrees of freedom are notoriously difficult to study trough
time series, even if multiple recordings are available. As an extreme example
take the dynamics of the human brain. The neurons and synapses are not only
enormous in number but they are also highly connected. The connection structure
can moreover change slowly with time. The system is quite inhomogeneous and has
to carry out many different tasks at different times. Certainly, only very
specific questions can be hoped to be answered on the basis of time series
recordings with a few channels. But even time-resolved imaging techniques can
only give a coarsened picture and do not adequately represent the connection
structure. Fortunately there are some interesting intermediate problems that
carry more promise to be tractable with dynamical methods. Spatial homogeneity
and local coupling leads to a class of systems which can show interesting but
still understandable dynamics. Apart from steady states and static patterns,
they can exhibit phenomena which are summarised under the term {\em weak
turbulence}. Neither this term, nor the notions of {\em spatio-temporal chaos}
or, more fashionably, {\em extensive chaos} have been clearly defined so
far. This is a direct consequence of the lack of a unifying framework for the
study of these systems. Extensivity in this context means that quantities like
the number of degrees of freedom, attractor dimension, entropy, etc.,
asymptotically grow linearly with the volume of the system. In the large system
limit, one can then define {\em intensive} quantities like a {\em dimension
density}.

This paper is certainly not the place to review the huge literature on
nonlinear, spatially extended dynamics, in particular since few of the
approaches have been shown to be useful when analysing observational data. The
reader may find interesting material and additional pointers to the literature
in the proceedings volumes Refs.~\cite{Busse,Bryn}, as well as in
Manneville~\cite{manneville}, Kapral and Showalter~\cite{Kapral}, and Cross and
Hohenberg~\cite{cross}. Probably the most immediate problem when analysing
spatio-temporal data is the choice of a useful representation of the system
states and dynamics. A high frequency sequence of images contains an amount of
information that is hardly managable, even with a powerful computer. The other
extreme, a small number of local probes, causes severe problems since the time
delay embedding technique is of very limited use with high dimensional
data~\cite{olbrich}.  Popular schemes to reduce the spatial information to a
few modes (like for example the Karhunen-Lo\'eve decomposition, see
Ref.~\cite{Zoldi1} for a recent application) are most often linear in nature
and therefore not quite appropriate for nonlinear systems. In certain
situations~\cite{Haken}, nonlinear mode dynamics have been used successfully to
describe spatio-temporal phenomena. A different approach that carries promise
in this respect is the representation by temporally periodic recurrent
patterns, or unstable periodic orbits, see the works by Christiansen and
coworkers~\cite{Hopf}, and by Zoldi and Greenside~\cite{Zoldi2}.  Despite these
efforts, the expectation raised in 1991 at the Santa Fe Institue Time Series
Contest that within five years we may have enough experience to enter a second
contest, this time on spatio-temporal data, has not been substantiated and it
seems that more than a slight relaxation of the time frame will be necessary.

High dimensional signals can also be produced by systems with only a few
components when a delayed feedback is involved. In biology, delayed feedback
loops are quite common due to the retarded response of subsystems to changes in
other parts of the system. In other fields, delayed feedback can be realised
for example when part of the output of a device is reflected back from a finite
distance, as it sometimes happens in laser or radar equipment, but also with
seismic waves. Delayed feedback is also often used for the control of chaotic
systems. There has been recent progress in the analysis of such systems, in
particular if some knowledge about the feedback structure is
available. B\"unner and coworkers~\cite{Buenner} have been able to extract
relatively simple dynamical equations from scalar time delay systems on the
basis of time series data, despite the high dimensionality of the signals. It
should be possible in principle also to infer the delay structure from
observations. The recovery of dynamical equations from data could then provide
a better understanding of many systems in nature.

\section*{Acknowledgments}
Let me first thank Peter Grassberger who has accompanied my work since I
started doing science. My work on nonlinear time series has again and again
led to close and enjoyable collaboration with Holger Kantz.  Among the people
who had impact on the research leading to this paper let me name James
Theiler, Daniel Kaplan, Leon Glass, Martin Casdagli, Tim Sauer, Rainer Hegger,
and Lenny Smith. I am grateful to Petr Saparin, John F. Hofmeister, Klaus
Lehnertz, and Thomas Sch\"urmann for letting me use their time series data in
this publication.  This work was supported by the SFB 237 of the Deutsche
Forschungsgemeinschaft.

\BIB
\addcontentsline{toc}{section}{References}
\bibitem{gss}
P. Grassberger, T. Schreiber, and C. Schaffrath,
   Nonlinear time sequence analysis,
   Int.\ J.\ Bifurcation and Chaos 1 (1991) 521.

\bibitem{abarbanel}
H.~D.~I. Abarbanel, R. Brown, J.~J. Sidorowich, and L.~Sh. Tsimring,
   The analysis of observed chaotic data in physical systems,
   Rev.\ Mod.\ Phys.\ 65 (1993) 1331.

\bibitem{kugiumtzis_rev1}
D. Kugiumtzis, B. Lillekjendlie, N. Christophersen,
   Chaotic time series I,
   Modeling, Identification and Control 15 (1994) 205.

\bibitem{kugiumtzis_rev2}
D. Kugiumtzis, B. Lillekjendlie, N. Christophersen,
   Chaotic time series II,
   Modeling, Identification and Control 15 (1994) 225.

\bibitem{coping}
E. Ott, T. Sauer, and J.~A. Yorke,
   Coping with chaos,
   Wiley, New York, 1994.

\bibitem{abarbook}
H.~D.~I. Abarbanel,
   Analysis of observed chaotic data,
   Springer, New York, 1996.

\bibitem{ourbook}
H. Kantz and T. Schreiber,
   Nonlinear time series analysis,
   Cambridge University Press, Cambridge, 1997.

\bibitem{Mayer-Kress}
G. Mayer-Kress, ed.,
   Dimensions and entropies in chaotic systems,
   Springer, Berlin, 1986.

\bibitem{casdagli}
M. Casdagli and S. Eubank, eds.,
   Nonlinear modeling and forecasting,
   Santa Fe Institute Studies in the Science of Complexity, Proc.~Vol.~XII,
   Addison-Wesley, Reading, MA, 1992.

\bibitem{SFI}
A.~S. Weigend and N.~A. Gershenfeld, eds.,
   Time series prediction: Forecasting the future and understanding the past,
   Santa Fe Institute Studies in the Science of Complexity, Proc.~Vol.~XV,
   Addison-Wesley, Reading, MA, 1993.

\bibitem{dyndis}
J. B\'elair, L. Glass, U. an der Heiden, and J. Milton, eds.,
   Dynamical disease,
   AIP Press, 1995.

\bibitem{freital}
H. Kantz, J. Kurths, and G. Mayer-Kress, eds.,
   Nonlinear techniques in physiological time series analysis,
   Springer series in synergetics, Springer, Heidelberg, 1998.

\bibitem{tong}
H. Tong,
   Non-linear time series analysis,
   Oxford University Press, Oxford, 1990.

\bibitem{priestley}
M.~B. Priestley,
   Non-linear and non-stationary time series analysis,
   Academic Press, London, 1988.

\bibitem{nico}
C. Nicolis and G. Nicolis,
   Is there a climatic attractor?
   Nature 326 (1987) 523.

\bibitem{fraedrich}
K. Fraedrich,
   Estimating the dimensions of weather and climate attractors,
   J.\ Atmos.\ Sci.\ 43 (1986) 419.

\bibitem{essex}
C. Essex, T. Lockman, and M.~A.~H. Nerenberg,
   The climate attractor over short time scales,
   Nature 326 (1987) 64.

\bibitem{hsieh}
D.~A. Hsieh,
   Chaos and nonlinear dynamics: Applications to financial markets,
   J.\ Finance 46 (1991) 1839.

\bibitem{peters}
E.~E. Peters,
   A chaotic attractor for the s+p 500,
   Financial Analysts J.\ 3 (1991) 55.

\bibitem{decoster}
G. DeCoster, W. Labys, and D. Mitchell,
   Evidence of chaos in commodity future prices,
   J.\ Futures Markets 12 (1992) 291.

\bibitem{goldberger}
A.~L. Goldberger, D.~R. Rigney, J. Mietus, E.~M. Antman, and S. Greenwald,
   Nonlinear dynamics in sudden cardiac death syndrome: Heart rate 
   oscillations and bifurcations,
   Experientia 44 (1988) 983.

\bibitem{chialvo}
D.~R. Chialvo and J. Jalife,
   Non-linear dynamics in cardiac excitation and impulse propagation,
   Nature 330 (1987) 749.

\bibitem{bablo}
A. Babloyantz,
   Strange attractors in the dynamics of brain activity,
   H. Haken, ed., Complex systems,
   Springer, Berlin, 1985.

\bibitem{prit}
W.~S. Pritchard,
   Electroencephalographic effects of cigarette smoking,
   Psychopharmacology 104 (1991) 485.

\bibitem{tisean}
T. Schreiber, R. Hegger, and H. Kantz,
   Practical implementation of nonlinear time series methods,
   to be published, 1998.

\bibitem{Ott}
E. Ott,
   Chaos in dynamical systems,
   Cambridge University Press, Cambridge, 1993.

\bibitem{Berge}
P. Berg\'e, Y. Pomeau, and C. Vidal,
   Order within chaos: Towards a deterministic approach to turbulence,
   Wiley, New York, 1986.

\bibitem{Schuster}
H.-G. Schuster,
   Deterministic chaos: An introduction,
   Physik Verlag, Weinheim, 1988.

\bibitem{KatokHasselblatt}
A. Katok and B. Hasselblatt,
   Introduction to the modern theory of dynamical systems,
   Cambridge University Press, Cambridge, 1996.

\bibitem{KaplanGlass}
D. Kaplan and L. Glass,
   Understanding nonlinear dynamics,
   Springer, New York, 1995.

\bibitem{tsonis}
A.~A. Tsonis,
   Chaos: From theory to applications,
   Plenum, New York, 1992.

\bibitem{lop2}
T. Schreiber and H. Kantz,
   Noise in chaotic data: Diagnosis and treatment,
   CHAOS 5 (1995) 133; Reprinted in~\cite{dyndis}.

\bibitem{jk}
L. Jaeger and H. Kantz, 
   Effective deterministic models for chaotic dynamics perturbed by noise,
   Phys.\ Rev.\ E 55 (1997) 5234.

\bibitem{fsid0}
J.~D. Farmer and J. Sidorowich,
   Predicting chaotic time series,
   Phys.\ Rev.\ Lett.\ 59 (1987) 845; Reprinted in~\cite{coping}.

\bibitem{Casdagli_pred}
M. Casdagli,
   Nonlinear prediction of chaotic time series,
   Physica D 35 (1989) 335; Reprinted in~\cite{coping}.

\bibitem{kost} % errors in variables
E.~J. Kostelich,
   Problems in estimating dynamics from data,
   Physica D 58 (1992) 138.

\bibitem{jaeger}
L. Jaeger and H. Kantz,
   Unbiased reconstruction underlying a noisy chaotic time series,
   CHAOS 6 (1996) 440.

\bibitem{brown_sync}
R. Brown, E.~R. Rulkov, and N.~F. Tracy,
   Modeling and synchronizing chaotic systems from time-series data,
   Phys.\ Rev.\ E 49 (1994) 3784.

\bibitem{Takens}
F. Takens,
   Detecting strange attractors in turbulence,
   in D.~A. Rand and L.-S. Young, eds.,
   Dynamical systems and turbulence,
   Lecture notes in mathematics Vol.~898, Springer, New York, 1981.

\bibitem{embed}
T. Sauer, J. Yorke, and M. Casdagli,
   Embedology,
   J.\ Stat.\ Phys.\ 65 (1991) 579.

\bibitem{stark}
J. Stark, D.S. Broomhead, M.E. Davies, and J.Huke,
   Takens embedding theorems for forced and stochastic systems,
   Proc. 2nd World Congress of Nonlinear Analysts,
   Athens, Greece, to appear, 1996.

\bibitem{saueryorke}
T. Sauer and J. Yorke,
   How many delay coordinates do you need?
   Int.\ J.\ Bifurcation and Chaos 3 (1993) 737.

\bibitem{numrec}
W.~H. Press, B.~P. Flannery, S.~A. Teukolsky, and W.~T. Vetterling,
   Numerical recipes, 2nd edition,
   Cambridge University Press, Cambridge, 1992.

\bibitem{rosenstein}
M.~T. Rosenstein, J.~J. Collins, C.~J. De~Luca,
   Reconstruction expansion as a geometry-based framework for choosing proper 
   delay times,
   Physica D 65 (1993) 117.

\bibitem{Holger} % lyap
H. Kantz,
   A robust method to estimate the maximal Lyapunov exponent of a time series,
   Phys.\ Lett.\ A 185 (1994) 77.

\bibitem{Eckmann}
J.-P. Eckmann, S. Oliffson Kamphorst, D. Ruelle, and S. Ciliberto,
   Lyapunov exponents from a time series,
   Phys.\ Rev.~A 34 (1986) 4971; Reprinted in~\cite{coping}.

\bibitem{sano}
M. Sano and Y. Sawada,
   Measurement of the Lyapunov spectrum from a chaotic time series,
   Phys.\ Rev.\ Lett.\ 55 (1985) 1082.

\bibitem{Ruedi1}
R. Stoop and P.~F. Meier,
   Evaluation of Lyapunov exponents and scaling functions from time series,
   J.\ Opt.\ Soc.\ Am.\ B 5 (1988) 1037.

\bibitem{Ruedi2}
R. Stoop and J. Parisi,
   Calculation of Lyapunov exponents avoiding spurious elements,
   Physica D 50 (1991) 89.

\bibitem{abar_locallyap}
H.~D.~I. Abarbanel, R. Brown, and M.~B. Kennel,
   Variation of Lyapunov exponents on a strange attractor,
   J.\ Nonlinear Sci.\ 1 (1991) 175.

\bibitem{lennylyap}
L. A. Smith,
   Local optimal prediction: exploiting strangeness and the variation of
   sensitiy to initial condition,
   Philos.\ Trans.\ Roy.\ Soc.\ A 348 (1994) 371.

\bibitem{abar_locallyap2}
H.~D.~I. Abarbanel, R. Brown, and M.~B. Kennel,
   Local Lyapunov exponents from observed data,
   J.\ Nonlinear Sci.\ 2 (1992) 343.

\bibitem{bruno_locallyap}
B. Eckhardt and D. Yao,
   Local Lyapunov exponents in chaotic systems,
   Physica D 65 (1993) 100.

\bibitem{Ellner}
B.~A. Bailey, S. Ellner, and D.~W. Nychka,
   Chaos with confidence: Asymptotics and applications of local Lyapunov
   exponents,
   Fields Inst.\ Comm.\ 11 (1997) 115.

\bibitem{GP}
P. Grassberger and I. Procaccia,
   Measuring the strangeness of strange attractors,
   Physica D 9 (1983) 189.

\bibitem{frank}
M. Frank, H.-R. Blank, J. Heindl, M. Kaltenh{\"a}user, H. K{\"o}chner,
N. M{\"u}ller, S. Pocher, R. Sporer, and T. Wagner,
   Improvement of $K_2$-entropy calculations by means of dimension scaled
   distances,
   Physica D 65 (1993) 359.

\bibitem{dimi_norm}
D. Kugiumtzis,
   Assessing different norms in nonlinear analysis of noisy time series,
   Physica D 105 (1997) 62.

\bibitem{pt}
D. Prichard and J. Theiler,
   Generalized redundancies for time series analysis,
   Physica D 84 (1995) 476.

\bibitem{CP}
A. Cohen and I. Procaccia,
   Computing the {K}olmogorov entropy from time signals of dissipative and
   conservative dynamical systems, 
   Phys.\ Rev.\ A 31 (1985) 1872.

\bibitem{ghez1}
J.~M. Ghez and S. Vaienti,
   Integrated wavelets on fractal sets I: The correlation dimension,
   Nonlinearity 5 (1992) 777.

\bibitem{ghez2}
J.~M. Ghez and S. Vaienti,
   Integrated wavelets on fractal sets II: The generalized dimensions,
   Nonlinearity 5 (1992) 791.

\bibitem{EEG}
   J.~L. Hern\'andez, R. Biscay, J.~C. Jimenez, P. Valdes, and R. Grave de
   Peralta,
   Measuring the dissimilarity between EEG recordings through a non-linear
   dynamical system approach,
   Int.\ J.\ Bio-Med.\ Comp.\ 38 (1995) 121.

\bibitem{clust}
T. Schreiber and A. Schmitz,
   Classification of time series data with nonlinear similarity measures,
   Phys.\ Rev.\ Lett.\ 79 (1997) 1475.

\bibitem{B1}
R. Manuca and R. Savit,
   Stationarity and nonstationarity in time series analysis,
   Physica D 99 (1996) 134.

\bibitem{casEEG}
M.~C. Casdagli, L.~D. Iasemidis, R.~S. Savit, R.~L. Gilmore, S. Roper, and
   J.~C. Sackellares,
   Non-linearity in invasive EEG recordings from patients with temporal
   lobe epilepsy,
   Electroencephalogr.\ Clin.\ Neurophysiol.\ 102 (1997) 98.

\bibitem{statio}
T. Schreiber,
   Detecting and analysing nonstationarity in a time series using
   nonlinear cross predictions,
   Phys.\ Rev.\ Lett.\ 78 (1997) 843.

\bibitem{lou}
L.~M. Pecora, T.~L. Carroll, and J.~F. Heagy,
   Statistics for mathematical properties of maps between time series
   embeddings,
   Phys.\ Rev.\ E 52 (1995) 3420.

\bibitem{koc1} % Generalized synchronization
L. Kocarev and U. Parlitz,
   General approach for chaotic synchronization with applications to
   communication,
   Phys.\ Rev.\ Lett.\ 74 (1995) 5028.

\bibitem{koc2} % Generalized synchronization
L. Kocarev and U. Parlitz,
   Generalized synchronization, predictability, and equivalence of
   unidirectionally coupled dynamical systems,
   Phys.\ Rev.\ Lett.\ 76 (1996) 1816.

\bibitem{rulkov}
N.~F. Rulkov, M.~M. Sushchik, L.~S. Tsimring, and H.~D.~I. Abarbanel,
   Generalized synchronization of chaos in directionally coupled chaotic
   systems,
   Phys.\ Rev.\ E 51 (1995) 980.

\bibitem{rosenblum}
M.~G. Rosenblum, A.~S. Pikovsky, and J. Kurths,
   Phase synchronisation of chaotic attractors,
   Phys.\ Rev.\ Lett. 76 (1996) 1804.

\bibitem{Kull}
S. Kullback,
   Information theory and statistics,
   Wiley, New York, 1959.

\bibitem{holger}
H. Kantz,
   Quantifying the closeness of factal measures,
   Phys.\ Rev.\ E 49 (1994) 5091.

\bibitem{diks_dist}
C. Diks, W.~R. van Zwet, F. Takens, and J. DeGoede,
   Detecting differences between delay vector distributions,
   Phys.\ Rev.\ E 53, (1996) 2169.

\bibitem{moeckel}  % distance between attractors
R. Moeckel and B. Murray,
   Measuring the distance between time series,
   Physica D 102 (1997) 187.

\bibitem{Kadtke}
J. Kadtke,
   Classification of highly noisy signals using global dynamical models,
   Phys.\ Lett.\ A 203 (1995) 196.

\bibitem{reconstruct}
M. Casdagli, S. Eubank, J.~D. Farmer, and J. Gibson,
   State space reconstruction in the presence of noise,
   Physica D 51 (1991) 52.

\bibitem{Casdagli-inout}
M. Casdagli,
   A dynamical systems approach to modeling input-output systems,
   in~\cite{casdagli}.

\bibitem{HeggusMuldoon}
M.~R. Muldoon, D. S. Broomhead, J. P. Huke, and R. Hegger,
   Delay embedding in the presence of dynamical noise,
   Dynamics and Stability of Systems, to appear, 1997.

\bibitem{ding}
M. Ding, C. Grebogi, E. Ott, T. Sauer, and J.~A. Yorke,
   Plateau onset for correlation dimension: When does it occur?
   Phys.\ Rev.\ Lett.\ 70 (1993) 3872; Reprinted in~\cite{coping}.

\bibitem{malinetskij}
G.~G. Malinetskii, A.~B. Potapov, A.~I. Rakhmanov, and E.~B. Rodichev,
   Limitations of delay reconstruction for chaotic systems with a broad
   spectrum,
   Phys.\ Lett. A 179 (1993) 15.

\bibitem{fraser}
A.~M. Fraser and H.~L. Swinney,
   Independant coordinates for strange attractors from mutual
   information,
   Phys.\ Rev.\ A 33 (1986) 1134.

\bibitem{ls}
W. Liebert, H.~G. Schuster,
   Proper choice of the time delays for the analysis of chaotic time
   series,
   Phys.\ Lett.\ A 142 (1989) 107.

\bibitem{lps}
W. Liebert, K. Pawelzik, and H.~G. Schuster,
   Optimal embeddings of chaotic attractors from topological
   considerations,
   Europhys.\ Lett.\ 14 (1991) 521.

\bibitem{kennelisabelle}
M.~B. Kennel and S. Isabelle,
   Method to distinguish possible chaos from colored noise and to
   determine embedding parameters,
   Phys.\ Rev.\ A 46 (1992) 3111.

\bibitem{buzpfist}
T. Buzug and G. Pfister,
   Comparison of algorithms calculating optimal parameters for delay time
   coordinates,
   Physica D 58 (1992) 127.

\bibitem{buzpfist2}
T. Buzug, T. Reimers, and G. Pfister
   Optimal reconstruction of strange attractors from purely geometrical
   arguments,
   Europhys.\ Lett.\ 13 (1990) 605.

\bibitem{kugi} % embedding
D. Kugiumtzis,
   State space reconstruction parameters in the analysis of chaotic time
   series -- the role of the time window length,
   Physica D 95 (1996) 13.

\bibitem{Lop}
T. Schreiber and H. Kantz,
   Observing and predicting chaotic signals: Is 2\% noise too much?
   Y. Kravtsov and J. Kadtke, eds.,
   Predictability of complex dynamical systems,
   Springer, New York, 1996.

\bibitem{kugiLL}
D. Kugiumtzis, O.~C. Lingj{\ae}rde, and N. Christophersen,
   Regularized local linear prediction of chaotic time series,
   Physica D 112 (1998) 344.

\bibitem{Sauer}
T. Sauer,
   Times series prediction using delay coordinate embedding,
   in~\cite{SFI}.

\bibitem{cas_stoch}
M. Casdagli,
   Chaos and deterministic versus stochastic nonlinear modeling,
   J.\ Roy.\ Stat.\ Soc.\ 54 (1991) 303.

\bibitem{analog}
E.~N. Lorenz,
   Atmospheric predictability as revealed by naturally occurring analogues,
   J.\ Atmos.\ Sci.\ 26 (1969) 636.

\bibitem{arkady}
A. Pikovsky,
   Discrete-time dynamic noise filtering,
   Sov.\ J.\ Commun.\ Technol.\ Electron.\ 31 (1986) 81.

\bibitem{SM}
G. Sugihara and R. May,
   Nonlinear forecasting as a way of distinguishing chaos from
   measurement error in time series,
   Nature 344 (1990) 734; Reprinted in~\cite{coping}.

\bibitem{KI}
M.~B. Kennel and S. Isabelle, 
   Method to distinguish possible chaos from colored noise and to determine 
   embedding parameters,
   Phys.\ Rev. A 46 (1992) 3111.

\bibitem{MARS}      % MARS
J.~H. Friedman,
   Multivariate adaptive regression splines (with discussion),
   Ann.\ Stat.\ 19 (1991) 1.

\bibitem{powell}
M.~J.~D. Powell,
   Radial basis functions for multivariable interpolation: A review,
   Proc.\ IMA Conf. Algorithms for the approximation of functions and data,
   RMCS, Shrivenham, 1985.

\bibitem{rbf}    % radial basis functions
D. Broomhead and D. Lowe,
   Multivariable function interpolation and adaptive networks,
   Complex Syst.\ 2 (1988) 321.

\bibitem{lenny_rbf}
L.~A. Smith,
   Identification and prediction of low-dimensional dynamics,
   Physica D 58 (1992) 50.

\bibitem{brei}
L. Breiman, J.~H. Friedman, R.~A. Olshen, and C.~J. Stone,
   Classification and regression trees,
   Chapman and Hall, New York, 1993.

\bibitem{akaike}
H. Akaike,
   A new look at the statistical model identification,
   IEEE Trans.\ Automat.\ Contr.\ AC-19 (1974) 716.

\bibitem{rissanen}
J. Rissanen,
   Consistent order estimates of autoregressive processes by shortest
   description of data,
   O. Jacobs et al., eds.,
   Analysis and optimisation of stochastic systems,
   Academic Press, New York, 1980.

\bibitem{fields}
J. Theiler and D. Prichard,
   Using `Surrogate Surrogate Data' to calibrate the actual rate of
   false positives in tests for nonlinearity in time series,
   Fields Inst.\ Comm.\ 11 (1997) 99.

\bibitem{Wolf}
A. Wolf, J.~B. Swift, H.~L.  Swinney, and J.~A. Vastano,
   Determining Lyapunov exponents from a time series,
   Physica D 16 (1985) 285.

\bibitem{geist}
K. Geist, U. Parlitz, and W. Lauterborn,
   Comparison of different methods for computing Lyapunov exponents,
   Prog.\ Thoeor.\ Phys.\ 83 (1990) 875.

\bibitem{dimerr}
J. Theiler,
   Statistical precision of dimension estimators,
   Phys.\ Rev. A 41 (1990) 3038.

\bibitem{smith}
R. L. Smith,
   Estimating dimension in noisy chaotic time-series,
   J.\ R.\ Statist.\ Soc.\ B 54 (1992) 329.

\bibitem{takens_est}
F. Takens,
   On the numerical determination of the dimension of an attractor,
   B.~L.~J. Braaksma, H.~W. Broer, and F. Takens, eds.,
   Dynamical systems and bifurcations,
   Lecture notes in mathematics Vol.~1125, Springer, Heidelberg, 1985.

\bibitem{cutler1}
C.~D. Cutler,
   Some results on the behavior and estimation of the fractal dimensions
   of distributions on attractors,
   J.\ Stat.\ Phys.\ 62 (1991) 651.

\bibitem{cutler2}
C.~D. Cutler,
   A theory of correlation dimension for stationary time series,
   Philosoph.\ Trans.\ Royal Soc.\ London A 348 (1995) 343.

\bibitem{oneoverf}
A.~R. Osborne and A. Provenzale,
   Finite correlation dimension for stochastic systems with power-law
   spectra,
   Physica D 35 (1989) 357.

\bibitem{oneoverft}
J. Theiler,
   Some comments on the correlation dimension of $1/f^\alpha$ noise,
   Phys.\ Lett.\ A 155 (1991) 480.

\bibitem{grass_nature1}
P. Grassberger,
   Do climatic attractors exist?
   Nature 323 (1986) 609.

\bibitem{grass_nature2}
P. Grassberger,
   Evidence or climatic attractors: Grassberger replies,
   Nature 326 (1987) 524.

\bibitem{ruelle_fiction}
D. Ruelle,
   Deterministic chaos: The science and the fiction,
   Proc.\ R.\ Soc.\ London A 427 (1990) 241.

\bibitem{theiler_dim}
J. Theiler,
   Estimating fractal dimension,
   J.\ Opt.\ Soc.\ Amer.\ A 7 (1990) 1055.

\bibitem{dim}
H. Kantz and T. Schreiber,
   Dimension estimates and physiological data,
   CHAOS 5 (1995) 143; Reprinted in~\cite{dyndis}.

\bibitem{grass_finite}
P. Grassberger,
   Finite sample corrections to entropy and dimension estimates,
   Phys.\ Lett.\ A 128 (1988) 369.

\bibitem{provenzale}
A. Provenzale, L.~A. Smith, R. Vio, and G. Murante,
   Distinguishing between low-dimensional dynamics and randomness in
   measured time series,
   Physica D 58 (1992) 31.

\bibitem{FNN}
M.~B. Kennel, R. Brown, and H.~D.~I. Abarbanel,
   Determining embedding dimension for phase-space reconstruction using
   a geometrical construction,
   Phys.\ Rev.\ A 45 (1992) 3403; Reprinted in~\cite{coping}.

\bibitem{kennel-unpub}
M.~B. Kennel and H.~D.~I. Abarbanel,
   False neighbors and false strands: A reliable
   minimum embedding dimension algorithm,
   INLS Preprint, 1994.

\bibitem{takens_theiler}
J. Theiler,
   Lacunarity in a best estimator of fractal dimension,
   Phys.\ Lett.\ A 135 (1988) 195.

\bibitem{diks}
C. Diks,
   Estimating invariants of noisy attractors,
   Phys.\ Rev.\ E 53 (1996) 4263.

\bibitem{kugig} % Gaussian noise
D. Kugiumtzis,
   Correction of the correlation dimension for noisy time series,
   Int.\ J.\ Bifurcation and Chaos, to appear, 1997.

\bibitem{olt}
H. Oltmans and P.~J.~T. Verheijen,
   The influence of noise on power law scaling functions and an
   algorithm for dimension estimations,
   Phys.\ Rev.\ E 56 (1997) 1160.

\bibitem{level}
T. Schreiber,
   Determination of the noise level of chaotic time series,
   Phys.\ Rev.\ E 48 (1993) R13.

\bibitem{comment}
T. Schreiber,
   Influence of Gaussian noise on the correlation exponent,
   Phys.\ Rev.\ E 56 (1997) 274.

\bibitem{olof}
E. Olofsen, J. Degoede, and R. Heijungs,
   A maximum likelihood approach to correlation dimension and entropy
   estimation,
   Bull.\ Math.\ Biol. 54 (1992) 45.

\bibitem{schouten}
J.~C. Schouten, F. Takens, and C.~M. van den Bleek,
   Maximum likelihood estimation of the entropy of an attractor,
   Phys.\ Rev.\ E 49 (1994) 126.

\bibitem{elger}
C. Elger and K. Lehnertz,
   Seizure prediction by nonlinear time series analysis of brain
   electrical activity
   European J.\ Neuroscience, to appear, 1997.

\bibitem{bowen} % symbolic dynamics
R. Bowen,
   Symbolic dynamics for hyperbolic flows,
   Amer.\ J.\ Math.\ 95 (1973) 429.

\bibitem{freddy}
F. Christiansen and A. Politi,
   Symbolic encoding in symplectic maps,
   Nonlinearity 9 (1996) 1623.

\bibitem{GK}
P. Grassberger and H. Kantz,
   Generating partitions for the dissipative H\'enon map,
   Phys.\ Lett.\ A 113 (1985) 235.

\bibitem{herzel} % symbol seq.
H. Herzel,
   Compelxity of symbol sequences,
   Syst.\ Anal.\ Modl.\ Simul.\ 5 (1988) 435.

\bibitem{ebeling} % symbol
W. Ebeling, Th. P{\"o}schel, and K.-F. Albrecht,
   Entropy, transinformation and word distribution of information-carrying
   sequences,
   Int.\ J.\ Bifurcation Chaos 5 (1995) 51.

\bibitem{symbol}
T. Sch\"urmann and P. Grassberger,
   Entropy estimation of symbol sequences,
   CHAOS 6 (1996) 414.

\bibitem{hao}
B.-L.~Hao,
   Elementary symbolic dynamics,
   World Scientific, Singapore, 1989.

\bibitem{wacker}
R. Wackerbauer, A. Witt, H. Atmanspacher, J. Kurths, and H. Scheingraber,
   A comparative classification of complexity measures,
   Chaos Solitons Fractals 4 (1994) 133.

\bibitem{k}
J. Kurths, A. Voss, P. Saparin, A. Witt, H.~J. Kleiner, and N. Wessel,
   Complexity measures for the analysis of heart rate variability,
   CHAOS 5 (1995) 88.

\bibitem{brockpaper1}
W.~A. Brock, W.~D. Dechert, J.~A. Scheinkman, and B. LeBaron,
   A test for independence based on the correlation dimension,
   University of Wisconsin Press, Madison, 1988.

\bibitem{brockpaper2}
W.~A. Brock, D.~A. Hseih, and B. LeBaron,
   Nonlinear dynamics, chaos, and instability: Statistical theory and
   economic evidence,
   MIT Press, Cambridge, MA, 1991.

\bibitem{B1b}
G. Sugihara, B. Grenfell, and R.~M. May,
   Distinguishing error from chaos in ecological time series,
   Phil.\ Trans.\ R.\ Soc.\ Lond.\ B 330 (1990) 235.

\bibitem{volterra}
M. Barahona and C.-S. Poon,
   Detection of nonlinear dynamics in short, noisy time series,
   Nature 381 (1996) 215.

\bibitem{power}
T. Schreiber and A. Schmitz,
   Discrimination power of measures for nonlinearity in a time series,
   Phys.\ Rev.\ E 55 (1997) 5443.

\bibitem{KG}
D.~T. Kaplan and L. Glass,
   Direct test for determinism in a time series,
   Phys.\ Rev.\ Lett.\ 68 (1992) 427; Reprinted in~\cite{coping}.

\bibitem{pompe}
B. Pompe,
   Measuring statistical dependencies in a time series,
   J.\ Stat.\ Phys.\ 73 (1993) 587.

\bibitem{milan2}
M. Palu\v{s},
   Testing for nonlinearity using redundancies: Quantitative and
   qualitative aspects,
   Physica D 80 (1995) 186.

\bibitem{milan1}
M. Palu\v{s},
   On entropy rates of dynamical systems and Gaussian processes,
   Phys.\ Lett.\ A 227 (1997) 301.

\bibitem{A0}
D. Auerbach, P. Cvitanovi\'c, J.-P. Eckmann, G. Gunaratne, and I. Procaccia,
   Exploring chaotic motion through periodic orbits,
   Phys.\ Rev.\ Lett.\ 58 (1987) 2387.

\bibitem{AAC}
R. Artuso, E. Aurell, and P. Cvitanovi\'c,
   Recycling of strange sets I, 
   Nonlinearity 3 (1990) 325.

\bibitem{AAC2}
R. Artuso, E. Aurell, and P. Cvitanovi\'c,
   Recycling of strange sets II, 
   Nonlinearity 3 (1990) 361.

\bibitem{PObook}
P. Cvitanovi\'c, R. Artuso, R. Mainieri, and G. Vattay, 
   Classical and quantum chaos: A cyclist treatise,
   Web-book in progress, available from {\tt http://www.nbi.dk/ChaosBook}
   (1998).

\bibitem{NMR-RMP}
R. Badii, E. Brun, M. Finardi, L. Flepp, R. Holzner, J. Parisi, C. Reyl,
   and J. Simonet,
   Progress in the analysis of experimental chaos through periodic orbits,
   Rev.\ Mod.\ Phys.\ 66 (1994) 1389.

\bibitem{NBT}
G.~B. Mindlin, X.-J. Hou, H.~G. Solari, R. Gilmore, and N.~B. Tufillaro, 
   Classification of strange attractors by integers, 
   Phys.\ Rev.\ Lett.\ 64 (1990) 2350.

\bibitem{OGY}
E. Ott, C. Grebogi, and J.~A. Yorke,
   Controlling chaos,
   Phys.\ Rev.\ Lett.\ 64 (1990) 1196; Reprinted in~\cite{coping}.

\bibitem{PM}
D. Pierson and F. Moss, 
   Detecting periodic unstable points in noisy chaotic and limit cycle
   attractors with applications to biology,
   Phys.\ Rev.\ Lett.\ 75 (1995) 2124.

\bibitem{soso}
P. So, E. Ott, S.~J. Schiff, D.~T. Kaplan, T. Sauer, and C. Grebogi,
   Detecting unstable periodic orbits in chaotic experimental data,
   Phys.\ Rev.\ Lett.\ 76 (1996) 4705.

\bibitem{sonicht}
D.~J. Christini and  J.~J. Collins,
   Controlling nonchaotic neuronal noise using chaos control techniques,
   Phys.\ Rev.\ Lett.\ 75 (1995) 2782.

\bibitem{albano}
A.~M. Albano, P.~E. Rapp, and A. Passamante,
   Kolmogorov-Smirnov test distinguishes attractors with similar
   dimensions,
   Phys.\ Rev. E 52 (1995) 196.

\bibitem{kurths}
H. Isliker and J. Kurths,
   A test for stationarity: Finding parts in a time series apt for
   correlation dimension estimates,
   Int.\ J.\ Bifurcation and Chaos 3 (1993) 1573.

\bibitem{kennel}
M.~B. Kennel,
   Statistical test for dynamical nonstationarity in observed
   time-series data,
   Phys.\ Rev.\ E 56 (1997) 316.

\bibitem{recurr}
J.~P. Eckmann, S. Oliffson Kamphorst, and D. Ruelle,
   Recurrence plots of dynamical systems,
   Europhys.\ Lett.\ 4 (1987) 973.

\bibitem{Koebbe}
M. Koebbe and G. Mayer-Kress,
   Use of the recurrence plots in the analysis of time-series data,
   in~\cite{casdagli}.

\bibitem{mcguire}
G. Mc Guire, N.~B. Azar, and M. Shelhamer,
   Recurrence matrices and the preservation of dynamical properties,
   Phys.\ Lett.\ A 237 (1997) 43.

\bibitem{zbilut}
C.~L. Webber and J.~P. Zbilut,
   Dynamical assessment of physiological systems and states using
   recurrence plot strategies,
   J.\ Appl.\ Physiol.\ 76 (1994) 965.

\bibitem{zbilut2}
J.~P. Zbilut, A. Giuliani, and C.~L. Webber,
   Recurrence quantification analysis and principal components in the
   detection of short complex signals,
   Phys.\ Lett.\ A 237 (1998) 131.

\bibitem{cas_recurr}
M. Casdagli,
   Recurrence plots revisited,
   Physica D 108 (1997) 206.

\bibitem{Bablo}
A. Babloyantz and A. Destexhe,
   Low-dimensional chaos in an instance of epilepsy,
   Proc.\ Natl.\ Acad.\ Sci.\ USA 83 (1986) 3513.

\bibitem{FEEG}
G.~W. Frank, T. Lookman, M.~A.~H. Nerenberg, C. Essex, J. Lemieux,
and W. Blume,
   Chaotic time series analyses of epileptic seizures,
   Physica D 46 (1990) 427.

\bibitem{TEEG}
J. Theiler,
   On the evidence for low--dimensional chaos in an epileptic
   electroencephalogram,
   Phys.\ Lett.\ A 196 (1995) 335.

\bibitem{TEEG2}
J. Theiler and P.~E. Rapp,
   Re-examination of the evidence for low-dimensional, nonlinear structure in
   the human electroencephalogram, 
   Electroencephalogr.\ Clin.\ Neurophysiol.\ 98 (1996) 213.

\bibitem{lerner}
D.~E. Lerner,
   Monitoring changing dynamics with correlation integrals: Case study of
   an epileptic seizure,
   Physica D 97 (1996) 563.

\bibitem{Pijn}
J.~P. Pijn, J. Van Neerven, A. Noest, and F.~H. Lopes da Silva,
   Chaos or noise in EEG signals; dependence on state and brain site,
   Electroencephalogr.\ Clin.\ Neurophysiol.\ 79 (1991) 371.

\bibitem{lehnertz}
K. Lehnertz, C.E. Elger,
   Spatio-Temporal dynamics of the primary epileptogenic area in temporal
   lobe epilepsy characterized by neuronal complexity loss,
   Electroencephalogr.\ Clin.\ Neurophysiol.\ 95 (1995) 108.

\bibitem{Pezard}
L. P\'ezard, J. Martinerie, J. M\"uller-Gerking, F.~J. Varela, and B. Renault,
   Entropy quantification of human brain spatio-temporal dynamics,
   Physica D 96 (1996) 344.

\bibitem{Rogo}
Z. Rogovski, I. Gath, and E. Bental,
   On the prediction of epileptic seizures,
   Biol.\ Cybern.\ 42 (1981) 9.

\bibitem{Dvo}
I. Dvor\'ak,
   Takens versus multichannel reconstruction in EEG
   correlation exponent estimates,
   Phys.\ Lett.\ A 151 (1990) 225.

\bibitem{hko}
R. Hegger, H. Kantz, and E. Olbrich, 
   Correlation dimension of intermittent signals,
   Phys.\ Rev.\ E 56 (1997) 199.

\bibitem{theiler1}
J. Theiler, S. Eubank, A. Longtin, B. Galdrikian, and J.D. Farmer,
   Testing for nonlinearity in time series: The method of surrogate data,
   Physica D 58 (1992) 77; Reprinted in~\cite{coping}.

\bibitem{theiler2}
J. Theiler, B. Galdrikian, A. Longtin, S. Eubank and J.D. Farmer,
   Using surrogate data to detect nonlinearity in time series,
   in~\cite{casdagli}.

\bibitem{BI}
T. Subba Rao and M.~M. Gabr,
   An introduction to bispectral analysis and bilinear time series models,
   Lecture notes in statistics Vol.~24, Springer, New York, 1984.

\bibitem{diks2} % time reversibility
C. Diks, J.~C. van Houwelingen, F. Takens, and J. DeGoede,
   Reversibility as a criterion for discriminating time series,
   Phys.\ Lett.\ A 201 (1995) 221.

\bibitem{skinner}
J.~E. Skinner, M. Molnar, and C. Tomberg,
   The point correlation dimension: Performance with nonstationary
   surrogate data and noise,
   Integrative Physiological and Behavioral Science 29 (1994) 217.

\bibitem{roulston}
M.~S. Roulston,
   Significance testing on information theoretic functionals,
   Physica D 110 (1997) 62.

\bibitem{tp}
J. Theiler and D. Prichard,
   Constrained-realization Monte-Carlo method for hypothesis testing,
   Physica D 94 (1996) 221.

\bibitem{anneal}
T. Schreiber,
   Constrained randomization of time series data,
   Phys.\ Rev.\ Lett. 80 (1998) 2105.

\bibitem{efron}
B. Efron,
   The jackknife, the bootstrap and other resampling plans,
   SIAM, Philadelphia, PA, 1982.

\bibitem{surrowe}
T. Schreiber and A. Schmitz,
   Improved surrogate data for nonlinearity tests,
   Phys.\ Rev.\ Lett.\ 77 (1996) 635.

\bibitem{theiler_sfi}
J. Theiler, P.~S. Linsay, and D.~M. Rubin,
   Detecting nonlinearity in data with long coherence times,
   in~\cite{SFI}.

\bibitem{metro}
N. Metropolis, A. Rosenbluth, M. Rosenbluth, A. Teller, and E. Teller,
   Equations of state calculations by fast computing machine,
   J.\ Chem.\ Phys.\ 21 (1953) 1097.

\bibitem{kirk}
S. Kirkpatrick, C.~D. Gelatt Jr., and M.~P. Vecchi,  
   Optimization by Simulated Annealing, 
   Science 220 (1983) 671.

\bibitem{annealbook}
R.~V.~V. Vidal, ed.,
   Applied simulated annealing,
   Lecture notes in economics and mathematical systems Vol.~396,
   Springer, Berlin, 1993.

\bibitem{gold}  %+
D.~R. Rigney, A.~L. Goldberger, W. Ocasio, Y. Ichimaru,  G.~B. Moody, and
   R. Mark,
   Multi-channel physiological data: Description and analysis,
   in~\cite{SFI}.

\bibitem{anneal_unpub}
T. Schreiber and A. Schmitz,
   unpublished, 1997.

\bibitem{garch}
T. Bollerslev,
   Generalized autoregressive conditional heteroscedasticity,
   J. Econometrics 31 (1986) 207.

\bibitem{dean}
D. Prichard,
   The correlation dimension of differenced data,
   Phys.\ Lett.\ A 191 (1994) 245.

\bibitem{fsid1}
J.~D. Farmer and J. Sidorowich,
   Exploiting chaos to predict the future and reduce noise,
   Y.C. Lee, ed., Evolution, learning and cognition,
   World Scientific, 1988.

\bibitem{ky2}
E.~J. Kostelich and J.~A. Yorke,
   Noise reduction in dynamical systems,
   Phys.\ Rev.\ A 38 (1988) 1649; Reprinted in~\cite{coping}.

\bibitem{sg}
T. Schreiber and P. Grassberger,
   A simple noise-reduction method for real data,
   Phys.\ Lett.\ A 160 (1991) 411.

\bibitem{on}
P. Grassberger, R. Hegger, H. Kantz, C. Schaffrath, and T. Schreiber,
   On noise reduction methods for chaotic data,
   CHAOS 3 (1993) 127; Reprinted in~\cite{coping}.

\bibitem{noiserev}
E.~J. Kostelich and T. Schreiber,
   Noise reduction in chaotic time series data: A survey of common methods,
   Phys.\ Rev.\ E 48 (1993) 1752.

\bibitem{processing}
T. Schreiber,
   Processing of physiological data,
   in~\cite{freital}.

\bibitem{PC}
I.~T. Jolliffe,
   Principal component analysis,
   Springer, New York, 1986.

\bibitem{BroomSVD}
D. Broomhead and G.~P. King,
   Extracting qualitative dynamics from experimental data,
   Physica D 20 (1986) 217.

\bibitem{ecg}
T. Schreiber and D.~T. Kaplan,
   Nonlinear noise reduction for electrocardiograms,
   CHAOS 6 (1996) 87.

\bibitem{case}
H. Kantz, T. Schreiber, I. Hoffmann, T. Buzug, G. Pfister, L.~G.~Flepp,
   J.~Simonet, R.~Badii, and E. Brun,
   Nonlinear noise reduction: A case study on experimental data,
   Phys.\ Rev.\ E 48 (1993) 1529.

\bibitem{neigh}
T. Schreiber,
   Efficient neighbor searching in nonlinear time series analysis,
   Int.\ J.\ Bifurcation and Chaos 5 (1995) 349.

\bibitem{stream}
T. Schreiber and M. Richter,
   Nonlinear projective filtering in a data stream,
   Wuppertal preprint WUB-98-8, 1998.

\bibitem{goldbook}
A.~L. Goldberger and E. Goldberger,
   Clinical electrocardiography,
   Mosby, St. Louis, 1977.

\bibitem{lazy}
T. Schreiber,
   Extremely simple nonlinear noise reduction method, 
   Phys.\ Rev.\ E 47 (1993) 2401.

\bibitem{unpub}
T. Schreiber,
   unpublished, 1997.

\bibitem{fetal1}
T. Schreiber and D.~T. Kaplan,
   Signal separation by nonlinear projections: The fetal electrocardiogram,
   Phys.\ Rev.\ E 53 (1996) 4326.

\bibitem{fetal2}
M. Richter, T. Schreiber, and D.~T. Kaplan,
   Fetal ECG extraction with nonlinear phase space projections,
   IEEE Trans.\ Bio-Med.\ Eng.\ 45 (1998) 133.

\bibitem{recording}
   J.~F Hofmeister, J.~C. Slocumb, L.~M. Kottmann, J.~B. Picchiottino, and
   D.~G. Ellis,
   A noninvasive method for recording the electrical activity of the human
   uterus in vivo,
   Biomed.\ Instr.\ Technol.\ (1994) 391.

\bibitem{Busse}
F.~H. Busse and L. Kramer, eds., 
   Nonlinear evolution of spatio-temporal structures in dissipative continuous
   systems,
   Plenum, New York, 1990.

\bibitem{Bryn}
A.~M. Albano, P.~E. Rapp, N.~B. Abraham, and A. Passamante, eds.,
   Measures of spatio-temporal dynamics, 
   Physica D 96 (1996).

\bibitem{manneville}
P. Manneville,
   Dissipative structures and weak turbulence,
   Academic Press, New York, 1989.

\bibitem{Kapral}
R. Kapral and K. Showalter, 
   Chemical waves and patterns, 
   Kluwer, Dordrecht, 1995.

\bibitem{cross}
M.~C. Cross and P.~C. Hohenberg,
   Pattern formation outside of equilibrium,
   Rev.\ Mod.\ Phys.\ 65 (1993) 851.

\bibitem{olbrich}
E. Olbrich and H. Kantz,
   Inferring chaotic dynamics from time-series: On which length scale
   determinism becomes visible,
   Phys.\ Lett.\ A 232 (1997) 63.

\bibitem{Zoldi1}
S.~M. Zoldi and H.~S. Greenside,
   Karhunen-Lo\'eve decomposition of extensive chaos,
   Phys.\ Rev.\ Lett.\ 78 (1997) 9.

\bibitem{Haken}
V.~K. Jirsa, R. Friedrich, and H. Haken,
   Reconstrution of the spatio-temporal dynamics of a human
   magnetoencephalogram,
   Physica D 89 (1995) 100.

\bibitem{Hopf}
F. Christiansen, P. Cvitanovi\'c, and V. Putkaradze,
   Hopf's last hope: spatiotemporal chaos in terms of unstable recurrent
   patterns,
   Nonlinearity 10 (1997) 1.

\bibitem{Zoldi2}
S.~M. Zoldi and H.~S. Greenside, 
   Spatially localized unstable periodic orbits of a high-dimensional chaotic
   system, 
   Phys.\ Rev.\ E 57 (1998) R2511.

\bibitem{Buenner}
M.~J. B\"unner, M. Popp, Th. Meyer, A. Kittel, and J. Parisi,
   Tool to recover scalar time-delay systems from experimental time series,
   Phys.\ Rev.\ E 54 (1996) 3082.
\end{thebibliography}
\end{document}